\documentclass{CSML}

\def\dOi{11(4:22)2015}
\lmcsheading%
{\dOi}
{1--43}
{}
{}
{Apr.~15, 2015}
{Dec.~31, 2015}
{}

\ACMCCS{[{\bf Theory of computation}]: Logic---Constructive mathematics}

\usepackage[T2A,T1]{fontenc}
\usepackage{amsmath}
\renewcommand{\ldots}{...}
\usepackage{amsthm}

\newtheorem*{definition*}{Definition}

\newtheorem*{lemma*}{Lemma}

\newtheorem*{theorem*}{Theorem}
\usepackage{amssymb}
\usepackage{cmll}
\usepackage{bm}
\usepackage{stmaryrd}
\usepackage[all]{xy}
\usepackage{bussproofs}
\usepackage{expl3}
\usepackage{hyperref}
\ExplSyntaxOn
\cs_generate_variant:Nn\tl_if_empty:nTF{f}
\newcommand*\ifpresent[3]{\tl_if_empty:fTF{#1}{#3}{#2}}
\ExplSyntaxOff
\newcommand*\AXM[1]{\AxiomC{$#1$}}
\newcommand*\UIM[1]{\UnaryInfC{$#1$}}
\newcommand*\BIM[1]{\BinaryInfC{$#1$}}
\newcommand*\TIM[1]{\TrinaryInfC{$#1$}}
\newcommand*\RLM[1]{\RightLabel{$\scriptstyle#1$}}
\newcommand*\DP\DisplayProof
\newcommand*\Def{\mathrel{\overset{\Delta}{=}}}
\newcommand*\GramDef{\quad\mathrel{::=}\quad}
\newcommand*\Entails{\mathrel{\vdash}}
\newcommand*\Derives{\mathrel{|\mkern-8.75mu\sim}}
\newcommand*\BarSep{\mathrel{|}}
\newcommand*\Sequent[3]{#1\Entails#2\ifpresent{#3}{\BarSep}{}#3}
\newcommand*\FV[1]{\text{FV}\left(#1\right)}
\newcommand*\SetSuch[2]{{\left\{#1\ifpresent{#2}{\!\;\middle|\;#2}{}\right\}}}
\newcommand*\SortBase\iota
\newcommand*\SortTo\to

\newcommand*\SortA{\sigma}
\newcommand*\SortB{\tau}
\newcommand*\SortC{\nu}

\newcommand*\LogSortedTerm[2]{#1^{#2}}
\newcommand*\LogTermA{t}
\newcommand*\LogTermB{u}
\newcommand*\LogTermC{v}

\newcommand*\LogVarA{x}
\newcommand*\LogVarB{y}
\newcommand*\LogVarC{z}
\newcommand*\LogVarD{u}
\newcommand*\LogVarE{v}
\newcommand*\LogVarF{w}
\newcommand*\LogConst[1]{\mathsf{#1}}
\newcommand*\LogConstA{\LogConst{c}}

\newcommand*\LogNeg[1]{{#1^-}}
\newcommand*\LogPos[1]{{#1^+}}
\newcommand*\LogImp{\mathbin{\Rightarrow}}
\newcommand*\LogAnd{\mathbin{\wedge}}
\newcommand*\LogBot\bot
\newcommand*\LogRel[1]{\llparenthesis#1\rrparenthesis}
\newcommand*\LogRelForm[1]{{#1^\mathrm{r}}}
\newcommand*\LogForallRel{\forall^\mathrm{r}}
\newcommand*\LogExistsRel{\exists^\mathrm{r}}
\newcommand*\LogFormA{A}
\newcommand*\LogFormB{B}
\newcommand*\LogFormC{C}

\newcommand*\LogAxioms{\mathcal{A}x}
\newcommand*\LogPredA{P}

\newcommand*\LogProofA{\mathfrak{p}}
\newcommand*\LogProofB{\mathfrak{q}}
\newcommand*\LogProofC{\mathfrak{r}}

\newcommand*\LogSubst[1]{\left\{#1\right\}}
\newcommand*\LogRuleAxConcl[3]{\Sequent{#1\ifpresent{#1}{,}{}#3}{#3}{#2}}
\newcommand*\LogRuleAx[3]{\AXM{}\UIM{\LogRuleAxConcl{#1}{#2}{#3}}\DP}
\newcommand*\LogRuleAxSigConcl[3]{\Sequent{#1}{#3}{#2}}
\newcommand*\LogRuleAxSig[3]{\AXM{}\RLM{\left(#3\in\LogAxioms\right)}\UIM{\LogRuleAxSigConcl{#1}{#2}{#3}}\DP}
\newcommand*\LogRuleImpIntroFirst[4]{\Sequent{#1\ifpresent{#1}{,}{}#3}{#4}{#2}}
\newcommand*\LogRuleImpIntroConcl[4]{\Sequent{#1}{#3\LogImp#4}{#2}}
\newcommand*\LogRuleImpIntro[4]{\AXM{\LogRuleImpIntroFirst{#1}{#2}{#3}{#4}}\UIM{\LogRuleImpIntroConcl{#1}{#2}{#3}{#4}}\DP}

\newcommand*\LogRuleImpElimFirst[4]{\Sequent{#1}{#3\LogImp#4}{#2}}
\newcommand*\LogRuleImpElimSecond[4]{\Sequent{#1}{#3}{#2}}
\newcommand*\LogRuleImpElimConcl[4]{\Sequent{#1}{#4}{#2}}
\newcommand*\LogRuleImpElim[4]{\AXM{\LogRuleImpElimFirst{#1}{#2}{#3}{#4}}\AXM{\LogRuleImpElimSecond{#1}{#2}{#3}{#4}}\BIM{\LogRuleImpElimConcl{#1}{#2}{#3}{#4}}\DP}

\newcommand*\LogRuleAndIntroFirst[4]{\Sequent{#1}{#3}{#2}}
\newcommand*\LogRuleAndIntroSecond[4]{\Sequent{#1}{#4}{#2}}
\newcommand*\LogRuleAndIntroConcl[4]{\Sequent{#1}{#3\LogAnd#4}{#2}}
\newcommand*\LogRuleAndIntro[4]{\AXM{\LogRuleAndIntroFirst{#1}{#2}{#3}{#4}}\AXM{\LogRuleAndIntroSecond{#1}{#2}{#3}{#4}}\BIM{\LogRuleAndIntroConcl{#1}{#2}{#3}{#4}}\DP}

\newcommand*\LogRuleAndElimFirst[3]{\Sequent{#1}{#3_1\LogAnd#3_2}{#2}}
\newcommand*\LogRuleAndElimConcl[3]{\Sequent{#1}{#3_i}{#2}}
\newcommand*\LogRuleAndElim[3]{\AXM{\LogRuleAndElimFirst{#1}{#2}{#3}}\UIM{\LogRuleAndElimConcl{#1}{#2}{#3}}\DP}

\newcommand*\LogRuleForallIntroFirst[4]{\Sequent{#1}{#4}{#2}}
\newcommand*\LogRuleForallIntroConcl[4]{\Sequent{#1}{\forall#3#4}{#2}}
\newcommand*\LogRuleForallIntro[4]{\AXM{\LogRuleForallIntroFirst{#1}{#2}{#3}{#4}}\RLM{\left(#3\notin\FV{#1,#2}\right)}\UIM{\LogRuleForallIntroConcl{#1}{#2}{#3}{#4}}\DP}

\newcommand*\LogRuleForallElimFirst[5]{\Sequent{#1}{\forall#3#5}{#2}}
\newcommand*\LogRuleForallElimConcl[5]{\Sequent{#1}{#5\LogSubst{#4/#3}}{#2}}
\newcommand*\LogRuleForallElim[5]{\AXM{\LogRuleForallElimFirst{#1}{#2}{#3}{#4}{#5}}\UIM{\LogRuleForallElimConcl{#1}{#2}{#3}{#4}{#5}}\DP}

\newcommand*\LogRuleBotIntroFirst[3]{\Sequent{#1}{#3}{#3\ifpresent{#2}{,}{}#2}}
\newcommand*\LogRuleBotIntroConcl[3]{\Sequent{#1}{\LogBot}{#3\ifpresent{#2}{,}{}#2}}
\newcommand*\LogRuleBotIntro[3]{\AXM{\LogRuleBotIntroFirst{#1}{#2}{#3}}\UIM{\LogRuleBotIntroConcl{#1}{#2}{#3}}\DP}

\newcommand*\LogRuleBotElimFirst[3]{\Sequent{#1}{\LogBot}{#3\ifpresent{#2}{,}{}#2}}
\newcommand*\LogRuleBotElimConcl[3]{\Sequent{#1}{#3}{#2}}
\newcommand*\LogRuleBotElim[3]{\AXM{\LogRuleBotElimFirst{#1}{#2}{#3}}\UIM{\LogRuleBotElimConcl{#1}{#2}{#3}}\DP}

\newcommand*\ModM{\mathcal{M}}
\newcommand*\ModElemA{a}
\newcommand*\ModElemB{b}
\newcommand*\ModElemC{c}

\newcommand*\ModMInterp[1]{{#1}^\ModM}
\newcommand*\Models\vDash
\newcommand*\LmSortBot0
\newcommand*\LmSortBase{X}
\newcommand*\LmSortExtract{Z}
\newcommand*\LmSortTimes\times
\newcommand*\LmSortTo\to
\newcommand*\LmSortA{T}
\newcommand*\LmSortB{U}
\newcommand*\LmSortC{V}

\newcommand*\LmConsts{\mathcal{C}st}
\newcommand*\LmTerm[2]{#1\mathrel{:}#2}
\newcommand*\LmTermA{M}
\newcommand*\LmTermB{N}
\newcommand*\LmTermC{P}

\newcommand*\LmVarA{x}
\newcommand*\LmVarB{y}
\newcommand*\LmVarC{z}
\newcommand*\LmVarD{u}
\newcommand*\LmVarE{v}
\newcommand*\LmVarF{w}
\newcommand*\LmMVarA\alpha
\newcommand*\LmMVarB\beta
\newcommand*\LmMVarC\beta
\newcommand*\LmMVarD\delta
\newcommand*\LmConst[1]{\mathsf{#1}}
\newcommand*\LmConstA{\LmConst{c}}

\newcommand*\LmPair[2]{\left\langle#1,#2\right\rangle}
\newcommand*\LmProj{\pi}
\newcommand*\LmSubst[1]{\left\{#1\right\}}
\newcommand*\Lam{\lambda^{R\times+}}
\newcommand*\LamTypeTo\to
\newcommand*\LamTypeTimes\times
\newcommand*\LamTypePlus{+}
\newcommand*\LamTypeUnit{1}
\newcommand*\LamTypeR{R}
\newcommand*\LamTypeA{T}
\newcommand*\LamTypeB{U}

\newcommand*\LamTermA{M}
\newcommand*\LamTermB{N}

\newcommand*\LamVarA{k}
\newcommand*\LamVarB{l}

\newcommand*\LamSubst[1]{\left\{#1\right\}}
\newcommand*\LamCPS[1]{\raisebox{0pt}[.9\height]{$\widetilde{\raisebox{0pt}[.9\height]{$#1$}}$}}
\newcommand*\LamIn{\mathsf{in}}
\newcommand*\LamCase[4]{\mathsf{case}\,#1\left\{\LamIn_1\,#2\mapsto#3\BarSep\LamIn_2\,#2\mapsto#4\right\}}
\newcommand*\LamCaseBlock[4]{\mathsf{case}\,#1\left\{\begin{gathered}\LamIn_1\,#2\mapsto#3\\\LamIn_2\,#2\mapsto#4\end{gathered}\right\}}
\newcommand*\LamUnit{*}
\newcommand*\LmRuleAxConcl[4]{\Sequent{#1\ifpresent{#1}{,}{}\LmTerm{#3}{#4}}{\LmTerm{#3}{#4}}{#2}}
\newcommand*\LmRuleAx[4]{\AXM{}\UIM{\LmRuleAxConcl{#1}{#2}{#3}{#4}}\DP}
\newcommand*\LmRuleAxSigConcl[4]{\Sequent{#1}{\LmTerm{\LmConst{#3}}{#4}}{#2}}
\newcommand*\LmRuleAxSig[4]{\AXM{}\RLM{(\LmTerm{\LmConst{#3}}{#4}\,\in\,\LmConsts)}\UIM{\LmRuleAxSigConcl{#1}{#2}{#3}{#4}}\DP}
\newcommand*\LmRuleImpIntroFirst[6]{\Sequent{#1\ifpresent{#1}{,}{}\LmTerm{#3}{#4}}{\LmTerm{#5}{#6}}{#2}}
\newcommand*\LmRuleImpIntroConcl[6]{\Sequent{#1}{\LmTerm{\lambda#3.#5}{#4\LmSortTo#6}}{#2}}
\newcommand*\LmRuleImpIntro[6]{\AXM{\LmRuleImpIntroFirst{#1}{#2}{#3}{#4}{#5}{#6}}\UIM{\LmRuleImpIntroConcl{#1}{#2}{#3}{#4}{#5}{#6}}\DP}

\newcommand*\LmRuleImpElimFirst[6]{\Sequent{#1}{\LmTerm{#3}{#6\LmSortTo#4}}{#2}}
\newcommand*\LmRuleImpElimSecond[6]{\Sequent{#1}{\LmTerm{#5}{#6}}{#2}}
\newcommand*\LmRuleImpElimConcl[6]{\Sequent{#1}{\LmTerm{#3#5}{#4}}{#2}}
\newcommand*\LmRuleImpElim[6]{\AXM{\LmRuleImpElimFirst{#1}{#2}{#3}{#4}{#5}{#6}}\AXM{\LmRuleImpElimSecond{#1}{#2}{#3}{#4}{#5}{#6}}\BIM{\LmRuleImpElimConcl{#1}{#2}{#3}{#4}{#5}{#6}}\DP}

\newcommand*\LmRuleAndIntroFirst[6]{\Sequent{#1}{\LmTerm{#3}{#4}}{#2}}
\newcommand*\LmRuleAndIntroSecond[6]{\Sequent{#1}{\LmTerm{#5}{#6}}{#2}}
\newcommand*\LmRuleAndIntroConcl[6]{\Sequent{#1}{\LmTerm{\LmPair{#3}{#5}}{#4\times#6}}{#2}}
\newcommand*\LmRuleAndIntro[6]{\AXM{\LmRuleAndIntroFirst{#1}{#2}{#3}{#4}{#5}{#6}}\AXM{\LmRuleAndIntroSecond{#1}{#2}{#3}{#4}{#5}{#6}}\BIM{\LmRuleAndIntroConcl{#1}{#2}{#3}{#4}{#5}{#6}}\DP}

\newcommand*\LmRuleAndElimFirst[4]{\Sequent{#1}{\LmTerm{#3}{#4_1\LmSortTimes#4_2}}{#2}}
\newcommand*\LmRuleAndElimConcl[4]{\Sequent{#1}{\LmTerm{\LmProj_i\,#3}{#4_i}}{#2}}
\newcommand*\LmRuleAndElim[4]{\AXM{\LmRuleAndElimFirst{#1}{#2}{#3}{#4}}\UIM{\LmRuleAndElimConcl{#1}{#2}{#3}{#4}}\DP}

\newcommand*\LmRuleBotIntroFirst[5]{\Sequent{#1}{\LmTerm{#4}{#5}}{\LmTerm{#3}{#5}}\ifpresent{#2}{,}{}#2}
\newcommand*\LmRuleBotIntroConcl[5]{\Sequent{#1}{\LmTerm{[#3]#4}{\LmSortBot}}{\LmTerm{#3}{#5}}\ifpresent{#2}{,}{}#2}
\newcommand*\LmRuleBotIntro[5]{\AXM{\LmRuleBotIntroFirst{#1}{#2}{#3}{#4}{#5}}\UIM{\LmRuleBotIntroConcl{#1}{#2}{#3}{#4}{#5}}\DP}

\newcommand*\LmRuleBotElimFirst[5]{\Sequent{#1}{\LmTerm{#4}{\LmSortBot}}{\LmTerm{#3}{#5}\ifpresent{#2}{,}{}#2}}
\newcommand*\LmRuleBotElimConcl[5]{\Sequent{#1}{\LmTerm{\mu#3.#4}{#5}}{#2}}
\newcommand*\LmRuleBotElim[5]{\AXM{\LmRuleBotElimFirst{#1}{#2}{#3}{#4}{#5}}\UIM{\LmRuleBotElimConcl{#1}{#2}{#3}{#4}{#5}}\DP}

\newcommand*\LamRuleOrIntroFirst[3]{\Sequent{#1}{\LmTerm{#2}{#3_i}}{}}
\newcommand*\LamRuleOrIntroConcl[3]{\Sequent{#1}{\LmTerm{\LamIn_i\,#2}{#3_1\LamTypePlus#3_2}}{}}
\newcommand*\LamRuleOrIntro[3]{\AXM{\LamRuleOrIntroFirst{#1}{#2}{#3}}\UIM{\LamRuleOrIntroConcl{#1}{#2}{#3}}\DP}

\newcommand*\LamRuleOrElimFirst[6]{\Sequent{#1}{\LmTerm{#3}{#5_1\LamTypePlus#5_2}}{}}
\newcommand*\LamRuleOrElimSecond[6]{\Sequent{#1\ifpresent{#1}{,}{}\LmTerm{#2}{#5_1}}{\LmTerm{#4_1}{#6}}{}}
\newcommand*\LamRuleOrElimThird[6]{\Sequent{#1\ifpresent{#1}{,}{}\LmTerm{#2}{#5_2}}{\LmTerm{#4_2}{#6}}{}}
\newcommand*\LamRuleOrElimConcl[6]{\Sequent{#1}{\LmTerm{\LamCase{#3}{#2}{#4_1}{#4_2}}{#6}}{}}
\newcommand*\LamRuleOrElim[6]{\AXM{\LamRuleOrElimFirst{#1}{#2}{#3}{#4}{#5}{#6}}\AXM{\LamRuleOrElimSecond{#1}{#2}{#3}{#4}{#5}{#6}}\AXM{\LamRuleOrElimThird{#1}{#2}{#3}{#4}{#5}{#6}}\TIM{\LamRuleOrElimConcl{#1}{#2}{#3}{#4}{#5}{#6}}\DP}

\newcommand*\LamRuleTopConcl[1]{\Sequent{#1}{\LmTerm{\LamUnit}{\LamTypeUnit}}{}}
\newcommand*\LamRuleTop[1]{\AXM{}\UIM{\LamRuleTopConcl{#1}}\DP}
\newcommand*\LmInterpForm[1]{{#1}^*}
\newcommand*\LmInterpProof[1]{{#1}^*}
\newcommand*\LmInterpAxiom[1]{M_{#1}}
\newcommand*\CatC{\mathcal{C}}
\newcommand*\CatObj[1]{Ob\left(#1\right)}
\newcommand*\CatObjA{A}
\newcommand*\CatObjB{B}
\newcommand*\CatObjC{C}
\newcommand*\CatObjD{D}
\newcommand*\CatR{R}
\newcommand*\CatRC{\CatExp{\CatR}{\CatC}}
\newcommand*\CatTimes\times
\newcommand*\CatPlus{+}
\newcommand*\CatExp[2]{#1^{#2}}
\newcommand*\CatPar\parr
\newcommand*\CatRCHomA\phi
\newcommand*\CatRCHomB\psi
\newcommand*\CatRCHomC\zeta
\newcommand*\CatRCHomD\xi
\newcommand*\CatRCHomE\varphi
\newcommand*\CatCHomA\mcyrzh
\newcommand*\CatCHomB\mcyri
\newcommand*\CatCHomC\mcyrl
\newcommand*\CatCHomD\mcyrch
\newcommand*\CatCHomE\mcyrd
\newcommand*\CatId[1]{\mathbf{Id}_{#1}}
\newcommand*\CatPair[2]{{\mathbf{pair}\bm{\left(}#1,#2\bm{\right)}}}
\newcommand*\CatEval{\mathbf{ev}}
\newcommand*\CatLambda{\bm{\Lambda}}
\newcommand*\CatTerm{\mathbf{1}}
\newcommand*\CatTermMorph[1]{\mathbf{1}_{#1}}
\newcommand*\CatInit{\mathbf{0}}
\newcommand*\CatInterpSortNeg[1]{{\left\llbracket#1\right\rrbracket}}
\newcommand*\CatInterpSort[1]{{\left[#1\right]}}
\newcommand*\CatInterpTerm[1]{\left[#1\right]}
\newcommand*\RealValNeg[1]{\left\|#1\right\|}
\newcommand*\RealVal[1]{\left|#1\right|}
\newcommand*\RealBot{{\bot\mkern-11mu\bot}}
\newcommand*\HA{{H\!A}}

\newcommand*\PA{{P\!A}}
\newcommand*\PAom{{\PA^\omega}}
\newcommand*\CA{{C\!A}}
\newcommand*\CAom{{\CA^\omega}}
\newcommand*\CASort\iota
\newcommand*\CASortList[1]{{#1}^\diamond}
\newcommand*\CALogs{\LogConst{s}}
\newcommand*\CALogsSort[3]{\left(#1\SortTo#2\SortTo#3\right)\SortTo\left(#1\SortTo#2\right)\SortTo#1\SortTo#3}
\newcommand*\CALogk{\LogConst{k}}
\newcommand*\CALogkSort[2]{#1\SortTo#2\SortTo#1}
\newcommand*\CALogZ{\LogConst{0}}
\newcommand*\CALogZSort{\CASort}
\newcommand*\CALogS{\LogConst{S}}
\newcommand*\CALogSSort{\CASort\SortTo\CASort}
\newcommand*\CALogrec{\LogConst{rec}}
\newcommand*\CALogrecSort[1]{#1\SortTo\left(\CASort\SortTo#1\SortTo#1\right)\SortTo\CASort\SortTo#1}
\newcommand*\CAAxName[1]{{\bm{\scriptstyle(#1)}}}
\newcommand*\CAReflName{\CAAxName{refl}}
\newcommand*\CARefl[1]{\forall\LogSortedTerm{\LogVarA}{#1}\left(\LogVarA=_#1\LogVarA\right)}

\newcommand*\CAReflTerm{\lambda\LmVarA.\LmVarA}
\newcommand*\CAReflSort{\LmSortBot\LmSortTo\LmSortBot}
\newcommand*\CALeibName{\CAAxName{Leib}}
\newcommand*\CALeib[3]{\forall\LogSortedTerm{\vec{\LogVarC}}{\vec{#3}}\,\forall\LogSortedTerm{\LogVarA}{#1}\,\forall\LogSortedTerm{\LogVarB}{#1}\left(\neg#2\LogImp#2\LogSubst{\LogVarB/\LogVarA}\LogImp\LogVarA\neq_#1\LogVarB\right)}

\newcommand*\CALeibTerm{\lambda\LmVarA.\LmVarA}
\newcommand*\CALeibSort[1]{\left(\LmInterpForm{#1}\LmSortTo\LmSortBot\right)\LmSortTo\LmInterpForm{#1}\LmSortTo\LmSortBot}
\newcommand*\CAdefsName{\CAAxName{\Delta s}}
\newcommand*\CAdefs[3]{\forall\LogSortedTerm{\LogVarA}{#1\SortTo#2\SortTo#3}\forall\LogSortedTerm{\LogVarB}{#1\SortTo#2}\forall\LogSortedTerm{\LogVarC}{#1}\left(\CALogs\,\LogVarA\,\LogVarB\,\LogVarC=_#3\LogVarA\,\LogVarC\left(\LogVarB\,\LogVarC\right)\right)}
\newcommand*\CAdefsNoType{\forall\LogVarA\,\forall\LogVarB\,\forall\LogVarC\;\;\CALogs\,\LogVarA\,\LogVarB\,\LogVarC=\LogVarA\,\LogVarC\left(\LogVarB\,\LogVarC\right)}
\newcommand*\CAdefsTerm{\lambda\LmVarA.\LmVarA}
\newcommand*\CAdefsSort{\LmSortBot\LmSortTo\LmSortBot}
\newcommand*\CAdefkName{\CAAxName{\Delta k}}
\newcommand*\CAdefk[2]{\forall\LogSortedTerm{\LogVarA}{#1}\forall\LogSortedTerm{\LogVarB}{#2}\left(\CALogk\,\LogVarA\,\LogVarB=_#1\LogVarA\right)}
\newcommand*\CAdefkNoType{\forall\LogVarA\,\forall\LogVarB\;\;\CALogk\,\LogVarA\,\LogVarB=\LogVarA}
\newcommand*\CAdefkTerm{\lambda\LmVarA.\LmVarA}
\newcommand*\CAdefkSort{\LmSortBot\LmSortTo\LmSortBot}
\newcommand*\CAdefrecZName{\CAAxName{\Delta rec0}}
\newcommand*\CAdefrecZ[1]{\forall\LogSortedTerm{\LogVarA}{#1}\forall\LogSortedTerm{\LogVarB}{\CASort\SortTo#1\SortTo#1}\left(\CALogrec\,\LogVarA\,\LogVarB\,\CALogZ=_#1\LogVarA\right)}
\newcommand*\CAdefrecZNoType{\forall\LogVarA\,\forall\LogVarB\;\;\CALogrec\,\LogVarA\,\LogVarB\,\CALogZ=\LogVarA}
\newcommand*\CAdefrecZTerm{\lambda\LmVarA.\LmVarA}
\newcommand*\CAdefrecZSort{\LmSortBot\LmSortTo\LmSortBot}
\newcommand*\CAdefrecSName{\CAAxName{\Delta recS}}
\newcommand*\CAdefrecS[1]{\forall\LogSortedTerm{\LogVarA}{#1}\forall\LogSortedTerm{\LogVarB}{\CASort\SortTo#1\SortTo#1}\forall\LogSortedTerm{\LogVarC}{\CASort}\left(\CALogrec\,\LogVarA\,\LogVarB\left(\CALogS\,\LogVarC\right)=_#1\LogVarB\,\LogVarC\left(\CALogrec\,\LogVarA\,\LogVarB\,\LogVarC\right)\right)}
\newcommand*\CAdefrecSNoType{\forall\LogVarA\,\forall\LogVarB\,\forall\LogVarC\;\;\CALogrec\,\LogVarA\,\LogVarB\left(\CALogS\,\LogVarC\right)=\LogVarB\,\LogVarC\left(\CALogrec\,\LogVarA\,\LogVarB\,\LogVarC\right)}
\newcommand*\CAdefrecSTerm{\lambda\LmVarA.\LmVarA}
\newcommand*\CAdefrecSSort{\LmSortBot\LmSortTo\LmSortBot}
\newcommand*\CASnZName{\CAAxName{Snz}}
\newcommand*\CASnZ{\forall\LogSortedTerm{\LogVarA}{\CASort}\left(\CALogS\,\LogVarA\neq_\CASort\CALogZ\right)}
\newcommand*\CASnZNoType{\forall\LogVarA\;\;\CALogS\,\LogVarA\neq\CALogZ}
\newcommand*\CASnZTerm{\CALmom}
\newcommand*\CASnZSort{\LmSortBot}
\newcommand*\CAindUnrelName{\CAAxName{ind}}

\newcommand*\CAindUnrelNoType[1]{\forall\vec{\LogVarB}\left(#1\LogSubst{\CALogZ/\LogVarA}\LogImp\forall\LogVarA\left(#1\LogImp#1\LogSubst{\CALogS\,\LogVarA/\LogVarA}\right)\LogImp\forall\LogVarA\,#1\right)}
\newcommand*\CAindName{\CAAxName{ind^r}}
\newcommand*\CAind[2]{\forall\LogSortedTerm{\vec{\LogVarB}}{\vec{#2}}\left(#1\LogSubst{\CALogZ/\LogVarA}\LogImp\LogForallRel\LogSortedTerm{\LogVarA}{\CASort}\left(#1\LogImp#1\LogSubst{\CALogS\,\LogVarA/\LogVarA}\right)\LogImp\LogForallRel\LogSortedTerm{\LogVarA}{\CASort}#1\right)}
\newcommand*\CAindNoType[1]{\forall\vec{\LogVarB}\left(#1\LogSubst{\CALogZ/\LogVarA}\LogImp\LogForallRel\LogVarA\left(#1\LogImp#1\LogSubst{\CALogS\,\LogVarA/\LogVarA}\right)\LogImp\LogForallRel\LogVarA\,#1\right)}
\newcommand*\CAindTerm\CALmrec
\newcommand*\CAindSort[1]{\LmInterpForm{#1}\LmSortTo\left(\CALmnSort\LmSortTo\LmInterpForm{#1}\LmSortTo\LmInterpForm{#1}\right)\LmSortTo\CALmnSort\LmSortTo\LmInterpForm{#1}}
\newcommand*\CADCUnrelName{\CAAxName{DC}}
\newcommand*\CADCUnrel[3]{\forall\LogSortedTerm{\vec{\LogVarD}}{\vec{#3}}\left(\forall\LogSortedTerm{\LogVarA}{\CASort}\,\forall\LogSortedTerm{\LogVarB}{#1}\,\exists\LogSortedTerm{\LogVarC}{#1}\,#2\LogSubst{\LogVarA,\LogVarB,\LogVarC}\;\;\LogImp\;\;\exists\LogSortedTerm{\LogVarE}{\CASort\SortTo#1}\,\forall\LogSortedTerm{\LogVarA}{\CASort}\,#2\LogSubst{\LogVarA,\LogVarE\,\LogVarA,\LogVarE\left(\CALogS\,\LogVarA\right)}\right)}

\newcommand*\CADCName{\CAAxName{DC^r}}
\newcommand*\CADC[3]{\forall\LogSortedTerm{\vec{\LogVarD}}{\vec{#3}}\left(\LogForallRel\LogSortedTerm{\LogVarA}{\CASort}\,\LogForallRel\LogSortedTerm{\LogVarB}{#1}\left(\forall\LogSortedTerm{\LogVarC}{#1}\,\neg#2\LogSubst{\LogVarA,\LogVarB,\LogVarC}\LogImp\forall\LogSortedTerm{\LogVarA'}{\CASort}\,#2\LogSubst{\LogVarA',\LogVarB,\LogVarB}\right)\;\LogImp\;\exists\LogSortedTerm{\LogVarE}{\CASort\SortTo#1}\,\LogForallRel\LogSortedTerm{\LogVarA}{\CASort}\,#2\LogSubst{\LogVarA,\LogVarE\,\LogVarA,\LogVarE\left(\CALogS\,\LogVarA\right)}\right)}

\newcommand*\CADCTerm{\lambda\LmVarA\LmVarB.\CALmbarrec\left(\lambda\LmVarE.\LmVarA\,\CALmlen{\LmVarE}\left(\CALmifz\,\CALmlen{\LmVarE}\,\CALmn{0}\left(\LmProj_1\,\CALmind{\LmVarE}{\CALmpred\,\CALmlen{\LmVarE}}\right)\right)\right)\LmVarB\,\CALmnil}
\newcommand*\CADCSort[2]{\left(\CALmnSort\LmSortTo#1\LmSortTo\left(\LmInterpForm{#2}\LmSortTo\LmSortBot\right)\LmSortTo\LmInterpForm{#2}\right)\LmSortTo\left(\left(\CALmnSort\LmSortTo\LmInterpForm{#2}\right)\LmSortTo\LmSortBot\right)\LmSortTo\LmSortBot}

\newcommand*\CARelsNoType{\LogRel{\CALogs}}
\newcommand*\CARelsTerm{\lambda\LmVarA\LmVarB\LmVarC.\LmVarA\,\LmVarC\left(\LmVarB\,\LmVarC\right)}
\newcommand*\CARelsSort[3]{\left(#1\LmSortTo#2\LmSortTo#3\right)\LmSortTo\left(#1\LmSortTo#2\right)\LmSortTo#1\LmSortTo#3}

\newcommand*\CARelkNoType{\LogRel{\CALogk}}
\newcommand*\CARelkTerm{\lambda\LmVarA\LmVarB.\LmVarA}
\newcommand*\CARelkSort[2]{#1\LmSortTo#2\LmSortTo#1}
\newcommand*\CARelZ{\LogRel{\LogSortedTerm{\CALogZ}{\CALogZSort}}}
\newcommand*\CARelZNoType{\LogRel{\CALogZ}}
\newcommand*\CARelZTerm{\CALmn{0}}
\newcommand*\CARelZSort{\CALmnSort}
\newcommand*\CARelS{\LogRel{\LogSortedTerm{\CALogS}{\CALogSSort}}}
\newcommand*\CARelSNoType{\LogRel{\CALogS}}
\newcommand*\CARelSTerm{\CALmsucc}
\newcommand*\CARelSSort{\CALmnSort\LmSortTo\CALmnSort}

\newcommand*\CARelrecNoType{\LogRel{\CALogrec}}
\newcommand*\CARelrecTerm{\CALmrec}
\newcommand*\CARelrecSort[1]{#1\LmSortTo\left(\CALmnSort\LmSortTo#1\LmSortTo#1\right)\LmSortTo\CALmnSort\LmSortTo#1}
\newcommand*\CALmn[1]{\LmConst{\overline{#1}}}
\newcommand*\CALmnSort{I}
\newcommand*\CALmom{\LmConst{\Omega}}
\newcommand*\CALmomSort\LmSortBot
\newcommand*\CALmsucc{\LmConst{succ}}
\newcommand*\CALmsuccSort{\CALmnSort\LmSortTo\CALmnSort}
\newcommand*\CALmrec{\LmConst{rec}}
\newcommand*\CALmrecSort[1]{#1\LmSortTo\left(\CALmnSort\LmSortTo#1\LmSortTo#1\right)\LmSortTo\CALmnSort\LmSortTo#1}
\newcommand*\CALmfix{\LmConst{Y}}
\newcommand*\CALmfixSort[1]{\left(#1\LmSortTo#1\right)\LmSortTo#1}
\newcommand*\CALmbarrec{\LmConst{barrec}}
\newcommand*\CALmbarrecSort[2]{\left(\CASortList{#1}\LmSortTo\left(#1\LmSortTo#2\right)\LmSortTo#1\right)\LmSortTo\left(\left(\CALmnSort\LmSortTo#1\right)\LmSortTo#2\right)\LmSortTo\CASortList{#1}\LmSortTo#2}
\newcommand*\CALmifz{\LmConst{if_0}}
\newcommand*\CALmifzSort[1]{\CALmnSort\LmSortTo#1\LmSortTo#1}
\newcommand*\CALmpred{\LmConst{pred}}

\newcommand*\CALmsub{\LmConst{sub}}
\newcommand*\CALmsubSort{\CALmnSort\LmSortTo\CALmnSort\LmSortTo\CALmnSort}
\newcommand*\CALmifl{\LmConst{if_<}}
\newcommand*\CALmiflSort[1]{\CALmnSort\LmSortTo\CALmnSort\LmSortTo#1\LmSortTo#1\LmSortTo#1}
\newcommand*\CALmife{\LmConst{if_=}}
\newcommand*\CALmifeSort[1]{\CALmnSort\LmSortTo\CALmnSort\LmSortTo#1\LmSortTo#1\LmSortTo#1}
\newcommand*\CALmconcat{\mathbin{@}}
\newcommand*\CALmextend{\mathbin{*}}
\newcommand*\CALmnil{\epsilon}
\newcommand*\CALmlen[1]{\left|#1\right|}
\newcommand*\CALmind[2]{#1\,\downharpoonright\mkern-5mu#2\mkern-5mu\downharpoonleft}
\DeclareSymbolFont{cyrillic}{T2A}{cmr}{m}{n}
\DeclareMathSymbol{\MCYRA}{\mathalpha}{cyrillic}{192}
\DeclareMathSymbol{\mcyra}{\mathalpha}{cyrillic}{224}
\DeclareMathSymbol{\MCYRB}{\mathalpha}{cyrillic}{193}
\DeclareMathSymbol{\mcyrb}{\mathalpha}{cyrillic}{225}
\DeclareMathSymbol{\MCYRV}{\mathalpha}{cyrillic}{194}
\DeclareMathSymbol{\mcyrv}{\mathalpha}{cyrillic}{226}
\DeclareMathSymbol{\MCYRG}{\mathalpha}{cyrillic}{195}
\DeclareMathSymbol{\mcyrg}{\mathalpha}{cyrillic}{227}
\DeclareMathSymbol{\MCYRD}{\mathalpha}{cyrillic}{196}
\DeclareMathSymbol{\mcyrd}{\mathalpha}{cyrillic}{228}
\DeclareMathSymbol{\MCYRE}{\mathalpha}{cyrillic}{197}
\DeclareMathSymbol{\mcyre}{\mathalpha}{cyrillic}{229}
\DeclareMathSymbol{\MCYRZH}{\mathalpha}{cyrillic}{198}
\DeclareMathSymbol{\mcyrzh}{\mathalpha}{cyrillic}{230}
\DeclareMathSymbol{\MCYRZ}{\mathalpha}{cyrillic}{199}
\DeclareMathSymbol{\mcyrz}{\mathalpha}{cyrillic}{231}
\DeclareMathSymbol{\MCYRI}{\mathalpha}{cyrillic}{200}
\DeclareMathSymbol{\mcyri}{\mathalpha}{cyrillic}{232}
\DeclareMathSymbol{\MCYRISHRT}{\mathalpha}{cyrillic}{201}
\DeclareMathSymbol{\mcyrishrt}{\mathalpha}{cyrillic}{233}
\DeclareMathSymbol{\MCYRK}{\mathalpha}{cyrillic}{202}
\DeclareMathSymbol{\mcyrk}{\mathalpha}{cyrillic}{234}
\DeclareMathSymbol{\MCYRL}{\mathalpha}{cyrillic}{203}
\DeclareMathSymbol{\mcyrl}{\mathalpha}{cyrillic}{235}
\DeclareMathSymbol{\MCYRM}{\mathalpha}{cyrillic}{204}
\DeclareMathSymbol{\mcyrm}{\mathalpha}{cyrillic}{236}
\DeclareMathSymbol{\MCYRN}{\mathalpha}{cyrillic}{205}
\DeclareMathSymbol{\mcyrn}{\mathalpha}{cyrillic}{237}
\DeclareMathSymbol{\MCYRO}{\mathalpha}{cyrillic}{206}
\DeclareMathSymbol{\mcyro}{\mathalpha}{cyrillic}{238}
\DeclareMathSymbol{\MCYRP}{\mathalpha}{cyrillic}{207}
\DeclareMathSymbol{\mcyrp}{\mathalpha}{cyrillic}{239}
\DeclareMathSymbol{\MCYRR}{\mathalpha}{cyrillic}{208}
\DeclareMathSymbol{\mcyrr}{\mathalpha}{cyrillic}{240}
\DeclareMathSymbol{\MCYRS}{\mathalpha}{cyrillic}{209}
\DeclareMathSymbol{\mcyrs}{\mathalpha}{cyrillic}{241}
\DeclareMathSymbol{\MCYRT}{\mathalpha}{cyrillic}{210}
\DeclareMathSymbol{\mcyrt}{\mathalpha}{cyrillic}{242}
\DeclareMathSymbol{\MCYRU}{\mathalpha}{cyrillic}{211}
\DeclareMathSymbol{\mcyru}{\mathalpha}{cyrillic}{243}
\DeclareMathSymbol{\MCYRF}{\mathalpha}{cyrillic}{212}
\DeclareMathSymbol{\mcyrf}{\mathalpha}{cyrillic}{244}
\DeclareMathSymbol{\MCYRH}{\mathalpha}{cyrillic}{213}
\DeclareMathSymbol{\mcyrh}{\mathalpha}{cyrillic}{245}
\DeclareMathSymbol{\MCYRC}{\mathalpha}{cyrillic}{214}
\DeclareMathSymbol{\mcyrc}{\mathalpha}{cyrillic}{246}
\DeclareMathSymbol{\MCYRCH}{\mathalpha}{cyrillic}{215}
\DeclareMathSymbol{\mcyrch}{\mathalpha}{cyrillic}{247}
\DeclareMathSymbol{\MCYRSH}{\mathalpha}{cyrillic}{216}
\DeclareMathSymbol{\mcyrsh}{\mathalpha}{cyrillic}{248}
\DeclareMathSymbol{\MCYRSHCH}{\mathalpha}{cyrillic}{217}
\DeclareMathSymbol{\mcyrshch}{\mathalpha}{cyrillic}{249}
\DeclareMathSymbol{\MCYRHRDSN}{\mathalpha}{cyrillic}{218}
\DeclareMathSymbol{\mcyrhrdsn}{\mathalpha}{cyrillic}{250}
\DeclareMathSymbol{\MCYRERY}{\mathalpha}{cyrillic}{219}
\DeclareMathSymbol{\mcyrery}{\mathalpha}{cyrillic}{251}
\DeclareMathSymbol{\MCYRSFTSN}{\mathalpha}{cyrillic}{220}
\DeclareMathSymbol{\mcyrsftsn}{\mathalpha}{cyrillic}{252}
\DeclareMathSymbol{\MCYREREV}{\mathalpha}{cyrillic}{221}
\DeclareMathSymbol{\mcyrerev}{\mathalpha}{cyrillic}{253}
\DeclareMathSymbol{\MCYRYU}{\mathalpha}{cyrillic}{222}
\DeclareMathSymbol{\mcyryu}{\mathalpha}{cyrillic}{254}
\DeclareMathSymbol{\MCYRYA}{\mathalpha}{cyrillic}{223}
\DeclareMathSymbol{\mcyrya}{\mathalpha}{cyrillic}{255}
\DeclareMathSymbol{\MCYRGUP}{\mathalpha}{cyrillic}{128}
\DeclareMathSymbol{\mcyrgup}{\mathalpha}{cyrillic}{160}
\DeclareMathSymbol{\MCYRGHCRS}{\mathalpha}{cyrillic}{129}
\DeclareMathSymbol{\mcyrghcrs}{\mathalpha}{cyrillic}{161}
\DeclareMathSymbol{\MCYRDJE}{\mathalpha}{cyrillic}{130}
\DeclareMathSymbol{\mcyrdje}{\mathalpha}{cyrillic}{162}
\DeclareMathSymbol{\MCYRTSHE}{\mathalpha}{cyrillic}{131}
\DeclareMathSymbol{\mcyrtshe}{\mathalpha}{cyrillic}{163}
\DeclareMathSymbol{\MCYRSHHA}{\mathalpha}{cyrillic}{132}
\DeclareMathSymbol{\mcyrshha}{\mathalpha}{cyrillic}{164}
\DeclareMathSymbol{\MCYRZHDSC}{\mathalpha}{cyrillic}{133}
\DeclareMathSymbol{\mcyrzhdsc}{\mathalpha}{cyrillic}{165}
\DeclareMathSymbol{\MCYRZDSC}{\mathalpha}{cyrillic}{134}
\DeclareMathSymbol{\mcyrzdsc}{\mathalpha}{cyrillic}{166}
\DeclareMathSymbol{\MCYRLJE}{\mathalpha}{cyrillic}{135}
\DeclareMathSymbol{\mcyrlje}{\mathalpha}{cyrillic}{167}
\DeclareMathSymbol{\MCYRYI}{\mathalpha}{cyrillic}{136}
\DeclareMathSymbol{\mcyryi}{\mathalpha}{cyrillic}{168}
\DeclareMathSymbol{\MCYRKDSC}{\mathalpha}{cyrillic}{137}
\DeclareMathSymbol{\mcyrkdsc}{\mathalpha}{cyrillic}{169}
\DeclareMathSymbol{\MCYRKBEAK}{\mathalpha}{cyrillic}{138}
\DeclareMathSymbol{\mcyrkbeak}{\mathalpha}{cyrillic}{170}
\DeclareMathSymbol{\MCYRKVCRS}{\mathalpha}{cyrillic}{139}
\DeclareMathSymbol{\mcyrkvcrs}{\mathalpha}{cyrillic}{171}
\DeclareMathSymbol{\MCYRAE}{\mathalpha}{cyrillic}{140}
\DeclareMathSymbol{\mcyrae}{\mathalpha}{cyrillic}{172}
\DeclareMathSymbol{\MCYRNDSC}{\mathalpha}{cyrillic}{141}
\DeclareMathSymbol{\mcyrndsc}{\mathalpha}{cyrillic}{173}
\DeclareMathSymbol{\MCYRNG}{\mathalpha}{cyrillic}{142}
\DeclareMathSymbol{\mcyrng}{\mathalpha}{cyrillic}{174}
\DeclareMathSymbol{\MCYRDZE}{\mathalpha}{cyrillic}{143}
\DeclareMathSymbol{\mcyrdze}{\mathalpha}{cyrillic}{175}
\DeclareMathSymbol{\MCYROTLD}{\mathalpha}{cyrillic}{144}
\DeclareMathSymbol{\mcyrotld}{\mathalpha}{cyrillic}{176}
\DeclareMathSymbol{\MCYRSDSC}{\mathalpha}{cyrillic}{145}
\DeclareMathSymbol{\mcyrsdsc}{\mathalpha}{cyrillic}{177}
\DeclareMathSymbol{\MCYRUSHRT}{\mathalpha}{cyrillic}{146}
\DeclareMathSymbol{\mcyrushrt}{\mathalpha}{cyrillic}{178}
\DeclareMathSymbol{\MCYRY}{\mathalpha}{cyrillic}{147}
\DeclareMathSymbol{\mcyry}{\mathalpha}{cyrillic}{179}
\DeclareMathSymbol{\MCYRYHCRS}{\mathalpha}{cyrillic}{148}
\DeclareMathSymbol{\mcyryhcrs}{\mathalpha}{cyrillic}{180}
\DeclareMathSymbol{\MCYRHDSC}{\mathalpha}{cyrillic}{149}
\DeclareMathSymbol{\mcyrhdsc}{\mathalpha}{cyrillic}{181}
\DeclareMathSymbol{\MCYRDZHE}{\mathalpha}{cyrillic}{150}
\DeclareMathSymbol{\mcyrdzhe}{\mathalpha}{cyrillic}{182}
\DeclareMathSymbol{\MCYRCHVCRS}{\mathalpha}{cyrillic}{151}
\DeclareMathSymbol{\mcyrchvcrs}{\mathalpha}{cyrillic}{183}
\DeclareMathSymbol{\MCYRCHRDSC}{\mathalpha}{cyrillic}{152}
\DeclareMathSymbol{\mcyrchrdsc}{\mathalpha}{cyrillic}{184}
\DeclareMathSymbol{\MCYRIE}{\mathalpha}{cyrillic}{153}
\DeclareMathSymbol{\mcyrie}{\mathalpha}{cyrillic}{185}
\DeclareMathSymbol{\MCYRSCHWA}{\mathalpha}{cyrillic}{154}
\DeclareMathSymbol{\mcyrschwa}{\mathalpha}{cyrillic}{186}
\DeclareMathSymbol{\MCYRNJE}{\mathalpha}{cyrillic}{155}
\DeclareMathSymbol{\mcyrnje}{\mathalpha}{cyrillic}{187}
\DeclareMathSymbol{\MCYRYO}{\mathalpha}{cyrillic}{156}
\DeclareMathSymbol{\mcyryo}{\mathalpha}{cyrillic}{188}
\DeclareMathSymbol{\MCYRII}{\mathalpha}{cyrillic}{73}
\DeclareMathSymbol{\mcyrii}{\mathalpha}{cyrillic}{105}
\DeclareMathSymbol{\MCYRJE}{\mathalpha}{cyrillic}{74}
\DeclareMathSymbol{\mcyrje}{\mathalpha}{cyrillic}{106}
\DeclareMathSymbol{\MCYRQ}{\mathalpha}{cyrillic}{81}
\DeclareMathSymbol{\mcyrq}{\mathalpha}{cyrillic}{113}
\DeclareMathSymbol{\MCYRW}{\mathalpha}{cyrillic}{87}
\DeclareMathSymbol{\mcyrw}{\mathalpha}{cyrillic}{119}
\begin{document}
\title{Typed realizability for first-order classical analysis}
\author{Valentin Blot}
\address{Department of Computer Science, University of Bath, United Kingdom}
\email{v.blot@bath.ac.uk}
\thanks{Research supported by the UK EPSRC grant EP/K037633/1.}
\begin{abstract}
We describe a realizability framework for classical first-order logic in which realizers live in (a model of) typed $\lambda\mu$-calculus. This allows a direct interpretation of classical proofs, avoiding the usual negative translation to intuitionistic logic. We prove that the usual terms of G\"odel's system T realize the axioms of Peano arithmetic, and that under some assumptions on the computational model, the bar recursion operator realizes the axiom of dependent choice. We also perform a proper analysis of relativization, which allows for less technical proofs of adequacy. Extraction of algorithms from proofs of $\Pi^0_2$ formulas relies on a novel implementation of Friedman's trick exploiting the control possibilities of the language. This allows to have extracted programs with simpler types than in the case of negative translation followed by intuitionistic realizability.
\end{abstract}
\maketitle
\section*{Introduction}
Realizability is a mean of formalizing the Brouwer-Heyting-Kolmogorov constructive interpretation of logic. To each formula is associated a set of programs, its realizers, which contain computational information about the formula. While the Curry-Howard isomorphism draws a correspondence between proofs in a certain logical system and programs in a suitable typed programming language, realizability takes a different approach and defines a realizer as a program which behaves like a proof. First, this allows to give computational content to the axioms of a theory, and second this allows to choose more freely the programming language, independently from the logical system. It is even possible to consider untyped programming languages, as was the case in the first realizability model from Kleene~\cite{Kleene}, in which a realizer may be any recursive function. It is however still possible to consider a typed language, as did Kreisel in his modified realizability model~\cite{Kreisel}. Both models from Kleene and Kreisel gave computational interpretation to Heyting arithmetic, the intuitionistic variant of Peano arithmetic.\par
G\"odel's negative translation~\cite{GodelNegative} allowed for the first interpretations of classical logic. Indeed, this translation from classical to intuitionistic logic, when followed by an interpretation of intuitionistic proofs, gives a computational interpretation to classical logic. This is what we call the indirect interpretation. This interpretation can easily be extended to arithmetic, and therefore allows to get computational interpretations of Peano arithmetic from the models of Kleene and Kreisel. Much later, Griffin discovered in~\cite{GriffinControl} that the \texttt{call/cc} operator of the functional language Scheme could be typed with a classical principle: the law of Peirce. This opened the possibility to what we call a direct interpretation of classical logic, using programming languages with control features. Following this path, Parigot defined in~\cite{ParigotLambdaMu} the $\lambda\mu$-calculus, a language for Gentzen's classical sequent calculus extending the Curry-Howard isomorphism to classical logic. Selinger axiomatized the universal categorical model of $\lambda\mu$-calculus in~\cite{SelingerControl}. On another side, Krivine considered untyped $\lambda$-calculus extended with the \texttt{call/cc} operator to give a realizability interpretation to classical second-order Zermelo-Fr\ae nkel set theory~\cite{KrivineZF}, later extended to handle the axiom of dependent choice~\cite{KrivineDependent,KrivinePanoramas}.\par
In this paper we define a realizability model for first-order classical logic which, contrary to Krivine's and similarly to Kreisel's (but for classical logic), uses typed programs as realizers. We interpret our proofs in the language $\mu$PCF, which is a combination of the functional Turing-complete language PCF with the control features of call-by-name $\lambda\mu$-calculus.  As in Krivine's model, we associate to each formula a truth value and a falsity value which are orthogonal to each other, but we perform a fine analysis of relativization and introduce positive predicates for which only the truth value is required and prove it to be correct through a suitable restriction on the proofs in our logical system. We validate our model by proving that the usual terms of G\"odel's system T realize the axioms of Peano arithmetic. We also implement Friedman's trick through the use of an external $\mu$-variable rather than through the replacement of the $\LogBot$ formula by an existential statement, which allows for a simpler and more effective interpretation. This variable is also used to define the orthogonality relation between our truth and falsity values.\par
Interpreting the axiom of dependent choice in a classical setting is much more complicated than interpreting arithmetic. Spector defined in~\cite{Spector} the bar recursor and used it in G\"odel's Dialectica interpretation~\cite{GodelDialectica} (a computational interpretation similar to realizability) to interpret the axiom of countable choice, and therefore the axiom schema of specification. This operator was later studied in~\cite{KohlenbachThesis}, and a more uniform version was used in~\cite{BerardiBezemCoquand} to give an indirect realizability interpretation of countable choice. A version with an implicit termination condition was later defined in~\cite{BergerOlivaChoice} and used to interpret, still in an indirect realizability setting, the double-negation shift principle and therefore the axiom of dependent choice.\par
We prove here that under some assumptions on our model of $\mu$PCF, the bar recursion operator of~\cite{BergerOlivaChoice} realizes the axiom of dependent choice in our direct interpretation of classical logic. Our implementation of Friedman's trick then allows us to obtain an extraction result on $\Pi^0_2$ formulas provable in classical analysis (Peano arithmetic + the axiom of dependent choice).
\section{Logic}
\label{logic}
We define in this section the logical system under which we will work through the article. First, we define the general case of classical multisorted first-order logic (handling classical reasoning by the use of multi-conclusioned sequents), then we describe the case of logics with equality, the case of Peano arithmetic and its extension with the axiom of dependent choice, and finally we recall some basic definitions about models of classical logic.
\subsection{Classical multisorted first-order logic}
The logical framework we use is multisorted first-order logic, where the sorts are fixed to be the types of simply typed $\lambda$-calculus (see e.g.~\cite{TroelstraVanDalenConstructivism}). We build from a set of base sorts $\SortBase$ the set of sorts:
$$\SortA,\SortB\GramDef\SortBase\BarSep\SortA\SortTo\SortB$$
which are used for the individuals of the logic. We fix a set of sorted individual constants (ranged over by $\LogSortedTerm{\LogConstA}{\SortA}$) from which we build the set of individuals of the logic:
$$\LogSortedTerm{\LogTermA}{\SortA},\LogSortedTerm{\LogTermB}{\SortB}\GramDef\LogSortedTerm{\LogConstA}{\SortA}\BarSep\LogSortedTerm{\LogVarA}{\SortA}\BarSep\LogSortedTerm{\left(\LogSortedTerm{\LogTermA}{\SortA\SortTo\SortB}\LogSortedTerm{\LogTermB}{\SortA}\right)}{\SortB}$$
We also fix a set of sorted predicates (ranged over by $\LogPredA$) from which we define the formulas of the logic:
$$\LogFormA,\LogFormB\GramDef\LogPredA\left(\LogSortedTerm{\LogTermA_1}{\SortA_1},\ldots,\LogSortedTerm{\LogTermA_n}{\SortA_n}\right)\BarSep\LogBot\BarSep\LogFormA\LogImp\LogFormB\BarSep\LogFormA\LogAnd\LogFormB\BarSep\forall\LogSortedTerm{\LogVarA}{\SortA}\LogFormA$$
The set of sorts, individuals and formulas of the logic is parameterized by a signature:
\begin{defi}
A signature $\Sigma$ is a set of base sorts together with a set of sorted constant individuals and a set of sorted predicates.
\end{defi}
Negation is defined as $\neg\LogFormA\Def\LogFormA\LogImp\LogBot$. We choose to have only negative connectives, since the interpretation of our logic in categories of continuations defined in section~\ref{LmInterp} is based on a negative call-by-name continuation-passing-style translation. The positive connectives are defined from the negative ones: $\LogFormA\vee\LogFormB\ \Def\ \neg\left(\neg\LogFormA\LogAnd\neg\LogFormB\right)$ and $\exists\LogSortedTerm{\LogVarA}{\SortA}\LogFormA\ \Def\ \neg\forall\LogSortedTerm{\LogVarA}{\SortA}\neg\LogFormA$. It is well-known that with this coding of positive connectives with negative ones, a formula $\LogFormA$ is provable in our system if and only if the formula obtained by replacing every $\LogPredA\left(\LogTermA_1,\ldots,\LogTermA_n\right)$ with $\neg\neg\LogPredA\left(\LogTermA_1,\ldots,\LogTermA_n\right)$ in $\LogFormA$ is provable in its intuitionistic restriction. In section~\ref{RelPred} we will also introduce the notion of negative basic predicates, which are those for which $\neg\neg\LogPredA\LogImp\LogPredA$ is valid under the realizability interpretation. We perform a more detailed comparison between our system and the usual ones in section~\ref{usualtheories}.\par
Since one of the goals of realizability is to provide a computational interpretation of theories beyond pure first-order logic, our model is dependent upon the particular set of axioms under consideration:
\begin{defi}
A theory on a given signature $\Sigma$ is a set $\LogAxioms$ of closed formulas (axioms) written in the language defined by $\Sigma$.
\end{defi}
We work in a variant of natural deduction in sequent style, so the interpretation of classical proofs in $\lambda\mu$-calculus is as direct as possible. In this setting, a context $\Gamma$ or $\Delta$ is a finite unordered sequence of formulas and a sequent is of the form:
$$\Sequent{\Gamma}{\LogFormA}{\Delta}$$
The formula $\LogFormA$ on the right is the formula that is being worked on, and this presentation is again chosen to have an easy interpretation in $\lambda\mu$-calculus. The above sequent should be interpreted as: the conjunction of the formulas of $\Gamma$ implies the disjunction of $\LogFormA$ and of the formulas of $\Delta$. If $\Delta$ is empty we simply write $\Sequent{\Gamma}{\LogFormA}{}$. The set of derivable sequents of a given theory is defined from $\LogAxioms$ using the rules of Figure~\ref{LogicRules}.
\begin{figure}
\begin{gather*}
\LogRuleAx{\Gamma}{\Delta}{\LogFormA}
\qquad
\LogRuleAxSig{\Gamma}{\Delta}{\LogFormA}
\qquad
\LogRuleBotIntro{\Gamma}{\Delta}{\LogFormA}
\qquad
\LogRuleBotElim{\Gamma}{\Delta}{\LogFormA}
\\[5pt]
\LogRuleImpIntro{\Gamma}{\Delta}{\LogFormA}{\LogFormB}
\qquad
\LogRuleImpElim{\Gamma}{\Delta}{\LogFormA}{\LogFormB}
\\[5pt]
\LogRuleAndIntro{\Gamma}{\Delta}{\LogFormA}{\LogFormB}
\qquad
\LogRuleAndElim{\Gamma}{\Delta}{\LogFormA}
\\[5pt]
\LogRuleForallIntro{\Gamma}{\Delta}{\LogSortedTerm{\LogVarA}{\SortA}}{\LogFormA}
\qquad
\LogRuleForallElim{\Gamma}{\Delta}{\LogSortedTerm{\LogVarA}{\SortA}}{\LogSortedTerm{\LogTermA}{\SortA}}{\LogFormA}
\end{gather*}
\caption{Deduction rules for first-order multisorted logic}
\label{LogicRules}
\end{figure}
In this system one can for example derive in any theory the ex-falso principle, the double-negation elimination and Peirce's law:
$$\Sequent{}{\LogBot\LogImp\LogFormA}{}\qquad\Sequent{}{\neg\left(\neg\LogFormA\right)\LogImp\LogFormA}{}\qquad\Sequent{}{\left(\left(\LogFormA\LogImp\LogFormB\right)\LogImp\LogFormA\right)\LogImp\LogFormA}{}$$
The weakening rule:
$$\AXM{\Sequent{\Gamma}{\LogFormA}{\Delta}}\RLM{\Gamma\subseteq\Gamma',\Delta\subseteq\Delta'}\UIM{\Sequent{\Gamma'}{\LogFormA}{\Delta'}}\DP$$
is admissible ($\alpha$-conversion may be necessary to preserve side condition of $\forall$-intro) and we use it without mentioning it. The left contraction rule is derivable and the right one is admissible:
$$\AXM{\Sequent{\Gamma,\LogFormA,\LogFormA}{\LogFormB}{\Delta}}\UIM{\Sequent{\Gamma,\LogFormA}{\LogFormA\LogImp\LogFormB}{\Delta}}\AXM{}\UIM{\Sequent{\Gamma,\LogFormA}{\LogFormA}{\Delta}}\BIM{\Sequent{\Gamma,\LogFormA}{\LogFormB}{\Delta}}\DP\qquad\qquad\qquad\AXM{\Sequent{\Gamma}{\LogFormA}{\LogFormB,\LogFormB,\Delta}}\RLM{\LogFormB,\LogFormB,\Delta\subseteq\LogFormA,\LogFormB,\LogFormB,\Delta}\UIM{\Sequent{\Gamma}{\LogFormA}{\LogFormA,\LogFormB,\LogFormB,\Delta}}\UIM{\Sequent{\Gamma}{\LogBot}{\LogFormA,\LogFormB,\LogFormB,\Delta}}\UIM{\Sequent{\Gamma}{\LogFormB}{\LogFormA,\LogFormB,\Delta}}\UIM{\Sequent{\Gamma}{\LogBot}{\LogFormA,\LogFormB,\Delta}}\UIM{\Sequent{\Gamma}{\LogFormA}{\LogFormB,\Delta}}\DP$$
so we also use them implicitly. If a sequent $\Sequent{}{\LogFormA}{}$ is derivable in a theory $\LogAxioms$, then since $\LogAxioms$ may contain infinitely many formulas we write to avoid confusion:
$$\LogAxioms\Derives\LogFormA$$
\subsection{Equational theories}
\label{equational}
Most of the theories used in mathematics involve equality, but there are several notions for it. Here we use a primitive Leibniz equality at each type. In fact, as in~\cite{KrivinePanoramas}, we use a primitive inequality rather than an equality, which allows us to realize the Leibniz scheme. Moreover, since we are in classical logic, the set of provable sequents is unchanged. In the intuitionistic case however, things are less clear and we discuss it in section~\ref{usualtheories}.\par
We say that a theory is an equational theory if it contains for each sort $\SortA$ an inequality predicate $\neq_\SortA$ between terms of sort $\SortA$ (for which we use infix notation), and the following axioms for reflexivity and the Leibniz scheme at all sorts:
$$\CAReflName\quad\CARefl{\SortA}\qquad\CALeibName\quad\CALeib{\SortA}{\LogFormA}{\SortB}$$
where $\LogTermA=\LogTermB\ \Def\ \neg\left(\LogTermA\neq\LogTermB\right)$. In this system, the following formulas are derivable:
$$\forall\LogVarA\,\forall\LogVarB\left(\LogVarA=\LogVarB\LogImp\forall\LogVarC\left(\LogVarC\,\LogVarA=\LogVarC\,\LogVarB\right)\right)\qquad\qquad\forall\LogVarA\,\forall\LogVarB\left(\LogVarA=\LogVarB\LogImp\forall\LogVarC\left(\LogVarA\,\LogVarC=\LogVarB\,\LogVarC\right)\right)$$
However, even though the reverse of the implication is provable in $\PAom$ (see next section~\ref{PeanoTheory}), the reverse of the second one doesn't hold in the general case.
\subsection{Peano arithmetic in finite types: \texorpdfstring{$\PAom$}{PAomega}}
\label{PeanoTheory}
Peano arithmetic in finite types, $\PAom$, is an extension of Peano arithmetic in which we can write any term of G\"odel's system T and quantify over functions of any type.\par
The signature $\Sigma_\PAom$ of Peano arithmetic in finite types contains a single base sort $\CASort$ and the sorted constants of Figure~\ref{PAomConsts}.
\begin{figure}
\begin{gather*}
\begin{aligned}
\CALogs&\quad\text{of sort}\quad\CALogsSort{\SortA}{\SortB}{\SortC}\quad&\quad\CALogk&\quad\text{of sort}\quad\CALogkSort{\SortA}{\SortB}\\
\CALogZ&\quad\text{of sort}\quad\CALogZSort\quad&\quad\CALogS&\quad\text{of sort}\quad\CALogSSort
\end{aligned}\\
\CALogrec\quad\text{of sort}\quad\CALogrecSort{\SortA}
\end{gather*}
\caption{Sorted constants of $\PAom$}
\label{PAomConsts}
\end{figure}
We will also write $\LogSortedTerm{\CALogZ}{\SortA}$ for the inductively defined term: $\LogSortedTerm{\CALogZ}{\SortA\SortTo\SortB}\Def\LogSortedTerm{\CALogk}{\CALogkSort{\SortB}{\SortA}}\,\LogSortedTerm{\CALogZ}{\SortB}$. $\PAom$ is an equational theory as defined in section~\ref{equational}, and inequality is the only predicate symbol:
$$\LogSortedTerm{\LogTermA}{\SortA}\neq\LogSortedTerm{\LogTermB}{\SortA}$$
The axioms of $\PAom$ are given in Figure~\ref{PAomAxioms}.
\begin{figure}
\begin{align*}
\CAReflName\;\;\;&\CARefl{\SortA}&\CALeibName\;\;\;&\CALeib{\SortA}{\LogFormA}{\SortB}\\
\CASnZName\;\;\;&\CASnZNoType&\CAindUnrelName\;\;\;&\CAindUnrelNoType{A}\\
\CAdefsName\;\;\;&\CAdefsNoType&\CAdefkName\;\;\;&\CAdefkNoType\\
\CAdefrecZName\;\;\;&\CAdefrecZNoType&\CAdefrecSName\;\;\;&\CAdefrecSNoType
\end{align*}
\caption{Axioms of $\PAom$}
\label{PAomAxioms}
\end{figure}
$\CAReflName$ and $\CALeibName$ are the axioms of equational theories, $\CAdefsName$, $\CAdefkName$, $\CAdefrecZName$ and $\CAdefrecSName$ are the definitional axioms for the individual constants $\CALogs$, $\CALogk$ and $\CALogrec$, and finally $\CASnZName$ and $\CAindUnrelName$ are the axiom stating that $0$ is not a successor and the axiom scheme of induction for every formula $\LogFormA$ with free variables among $\LogSortedTerm{\LogVarA}{\CASort},\LogSortedTerm{\vec{\LogVarB}}{\vec{\SortA}}$. Since the predecessor function is definable in system T, the injectivity of the successor:
$$\forall\LogVarA\,\forall\LogVarB\left(\CALogS\,\LogVarA=\CALogS\,\LogVarB\LogImp\LogVarA=\LogVarB\right)$$
is derivable in $\PAom$. Using the provable equality $\forall\LogVarA\left(\CALogS\,\CALogk\,\CALogk\,\LogVarA=\LogVarA\right)$, we can prove in $\PAom$:
$$\forall\LogVarA\,\forall\LogVarB\left(\forall\LogVarC\left(\LogVarC\,\LogVarA=\LogVarC\,\LogVarB\right)\LogImp\LogVarA=\LogVarB\right)$$
but the status of $\forall\LogVarA\,\forall\LogVarB\left(\forall\LogVarC\left(\LogVarA\,\LogVarC=\LogVarB\,\LogVarC\right)\LogImp\LogVarA=\LogVarB\right)$ is less clear and we won't consider it here.
\subsubsection{Relation to usual theories}
\label{usualtheories}
We discuss in this section about the relation between our version of $\PAom$, in which the inequality predicate is atomic, and other intuitionistic or classical systems. First, in this section we will consider the following systems:
$$
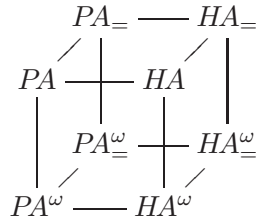
\xymatrix@!0{
&\PA_=\ar@{-}[rr]&&\HA_=\\
\PA\ar@{-}[ur]\ar@{-}[rr]&&\HA\ar@{-}[ur]\\
&\PA^\omega_=\ar@{-}[uu]\ar@{-}[rr]&&\HA^\omega_=\ar@{-}[uu]\\
\PA^\omega\ar@{-}[uu]\ar@{-}[ur]\ar@{-}[rr]&&\HA^\omega\ar@{-}[uu]\ar@{-}[ur]
}$$
where the $P/H$ part corresponds to Peano and Heyting arithmetic, the latter being the intuitionistic version of the former, the $\omega$ part corresponds to the possibility of having higher type variables, and the $=$ part corresponds to having an atomic predicate for equality while the absence of $=$ indicates a primitive inequality. The systems without higher type variables have addition and multiplication as individual constants and their defining equations as axioms, instead of $\CALogs$, $\CALogk$ and $\CALogrec$ and their defining equations. For the systems based on a primitive equality we refer to~\cite{TroelstraVanDalenConstructivism}. In the case of primitive equality, the Leibniz scheme is:
$$\forall\LogSortedTerm{\vec{\LogVarC}}{\vec{\SortB}}\forall\LogSortedTerm{\LogVarA}{\SortA}\forall\LogSortedTerm{\LogVarB}{\SortA}\left(\LogVarA=\LogVarB\LogImp\LogFormA\LogImp\LogFormA\LogSubst{\LogVarB/\LogVarA}\right)$$
while in the case of primitive inequality we define it is as in section~\ref{equational}:
$$\forall\LogSortedTerm{\vec{\LogVarC}}{\vec{\SortB}}\forall\LogSortedTerm{\LogVarA}{\SortA}\forall\LogSortedTerm{\LogVarB}{\SortA}\left(\neg\LogFormA\LogImp\LogFormA\LogSubst{\LogVarB/\LogVarA}\LogImp\LogVarA\neq\LogVarB\right)$$
Similarly, in $\PA_=$ and $\HA_=$ the injectivity of successor is:
$$\forall\LogSortedTerm{\LogVarA}{\CASort}\forall\LogSortedTerm{\LogVarB}{\CASort}\left(\CALogS\,\LogVarA=\CALogS\,\LogVarB\LogImp\LogVarA=\LogVarB\right)$$
which is derivable in $\PA^\omega_=$ and $\HA^\omega_=$, while in $\PA$ and $\HA$ we define it to be the following formula:
$$\forall\LogSortedTerm{\LogVarA}{\CASort}\forall\LogSortedTerm{\LogVarB}{\CASort}\left(\LogVarA\neq\LogVarB\LogImp\CALogS\,\LogVarA\neq\CALogS\,\LogVarB\right)$$
which is derivable in $\PA^\omega$ and $HA^\omega$.\par
It is well known that $\HA_=\Derives\forall\LogSortedTerm{\LogVarA}{\CASort}\forall\LogSortedTerm{\LogVarB}{\CASort}\left(\neg\neg\LogVarA=\LogVarB\LogImp\LogVarA=\LogVarB\right)$, so in our system with only negative connectives, $\HA_=$ and $\PA_=$ prove exactly the same formulas. It is quite easy to prove also that $\HA\Derives\forall\LogSortedTerm{\LogVarA}{\CASort}\forall\LogSortedTerm{\LogVarB}{\CASort}\left(\neg\neg\LogVarA\neq\LogVarB\LogImp\LogVarA\neq\LogVarB\right)$, so $\HA$ and $\PA$ also prove exactly the same formulas (again, in the particular case of our system with only negative connectives). However, despite the equivalence of provability between $\PA$ and $\HA$, the associated proof terms are very different. Indeed, the proof of $\HA\Derives\forall\LogSortedTerm{\LogVarA}{\CASort}\forall\LogSortedTerm{\LogVarB}{\CASort}\left(\neg\neg\LogVarA\neq\LogVarB\LogImp\LogVarA\neq\LogVarB\right)$ is obtained by a double induction on $\LogVarA$ and $\LogVarB$, while in $\PA$, a proof of $\PA\Derives\forall\LogSortedTerm{\LogVarA}{\CASort}\forall\LogSortedTerm{\LogVarB}{\CASort}\left(\neg\neg\LogVarA\neq\LogVarB\LogImp\LogVarA\neq\LogVarB\right)$ can also be obtained through classical logic, in which case its computational content relies on the control features of languages for classical logic (like $\lambda\mu$-calculus).\par
When we switch to higher type equalities, things become more complicated. Indeed, $\HA^\omega_=$ fails to prove $\forall\LogSortedTerm{\LogVarA}{\SortA}\forall\LogSortedTerm{\LogVarB}{\SortA}\left(\neg\neg\LogVarA=\LogVarB\LogImp\LogVarA=\LogVarB\right)$, since we cannot perform induction on higher type variables, so the systems $\HA^\omega_=$ and $\PA^\omega_=$ are not equivalent anymore, even at the level of provability.\par
For classical systems $\PA^\omega$ and $\PA^\omega_=$, a formula with equalities and inequalities is provable in $\PA^\omega$ (through the encoding $\left(\LogVarA=\LogVarB\right)\equiv\left(\LogVarA\neq\LogVarB\LogImp\LogBot\right)$) if and only if it is provable in $\PA^\omega_=$ (through the encoding $\left(\LogVarA\neq\LogVarB\right)\equiv\left(\LogVarA=\LogVarB\LogImp\LogBot\right)$). However, for intuitionistic systems $\HA^\omega$ and $\HA^\omega_=$ things are less clear since double negations cannot be eliminated on atomic formulas. In particular, we did not investigate the relation between the system $\HA^\omega$ with primitive inequalities and $\HA^\omega_=$ or $\PA^\omega/\PA^\omega_=$.
\subsection{The axiom of choice}
In this section, $\LogFormA$ is a formula over $\Sigma_\PAom$ with free variables among $\LogSortedTerm{\LogVarA}{\CASort},\LogSortedTerm{\LogVarB}{\SortA},\LogSortedTerm{\LogVarC}{\SortA},\LogSortedTerm{\vec{\LogVarD}}{\vec{\SortB}}$. For clarity we write $\LogFormA\LogSubst{\LogTermA,\LogTermB,\LogTermC}$ instead of $\LogFormA\LogSubst{\LogTermA/\LogVarA,\LogTermB/\LogVarB,\LogTermC/\LogVarC}$. Classical analysis $\CAom$ (in the sense of~\cite{KohlenbachProofTheory}) is defined to be $\PAom$ augmented with the following axiom scheme:
$$\CADCUnrelName\quad\CADCUnrel{\SortA}{\LogFormA}{\SortB}$$
As proved in~\cite{KohlenbachProofTheory}, $\CADCUnrelName$ implies in $\PAom$ both countable choice:
$$\forall\LogSortedTerm{\vec{\LogVarD}}{\vec{\SortB}}\left(\forall\LogSortedTerm{\LogVarA}{\CASort}\,\exists\LogSortedTerm{\LogVarB}{\SortA}\,\LogFormB\;\;\LogImp\;\;\exists\LogSortedTerm{\LogVarE}{\CASort\SortTo\SortA}\,\forall\LogSortedTerm{\LogVarA}{\CASort}\,\LogFormB\LogSubst{\LogVarE\,\LogVarA/\LogVarB}\right)$$
where $\LogFormB$ has free variables among $\LogSortedTerm{\LogVarA}{\CASort},\LogSortedTerm{\LogVarB}{\SortA},\LogSortedTerm{\vec{\LogVarD}}{\vec{\SortB}}$, and the more usual version of dependent choice:
$$\forall\LogSortedTerm{\vec{\LogVarD}}{\vec{\SortB}}\left(\forall\LogSortedTerm{\LogVarA}{\CASort}\,\forall\LogSortedTerm{\LogVarB}{\SortA}\,\exists\LogSortedTerm{\LogVarC}{\SortA}\,\LogFormA\LogSubst{\LogVarA,\LogVarB,\LogVarC}\;\;\LogImp\;\;\forall\LogSortedTerm{\LogVarF}{\SortA}\exists\LogSortedTerm{\LogVarE}{\CASort\SortTo\SortA}\left(\LogVarE\,\CALogZ=\LogVarF\LogAnd\forall\LogSortedTerm{\LogVarA}{\CASort}\,\LogFormA\LogSubst{\LogVarA,\LogVarE\,\LogVarA,\LogVarE\left(\CALogS\,\LogVarA\right)}\right)\right)$$
\subsection{Models of classical multisorted first-order logic}
\label{models}
In this section we recall what is a model of a given multisorted first-order theory. We fix a signature $\Sigma$.
\begin{defi}
A $\Sigma$-structure $\ModM$ is given by:
\begin{itemize}
\item a set $\ModMInterp{\SortA}$ for each sort $\SortA$ constructed from the base sorts of $\Sigma$
\item an application function from $\ModMInterp{\left(\SortA\SortTo\SortB\right)}\times\ModMInterp{\SortA}$ to $\ModMInterp{\SortB}$
\item an element $\ModMInterp{\LogConstA}\in\ModMInterp{\SortA}$ for each individual constant $\LogSortedTerm{\LogConstA}{\SortA}$ of $\Sigma$
\item a set $\ModMInterp{\LogPredA}\subseteq\ModMInterp{\SortA_1}\times\ldots\times\ModMInterp{\SortA_n}$ for each sorted predicate $\LogPredA\left(\LogSortedTerm{\LogTermA_1}{\SortA_1},\ldots,\LogSortedTerm{\LogTermA_n}{\SortA_n}\right)$ of $\Sigma$
\end{itemize}
\end{defi}
Using the application function, we can extend the interpretation to individuals with parameters: if $\LogTermA$ is an individual of sort $\SortA$ with free variables $\LogSortedTerm{\vec{\LogVarA}}{\vec{\SortB}}$ and if $\vec{\ModElemA}$ are elements of $\ModMInterp{\vec{\SortB}}$, then $\ModMInterp{\left(\LogTermA\LogSubst{\vec{\ModElemA}/\vec{\LogVarA}}\right)}$ is an element of $\ModMInterp{\SortA}$.\par
As usual in model theory, for any $\Sigma$-structure $\ModM$ and any closed formula $\LogFormA$ on $\Sigma$ we can define when $\ModM$ validates $\LogFormA$, written $\ModM\Models\LogFormA$. Now we fix a theory $\LogAxioms$ on $\Sigma$ and we define what is a model of $\LogAxioms$:
\begin{defi}
If $\ModM$ is a $\Sigma$-structure, then $\ModM$ is a model of $\LogAxioms$ if for any $\LogFormA\in\LogAxioms$:
$$\ModM\Models\LogFormA$$
\end{defi}
The soundness theorem states that if $\ModM$ is a model of $\LogAxioms$, then for any closed formula $\LogFormA$, $\LogAxioms\Derives\LogFormA$ implies $\ModM\Models\LogFormA$. In the particular case of equational theories we fix the interpretation of the $\neq_\SortA$ predicate to be the following set of pairs of elements of $\ModMInterp{\SortA}$:
$$\ModMInterp{\neq_\SortA}\quad\Def\quad\SetSuch{\left(\ModElemA,\ModElemB\right)\in\ModMInterp{\SortA}\times\ModMInterp{\SortA}}{\ModElemA\neq\ModElemB}$$
\section{Syntax and semantics of \texorpdfstring{$\lambda\mu$}{lambda-mu}-calculus}
\label{typedLanguages}
In this section, we define the programming language to which we map classical proofs, and define the categorical model of this language. First, we describe $\lambda\mu$-calculus as a typed language and give its equational theory under call-by-name semantics before defining how we map classical proofs in this language. Then we define the language $\mu$PCF (that we will use in order to realize Peano arithmetic and the axiom of choice) as a $\lambda\mu$-theory. Finally we describe categories of continuations as a model of call-by-name $\lambda\mu$-calculus and we present the connection between the interpretation of $\lambda\mu$-calculus in categories of continuations and that of its continuation-passing-style translation in the underlying cartesian category with exponentials of a fixed object.
\subsection{\texorpdfstring{$\lambda\mu$}{lambda-mu}-calculus}
The $\lambda\mu$-calculus is an extension of the $\lambda$-calculus introduced by Parigot in~\cite{ParigotLambdaMu} in order to represent and evaluate classical proofs. In $\lambda\mu$-calculus, there is another kind of variables along standard $\lambda$-variables: the $\mu$-variables. Just like the $\lambda$-variables are bound by the construct $\lambda\LmVarA.\LmTermA$, the $\mu$-variables (which will be written $\alpha,\beta,\ldots$) are bound by the new construct of $\lambda\mu$-calculus: $\mu\LmMVarA.\LmTermA$. The other new construct, $\left[\LmMVarA\right]\,\LmTermA$ should be understood as a mean for $\LmTermA$ to get arguments which are passed to the enclosing $\mu\LmMVarA.\LmTermB$.\par
$\lambda\mu$-calculus provides a direct interpretation of classical proofs just like $\lambda$-calculus is an interpretation of the intuitionistic ones. In classical sequent calculus, sequents are of the form $\Sequent{\LogFormA_1,\ldots,\LogFormA_n}{\LogFormB_1,\ldots,\LogFormB_m}{}$, and should be interpreted as the formula $\LogFormA_1\LogAnd\ldots\LogAnd\LogFormA_n\LogImp\LogFormB_1\vee\ldots\vee\LogFormB_m$. The interpretation of a proof of such a sequent in $\lambda\mu$-calculus is then a $\lambda\mu$-term with free $\lambda$-variables $\LmVarA_1,\ldots,\LmVarA_n$, and free $\mu$-variables $\LmMVarA_1,\ldots,\LmMVarA_m$, $\LmVarA_i$ representing the possibility to use the hypothesis $\LogFormA_i$, and $\LmMVarA_i$ representing the possibility to produce (part of) a proof of $\LogFormB_i$. Under this view, a $\lambda\mu$-term may provide many proofs at the same time, each proof being given step-by-step.\par
In the original version of~\cite{ParigotLambdaMu}, the terms were restricted to $\lambda$-abstractions, applications and $\mu\LmMVarA.\left[\LmMVarB\right]\,\LmTermA$ for some term $\LmTermA$. This syntax was extended in~\cite{DeGrooteLambdaMu} by allowing terms $\mu\LmMVarA.\LmTermA$ and $\left[\LmMVarA\right]\,\LmTermA$. This extension of syntax was used in~\cite{SaurinLambdaMu} together with a well-chosen set of reduction rules (this calculus being called $\Lambda\mu$-calculus) to recover the separation property that failed in Parigot's $\lambda\mu$-calculus, as shown in~\cite{DavidPyBohm}. Later on in~\cite{ArilaHerbelinSabryContinuations}, Parigot's version was related to minimal classical logic, and De Groote's to full classical logic. Here we use the version of~\cite{SelingerControl} (but without disjunction types), which is in the lines of~\cite{DeGrooteLambdaMu,SaurinLambdaMu}. We recommend~\cite{HerbelinSaurinLambdaMu} for an historical overview of the different versions of $\lambda\mu$-calculus and how they relate to each other.
\subsubsection{Type system}
\label{typeSystem}
The types of $\lambda\mu$-calculus are those of simply typed $\lambda$-calculus with a fixed set of base types (ranged over by $\LmSortBase$) together with a product type and a distinguished empty type $\LmSortBot$ used to type terms $\left[\LmMVarA\right]\,\LmTermA$:
$$\LmSortA,\LmSortB\GramDef\LmSortBase\BarSep\LmSortA\LmSortTo\LmSortB\BarSep\LmSortA\LmSortTimes\LmSortB\BarSep\LmSortBot$$
In addition to the base types, $\lambda\mu$-calculus is parameterized with a fixed set of typed constants: $\LmConsts=\SetSuch{\LmTerm{\LmConstA}{\LmSortA},\ldots}{}$. These two parameters form a signature of $\lambda\mu$-calculus:
\begin{defi}[$\lambda\mu$ signature]
A $\lambda\mu$ signature is a set of base types together with a set of typed constants.
\end{defi}
From a $\lambda\mu$ signature, we will now define the derivable typing judgments, which are presented as sequents of the form:
$$\Sequent{\LmTerm{\LmVarA_1}{\LmSortA_1},\ldots,\LmTerm{\LmVarA_n}{\LmSortA_n}}{\LmTerm{\LmTermA}{\LmSortA}}{\LmTerm{\LmMVarA_1}{\LmSortB_1},\ldots,\LmTerm{\LmMVarA_m}{\LmSortB_m}}$$
where the free $\lambda$-variables of $\LmTermA$ are among $\LmVarA_1,\ldots,\LmVarA_n$ and its free $\mu$-variables are among $\LmMVarA_1,\ldots,\LmMVarA_m$. The $\LmTerm{\LmVarA_1}{\LmSortA_1},\ldots,\LmTerm{\LmVarA_n}{\LmSortA_n}$ part will be called the $\lambda$-context, and $\LmTerm{\LmMVarA_1}{\LmSortB_1},\ldots,\LmTerm{\LmMVarA_m}{\LmSortB_m}$ the $\mu$-context. If the $\mu$-context is empty, then we will simply write:
$$\Sequent{\LmTerm{\LmVarA_1}{\LmSortA_1},\ldots,\LmTerm{\LmVarA_n}{\LmSortA_n}}{\LmTerm{\LmTermA}{\LmSortA}}{}$$
The derivable typing judgments are defined in Figure~\ref{TypingRules} for a given signature of $\lambda\mu$-calculus.
\begin{figure}
\begin{gather*}
\LmRuleAx{\LmTerm{\vec{\LmVarA}}{\vec{\LmSortA}}}{\LmTerm{\vec{\LmMVarA}}{\vec{\LmSortB}}}{\LmVarA}{\LmSortA}
\qquad
\LmRuleAxSig{\LmTerm{\vec{\LmVarA}}{\vec{\LmSortA}}}{\LmTerm{\vec{\LmMVarA}}{\vec{\LmSortB}}}{\LmConstA}{\LmSortA}
\\[5pt]
\LmRuleImpIntro{\LmTerm{\vec{\LmVarA}}{\vec{\LmSortA}}}{\LmTerm{\vec{\LmMVarA}}{\vec{\LmSortB}}}{\LmVarA}{\LmSortA}{\LmTermA}{\LmSortB}
\qquad
\LmRuleImpElim{\LmTerm{\vec{\LmVarA}}{\vec{\LmSortA}}}{\LmTerm{\vec{\LmMVarA}}{\vec{\LmSortB}}}{\LmTermA}{\LmSortB}{\LmTermB}{\LmSortA}
\\[5pt]
\LmRuleAndIntro{\LmTerm{\vec{\LmVarA}}{\vec{\LmSortA}}}{\LmTerm{\vec{\LmMVarA}}{\vec{\LmSortB}}}{\LmTermA}{\LmSortA}{\LmTermB}{\LmSortB}
\qquad
\LmRuleAndElim{\LmTerm{\vec{\LmVarA}}{\vec{\LmSortA}}}{\LmTerm{\vec{\LmMVarA}}{\vec{\LmSortB}}}{\LmTermA}{\LmSortA}
\\[5pt]
\LmRuleBotIntro{\LmTerm{\vec{\LmVarA}}{\vec{\LmSortA}}}{\LmTerm{\vec{\LmMVarA}}{\vec{\LmSortB}}}{\LmMVarA}{\LmTermA}{\LmSortB}
\qquad
\LmRuleBotElim{\LmTerm{\vec{\LmVarA}}{\vec{\LmSortA}}}{\LmTerm{\vec{\LmMVarA}}{\vec{\LmSortB}}}{\LmMVarA}{\LmTermA}{\LmSortB}
\end{gather*}
\caption{Typing rules of $\lambda\mu$-calculus}
\label{TypingRules}
\end{figure}
The left and right weakening rules are admissible in that type system, and we use them without explicitly mentioning it. Here are some examples of derivable typing judgments.
\begin{gather*}
\Sequent{}{\LmTerm{\lambda\LmVarA.\mu\LmMVarA.\LmVarA}{\LmSortBot\LmSortTo\LmSortA}}{}\qquad\Sequent{}{\LmTerm{\lambda\LmVarB.\mu\LmMVarA.\LmVarB\left(\lambda\LmVarA.[\LmMVarA]\,\LmVarA\right)}{\left(\left(\LmSortA\LmSortTo\LmSortBot\right)\LmSortTo\LmSortBot\right)\LmSortTo\LmSortA}}{}\\[5pt]
\Sequent{}{\LmTerm{\lambda\LmVarB.\mu\LmMVarA.[\LmMVarA]\,\LmVarB\left(\lambda\LmVarA.\mu\LmMVarB.[\LmMVarA]\,\LmVarA\right)}{\left(\left(\LmSortA\LmSortTo\LmSortB\right)\LmSortTo\LmSortA\right)\LmSortTo\LmSortA}}{}
\end{gather*}
The logical counterparts of these terms are respectively the ex falso formula $\LogBot\LogImp\LogFormA$, the double-negation elimination $\left(\left(\LogFormA\LogImp\LogBot\right)\LogImp\LogBot\right)\LogImp\LogFormA$ and the law of Peirce $\left(\left(\LogFormA\LogImp\LogFormB\right)\LogImp\LogFormA\right)\LogImp\LogFormA$, these last two being classical principles. Remark that in the first judgment we write $\mu\LmMVarA.\LmVarA$, and in the second example we write $\lambda\LmVarA.[\LmMVarA]\,\LmVarA$. This is a consequence of using the extended syntax of \cite{DeGrooteLambdaMu,SaurinLambdaMu}, where the empty type is handled as any other. The third term corresponds to the \texttt{call/cc} operator of the Scheme programming language which, as observed by Griffin in~\cite{GriffinControl}, can be typed with the law of Peirce.
\subsubsection{Interpreting classical logic in \texorpdfstring{$\lambda\mu$}{lambda-mu}-calculus}
\label{LmInterp}
Here we interpret every formula $\LogFormA$ as a type $\LmInterpForm{\LogFormA}$ of $\lambda\mu$-calculus, and every proof of a sequent $\Sequent{\Gamma}{\LogFormA}{\Delta}$ in a given theory as a typing derivation of a term $\Sequent{\LmInterpForm{\Gamma}}{\LmTerm{\LmTermA}{\LmInterpForm{\LogFormA}}}{\LmInterpForm{\Delta}}$. The interpratation of the axioms is for the time being a parameter of the interpretation of proofs, which is the first component of our realizability interpretation of classical logic.\par
Fix a first-order signature $\Sigma$ and a $\lambda\mu$ signature. Fix also for each predicate $\LogPredA$ of $\Sigma$ a type $\LmInterpForm{\LogPredA}$ of $\lambda\mu$-calculus. We extend this interpretation to every formula over $\Sigma$ the following way:
$$\LmInterpForm{\LogBot}=\LmSortBot\qquad\LmInterpForm{\left(\LogFormA\LogImp\LogFormB\right)}=\LmInterpForm{\LogFormA}\LmSortTo\LmInterpForm{\LogFormB}\qquad\LmInterpForm{\left(\LogFormA\LogAnd\LogFormB\right)}=\LmInterpForm{\LogFormA}\LmSortTimes\LmInterpForm{\LogFormB}\qquad\LmInterpForm{\left(\forall\LogSortedTerm{\LogVarA}{\SortA}\,\LogFormA\right)}=\LmInterpForm{\LogFormA}$$
An important fact is that during this interpretation we simply forget about first-order quantifications. This is a Curry-style interpretation (as opposed to a Church-style interpretation). The effect of this is that the term interpreting a proof of a universal statement must not depend on the particular individual. This is indeed true for the equational axioms, but when it comes to the axiom schemes of induction and choice it is no longer the case. For that reason we will introduce in section~\ref{RelPred} a relativization predicate. The interpretation of a formula as a type of $\lambda\mu$-calculus is then extended to contexts: $\LmInterpForm{\Gamma}$ (resp. $\LmInterpForm{\Delta}$) is a context of $\lambda$-variables (resp. $\mu$-variables) with types $\LmInterpForm{\LogFormB}$ for $\LogFormB$ in $\Gamma$ (resp. $\Delta$).\par
Fix now a first-order theory $\LogAxioms$ and for each $\LogFormA\in\LogAxioms$ a closed $\lambda\mu$-term $\LmInterpAxiom{\LogFormA}$ of type $\LmInterpForm{\LogFormA}$ which is a parameter of the interpretation. The interpretation of first-order proofs as typing derivations in $\lambda\mu$-calculus is given in Figure~\ref{ProofInterp}. In this interpretation, the structural rules on the right part of a sequent are interpreted with the $\mu$ constructs of $\lambda\mu$-calculus. Indeed, these rules are the ones which make our system a classical one, so it doesn't come as a surprise that they are interpreted using the control features of $\lambda\mu$-calculus.
\begin{figure}
\begin{align*}
\LmInterpProof{\left(\LogRuleAx{\Gamma}{\Delta}{\LogFormA}\right)}&=\LmRuleAx{\LmInterpForm{\Gamma}}{\LmInterpForm{\Delta}}{\LmVarA}{\LmInterpForm{\LogFormA}}\\[10pt]
\LmInterpProof{\left(\LogRuleAxSig{\Gamma}{\Delta}{\LogFormA}\right)}&=\Sequent{\LmInterpForm{\Gamma}}{\LmTerm{\LmInterpAxiom{\LogFormA}}{\LmInterpForm{\LogFormA}}}{\LmInterpForm{\Delta}}\\[10pt]
\LmInterpProof{\left(\LogRuleImpIntro{\Gamma}{\Delta}{\LogFormA}{\LogFormB}\right)}&=\LmRuleImpIntro{\LmInterpForm{\Gamma}}{\LmInterpForm{\Delta}}{\LmVarA}{\LmInterpForm{\LogFormA}}{\LmTermA}{\LmInterpForm{\LogFormB}}\\[10pt]
\LmInterpProof{\left(\LogRuleImpElim{\Gamma}{\Delta}{\LogFormA}{\LogFormB}\right)}&=\LmRuleImpElim{\LmInterpForm{\Gamma}}{\LmInterpForm{\Delta}}{\LmTermA}{\LmInterpForm{\LogFormB}}{\LmTermB}{\LmInterpForm{\LogFormA}}\\[10pt]
\LmInterpProof{\left(\LogRuleAndIntro{\Gamma}{\Delta}{\LogFormA}{\LogFormB}\right)}&=\LmRuleAndIntro{\LmInterpForm{\Gamma}}{\LmInterpForm{\Delta}}{\LmTermA}{\LmInterpForm{\LogFormA}}{\LmTermB}{\LmInterpForm{\LogFormB}}\\[10pt]
\LmInterpProof{\left(\LogRuleAndElim{\Gamma}{\Delta}{\LogFormA}\right)}&=\LmRuleAndElim{\LmInterpForm{\Gamma}}{\LmInterpForm{\Delta}}{\LmTermA}{\LmInterpForm{\LogFormA}}\\[10pt]
\LmInterpProof{\left(\LogRuleForallIntro{\Gamma}{\Delta}{\LogSortedTerm{\LogVarA}{\SortA}}{\LogFormA}\right)}&=\AXM{\Sequent{\LmInterpForm{\Gamma}}{\LmTerm{\LmTermA}{\LmInterpForm{\LogFormA}}}{\LmInterpForm{\Delta}}}\DP\\[10pt]
\LmInterpProof{\left(\LogRuleForallElim{\Gamma}{\Delta}{\LogSortedTerm{\LogVarA}{\SortA}}{\LogSortedTerm{\LogTermA}{\SortA}}{\LogFormA}\right)}&=\AXM{\Sequent{\LmInterpForm{\Gamma}}{\LmTerm{\LmTermA}{\LmInterpForm{\LogFormA}}}{\LmInterpForm{\Delta}}}\DP\\[10pt]
\LmInterpProof{\left(\LogRuleBotIntro{\Gamma}{\Delta}{\LogFormA}\right)}&=\LmRuleBotIntro{\LmInterpForm{\Gamma}}{\LmInterpForm{\Delta}}{\LmMVarA}{\LmTermA}{\LmInterpForm{\LogFormA}}\\[10pt]
\LmInterpProof{\left(\LogRuleBotElim{\Gamma}{\Delta}{\LogFormA}\right)}&=\LmRuleBotElim{\LmInterpForm{\Gamma}}{\LmInterpForm{\Delta}}{\LmMVarA}{\LmTermA}{\LmInterpForm{\LogFormA}}
\end{align*}
\caption{Interpretation of classical proofs in $\lambda\mu$-calculus}
\label{ProofInterp}
\end{figure}
\subsubsection{\texorpdfstring{$\lambda\mu$}{lambda-mu} theories}
\label{lmtheo}
We follow~\cite{SelingerControl} and define $\lambda\mu$-calculus as an equational theory. We only consider here the case of call-by-name semantics. The axioms of the call-by-name $\lambda\mu$-calculus are given in Figure~\ref{LambdaMuAxioms}, where in each equation the two terms are typed with the same type.
\begin{figure}
\begin{alignat*}{5}
&\left(\beta_\LmSortTo\right)\quad&\left(\lambda\LmVarA.\LmTermA\right)\LmTermB&=\LmTermA\LmSubst{\LmTermB/\LmVarA}&\qquad\qquad&\left(\eta_\LmSortTo\right)\quad&\lambda\LmVarA.\LmTermA\,\LmVarA&=\LmTermA&\qquad&\left(\LmVarA\notin\FV{\LmTermA}\right)\\
&\left(\beta_\LmSortTimes\right)\quad&\LmProj_i\LmPair{\LmTermA_1}{\LmTermA_2}&=\LmTermA_i&&\left(\eta_\LmSortTimes\right)\quad&\LmPair{\LmProj_1\,\LmTermA}{\LmProj_2\,\LmTermA}&=\LmTermA\\
&\left(\beta_\LmSortBot\right)\quad&\left[\LmMVarA\right]\,\mu\LmMVarB.\LmTermA&=\LmTermA\LmSubst{\LmMVarA/\LmMVarB}&&\left(\eta_\LmSortBot\right)\quad&\mu\LmMVarA.\left[\LmMVarA\right]\,\LmTermA&=\LmTermA&\qquad&\left(\LmMVarA\notin\FV{\LmTermA}\right)\\
\end{alignat*}
\begin{alignat*}{3}
&\left(\zeta_\LmSortTo\right)\quad&\left(\mu\LmMVarA.\LmTermA\right)\LmTermB&=\mu\LmMVarA.\LmTermA\LmSubst{\left[\LmMVarA\right]\,\underline{\ \ }\,\LmTermB/\left[\LmMVarA\right]\,\underline{\ \ }\,}\\
&\left(\zeta_\LmSortTimes\right)\quad&\LmProj_i\left(\mu\LmMVarA.\LmTermA\right)&=\mu\LmMVarA.\LmTermA\LmSubst{\left[\LmMVarA\right]\,\LmProj_i\,\underline{\ \ }\,/\left[\LmMVarA\right]\,\underline{\ \ }\,}\\
&\left(\zeta_\LmSortBot\right)\quad&\mu\LmMVarA.\LmTermA&=\LmTermA\LmSubst{\,\underline{\ \ }\,/\left[\LmMVarA\right]\,\underline{\ \ }\,}&\qquad&\left(\LmTerm{\LmMVarA}{\LmSortBot}\right)
\end{alignat*}
\caption{Axioms of the call-by-name $\lambda\mu$-calculus}
\label{LambdaMuAxioms}
\end{figure}
In the equations $\left(\zeta_\LmSortTo\right)$, $\left(\zeta_\LmSortTimes\right)$ and $\left(\zeta_\LmSortBot\right)$, $\underline{\ \ }$ is a placeholder for the term coming after $\left[\LmMVarA\right]$, i.e. $\LmTermA\LmSubst{\left[\LmMVarA\right]\,\underline{\ \ }\,\LmTermB/\left[\LmMVarA\right]\,\underline{\ \ }\,}$ is obtained by replacing in $\LmTermA$ all the subterms of the form $\left[\LmMVarA\right]\,\LmTermC$ with $\left[\LmMVarA\right]\,\LmTermC\,\LmTermB$. From these axioms  we define the notion of $\lambda\mu$ theory:
\begin{defi}[$\lambda\mu$ theory]
A $\lambda\mu$ theory is a set of equations between typed terms of the same type (with free variables of the same type) which contains the axioms of call-by-name $\lambda\mu$-calculus and is a congruence (contextually-closed equivalence relation).
\end{defi}
We use here a slightly different set of axioms from that of Selinger~\cite{SelingerControl}. Our $\beta_0$ and $\eta_0$ equations are the $\beta_\mu$ and $\eta_\mu$ equations of Selinger, and we replace his $\beta_\bot$ (which is $\left[\LmMVarA\right]\,\LmTermA=\LmTermA$ if $\LmTerm{\LmTermA}{\LmSortBot}$) with our $\zeta_\LmSortBot$. However, the two systems are equivalent:
\begin{lem}
Under the contextual closures of $\beta_\LmSortBot$ and $\eta_\LmSortBot$ (the $\beta_\mu$ and $\eta_\mu$ of Selinger), the equation $\beta_\bot$ of Selinger is equivalent to $\zeta_\LmSortBot$.
\end{lem}
\proof
Suppose $\beta_\bot$ holds, let $\LmTerm{\LmTermA}{\LmSortBot}$ and $\LmMVarA$ be a $\mu$-variable of type $\LmSortBot$. Using the contextual closure of $\beta_\bot$, we have $\LmTermA=\LmTermA\LmSubst{\,\underline{\ \ }\,/\left[\LmMVarA\right]\,\underline{\ \ }\,}$, and again by $\beta_\bot$ we get $\LmTermA=\left[\LmMVarA\right]\,\LmTermA\LmSubst{\,\underline{\ \ }\,/\left[\LmMVarA\right]\,\underline{\ \ }\,}$. Then by contextual closure $\mu\LmMVarA.\LmTermA=\mu\LmMVarA.\left[\LmMVarA\right]\,\LmTermA\LmSubst{\,\underline{\ \ }\,/\left[\LmMVarA\right]\,\underline{\ \ }\,}$, which is equal to $\LmTermA\LmSubst{\,\underline{\ \ }\,/\left[\LmMVarA\right]\,\underline{\ \ }\,}$ using $\eta_\LmSortBot$, since $\LmMVarA$ does not appear free in $\LmTermA\LmSubst{\,\underline{\ \ }\,/\left[\LmMVarA\right]\,\underline{\ \ }\,}$.\par
Conversely, suppose $\zeta_\LmSortBot$ holds, let $\LmTerm{\LmTermA}{\LmSortBot}$, $\LmMVarA$ be a $\mu$-variable of type $\LmSortBot$ and $\beta$ be a $\mu$-variable of type $\LmSortBot$ which does not appear free in $\LmTermA$. Using $\zeta_\LmSortBot$ we have $\LmTermA=\mu\beta.\LmTermA$, so by contextual closure $\left[\LmMVarA\right]\,\LmTermA=\left[\LmMVarA\right]\,\mu\LmMVarB.\LmTermA$, so by $\beta_\LmSortBot$ we get $\left[\LmMVarA\right]\,\LmTermA=\LmTermA\LmSubst{\LmMVarA/\LmMVarB}$, but $\LmTermA\LmSubst{\LmMVarA/\LmMVarB}=\LmTermA$ since $\beta$ does not appear free in $\LmTermA$, and we get finally $\left[\LmMVarA\right]\,\LmTermA=\LmTermA$.
\qed
Replacing Selinger's $\beta_\bot$ allows us to have a more symmetric calculus, with a set of three equations for each type constructor, the $\zeta$ equations representing the transmission of the context of a $\mu\LmMVarA.\LmTermA$ to the subterms $\left[\LmMVarA\right]\,\LmTermB$.
\subsubsection{\texorpdfstring{$\mu$}{mu}PCF}
\label{muPCF}
We give here an example of a $\lambda\mu$ theory called $\mu$PCF, which will also be the language in which we will interpret Peano arithmetic and the axiom of choice. Programming language for Computable Functions (PCF) is a functional programming language described by Plotkin in~\cite{PlotkinPCF}. It is based on Scott's Logic for Computable Functions (LCF), which was presented in~\cite{ScottLCF}. The language contains constants for natural numbers and general recursion. It is probably the simplest example of a Turing-complete higher-order language. Here we consider an extension of PCF to primitively handle control operators, by presenting PCF as a $\lambda\mu$ theory. In \cite{OngStewartControl} the authors define a call-by-value semantics for $\lambda\mu$-calculus, and they illustrate it with $\mu$PCF$_V$, a call-by-value version of PCF with control. Later on, Laird defines in \cite{LairdThesis} its call-by-name version, which is the version we use here. The choice of this language is justified by our will to get computational content directly from classical proofs.\par
Our version of $\mu$PCF has only one base type for natural numbers: $\CALmnSort$, and the constants are:
\begin{gather*}
\LmTerm{\CALmn{n}}{\CALmnSort}\qquad\LmTerm{\CALmsucc}{\CALmsuccSort}\qquad\LmTerm{\CALmpred}{\CALmsuccSort}\qquad\LmTerm{\CALmifz}{\CALmifzSort{\LmSortA}}\qquad\LmTerm{\CALmfix}{\CALmfixSort{\LmSortA}}
\end{gather*}
It will also be useful to have a canonical term on each type so we define:
$$\CALmom\Def\LmTerm{\CALmfix\left(\lambda x.x\right)}{\LmSortA}$$
This term represents non-termination, or ``undefined''. The equations of $\mu$PCF given in Figure~\ref{muPCFEq} are standard and include the interactions of the constants with the $\mu$ operator.
\begin{figure}
\begin{align*}
\CALmfix\,\LmTermA&=\LmTermA\left(\CALmfix\,\LmTermA\right)
&
\CALmsucc\,\CALmn{n}&=\CALmn{n+1}
&
\CALmsucc\left(\mu\LmMVarA.\LmTermA\right)&=\LmTermA\LmSubst{\CALmsucc\left(\underline{\ \ }\right)/\left[\LmMVarA\right]\,\underline{\ \ }\,}
\\
\CALmpred\,\CALmn{0}&=\CALmn{0}
&
\CALmpred\,\CALmn{n+1}&=\CALmn{n}
&
\CALmpred\left(\mu\LmMVarA.\LmTermA\right)&=\LmTermA\LmSubst{\CALmpred\left(\underline{\ \ }\right)/\left[\LmMVarA\right]\,\underline{\ \ }\,}
\\
\CALmifz\,\CALmn{0}\,\LmTermA\,\LmTermB&=\LmTermA
&
\CALmifz\,\CALmn{n+1}\,\LmTermA\,\LmTermB&=\LmTermB
&
\CALmifz\left(\mu\LmMVarA.\LmTermA\right)\,\LmTermB\,\LmTermC&=\LmTermA\LmSubst{\CALmifz\left(\underline{\ \ }\right)\,\LmTermB\,\LmTermC\,/\left[\LmMVarA\right]\,\underline{\ \ }\,}
\end{align*}
\caption{Equations of $\mu$PCF}
\label{muPCFEq}
\end{figure}
We can finally define $\mu$PCF as the $\lambda\mu$ theory generated by these equations:
\begin{defi}[$\mu$PCF]
The $\lambda\mu$ theory $\mu$PCF is the smallest $\lambda\mu$ theory containing the equations of Figure~\ref{muPCFEq}.
\end{defi}
\subsection{Categories of continuations}
\label{catcont}
Categories of continuations are to call-by-name $\lambda\mu$-calculus what cartesian closed categories are to $\lambda$-calculus, in the sense that if we fix a signature, there is a one-to-one correspondence between $\lambda\mu$ theories and categories of continuations together with an interpretation of the signature. Ong defines $\lambda\mu$ categories in~\cite{OngClassicalProofs} by reformulating the syntax of $\lambda\mu$-calculus in categorical terms. Later on, Hofmann and Streicher proved the soundness and completeness of categories of continuations with respect to $\lambda\mu$-calculus, providing the first abstract version of $\lambda\mu$-categories. Finally, Selinger axiomatized these in~\cite{SelingerControl} under the name of control categories, proving that they are equivalent (for a suitable notion of equivalence based on weak functors) to categories of continuations, and therefore sound and complete with respect to call-by-name $\lambda\mu$-calculus. Moreover, he proved that the categorical dual of control categories are sound and complete with respect to call-by-value $\lambda\mu$-calculus. Here we are only interested in the call-by-name version, and we use the model of categories of continuations.
\begin{defi}[Category of continuations]
Let $\CatC$ be a distributive category, that is, a category with finite products and coproducts such that the canonical distributivity morphisms from $\CatObjA\CatTimes\CatObjB\CatPlus\CatObjA\CatTimes\CatObjC$ to $\CatObjA\CatTimes\left(\CatObjB\CatPlus\CatObjC\right)$ are isomorphisms (which implies that the morphism from $\CatInit$ to $\CatObjA\CatTimes\CatInit$ is also an isomorphism), and let $\CatR\in\CatObj{\CatC}$ be a fixed object such that all exponentials $\CatExp{\CatR}{\CatObjA}$ for $\CatObjA\in\CatObj{\CatC}$ exist. Then the full subcategory $\CatRC$ of $\CatC$ consisting of the objects $\CatExp{\CatR}{\CatObjA}$ for $\CatObjA\in\CatC$ is called a category of continuations.
\end{defi}
We will differentiate morphisms in $\CatC$ and $\CatRC$ by writing $\CatCHomA$, $\CatCHomB$\footnote{Pronounced as in $\MCYRD\mcyro\mcyrk\mcyrt\mcyro\mcyrr$ $\MCYRZH\mcyri\mcyrv\mcyra\mcyrg\mcyro$ (Doctor Zhivago).}  for morphisms in $\CatC$ and $\CatRCHomA$, $\CatRCHomB$, $\CatRCHomC$, $\CatRCHomD$ for morphisms in $\CatRC$. As observed in~\cite{LafontReusStreicher}, a category of continuations $\CatRC$ is in particular a cartesian closed category:
\begin{lem}
If $\CatRC$ is a category of continuations, then $\CatExp{\CatR}{\CatInit}$ defines a terminal object and $\CatExp{\CatR}{\CatObjA\CatPlus\CatObjB}$ defines a cartesian product of $\CatExp{\CatR}{\CatObjA}$ and $\CatExp{\CatR}{\CatObjB}$, so $\CatRC$ is cartesian. Moreover, if $\CatObjA\in\CatObj{\CatC}$ and $\CatExp{\CatR}{\CatObjB}\in\CatObj{\CatRC}$, then $\CatExp{\CatR}{\CatObjA\CatTimes\CatObjB}$ defines an exponential in $\CatC$ of $\CatExp{\CatR}{\CatObjB}$ by $\CatObjA$. Consequently, $\CatRC$ is cartesian closed, the exponential of $\CatExp{\CatR}{\CatObjB}$ by $\CatExp{\CatR}{\CatObjA}$ being $\CatExp{\CatR}{\CatExp{\CatR}{\CatObjA}\CatTimes\CatObjB}$.
\end{lem}
Be careful that we have two (isomorphic) terminal objects, one in $\CatC$ and one in $\CatRC$, and two (isomorphic) products of $\CatExp{\CatR}{\CatObjA}$ and $\CatExp{\CatR}{\CatObjB}$, again one in $\CatC$ and one in $\CatRC$. To avoid confusion and without loss of generality we will suppose that they are equal: $\CatExp{\CatR}{\CatInit}=\CatTerm$ and $\CatExp{\CatR}{\CatObjA\CatPlus\CatObjB}=\CatExp{\CatR}{\CatObjA}\CatTimes\CatExp{\CatR}{\CatObjB}$. For the same reason, we also suppose $\CatR=\CatExp{\CatR}{\CatTerm}$.
\subsubsection{Classical disjunction in categories of continuations.} We could have added a primitive connective $\vee$ for the disjunction in the logic, with the following rules:
$$\AXM{\Sequent{\Gamma}{\LogFormA\vee\LogFormB}{\Delta,\LogFormA,\LogFormB}}\UIM{\Sequent{\Gamma}{\LogBot}{\Delta,\LogFormA,\LogFormB}}\DP\qquad\qquad\qquad\AXM{\Sequent{\Gamma}{\LogBot}{\Delta,\LogFormA,\LogFormB}}\UIM{\Sequent{\Gamma}{\LogFormA\vee\LogFormB}{\Delta}}\DP$$
and then interpret these logical rules by adding the following typing rules to $\lambda\mu$-calculus:
$$\AXM{\Sequent{\LmTerm{\vec{\LmVarA}}{\vec{\LmSortA}}}{\LmTerm{\LmTermA}{\LmSortB_1\parr\LmSortB_2}}{\LmTerm{\vec{\LmMVarA}}{\vec{\LmSortB}},\LmTerm{\LmMVarB_1}{\LmSortB_1},\LmTerm{\LmMVarB_2}{\LmSortB_2}}}\UIM{\Sequent{\LmTerm{\vec{\LmVarA}}{\vec{\LmSortA}}}{\LmTerm{\left[\LmMVarB_1,\LmMVarB_2\right]\,\LmTermA}{\LmSortBot}}{\LmTerm{\vec{\LmMVarA}}{\vec{\LmSortB}},\LmTerm{\LmMVarB_1}{\LmSortB_1},\LmTerm{\LmMVarB_2}{\LmSortB_2}}}\DP\qquad\AXM{\Sequent{\LmTerm{\vec{\LmVarA}}{\vec{\LmSortA}}}{\LmTerm{\LmTermA}{\LmSortBot}}{\LmTerm{\vec{\LmMVarA}}{\vec{\LmSortB}},\LmTerm{\LmMVarB_1}{\LmSortB_1},\LmTerm{\LmMVarB_2}{\LmSortB_2}}}\UIM{\Sequent{\LmTerm{\vec{\LmVarA}}{\vec{\LmSortA}}}{\LmTerm{\mu\left(\LmMVarB_1,\LmMVarB_2\right).\LmTermA}{\LmSortB_1\parr\LmSortB_2}}{\LmTerm{\vec{\LmMVarA}}{\vec{\LmSortB}}}}\DP$$
These rules are present in~\cite{SelingerControl}, however we choose here to keep things simple and stick to the usual $\lambda\mu$-calculus without disjunction types. Nevertheless we still use the binoidal functor $\CatPar$ in categories of continuations to interpret multi-conclusioned sequents. Following~\cite{SelingerControl} we write: $\CatExp{\CatR}{\CatObjA}\CatPar\CatExp{\CatR}{\CatObjB}\Def\CatExp{\CatR}{\CatObjA\CatTimes\CatObjB}$ for the interpretation of the classical disjunction between $\CatExp{\CatR}{\CatObjA}$ and $\CatExp{\CatR}{\CatObjB}$, $i:\CatR\to\CatExp{\CatR}{\CatObjA}$ for the interpretation of the right weakening rule and $\nabla:\CatExp{\CatR}{\CatObjA}\CatPar\CatExp{\CatR}{\CatObjA}\to\CatExp{\CatR}{\CatObjA}$ for the interpretation of the right contraction rule.\par
Another interesting fact about categories of continuations is that we can define a functor from $\CatC^{op}$ to $\CatRC$ which maps $\CatObjA\in\CatObj{\CatC}$ to $\CatExp{\CatR}{\CatObjA}\in\CatObj{\CatRC}$, and $\CatCHomA:\CatObjA\to\CatObjB$ to $\CatExp{\CatR}{\CatCHomA}:\CatExp{\CatR}{\CatObjB}\to\CatExp{\CatR}{\CatObjA}$ which is the currying of:
$$\xymatrix{\CatExp{\CatR}{\CatObjB}\CatTimes\CatObjA\ar[rr]^-{\CatId{\CatExp{\CatR}{\CatObjB}}\CatTimes\CatCHomA}&\hspace{10pt}&\CatExp{\CatR}{\CatObjB}\CatTimes\CatObjB\ar[r]^-{\CatEval}&\CatR}$$
The morphism $i:\CatR\to\CatExp{\CatR}{\CatObjA}$ corresponds to $\CatExp{\CatR}{\CatTermMorph{\CatObjA}}$ (where $\CatTermMorph{\CatObjA}$ is the unique morphism from $\CatObjA$ to the terminal object $\CatTerm$) and the morphism $\nabla:\CatExp{\CatR}{\CatObjA}\CatPar\CatExp{\CatR}{\CatObjA}\to\CatExp{\CatR}{\CatObjA}$ corresponds to $\CatExp{\CatR}{\CatPair{\CatId{\CatObjA}}{\CatId{\CatObjA}}}$ (remember that $\CatExp{\CatR}{\CatObjA}\CatPar\CatExp{\CatR}{\CatObjA}=\CatExp{\CatR}{\CatObjA\CatTimes\CatObjA}$). Also, in particular, if $\CatObjA\simeq\CatObjB$ in $\CatC$, then $\CatExp{\CatR}{\CatObjA}\simeq\CatExp{\CatR}{\CatObjB}$ in $\CatRC$ (and in $\CatC$). Through this functor, the cocartesian structure of $\CatC$ translates to the cartesian structure of $\CatRC$.
\subsubsection{Interpretation of \texorpdfstring{$\lambda\mu$}{lambda-mu}-calculus in categories of continuations}
\label{CatInterp}
We describe here the interpretation of call-by-name $\lambda\mu$-calculus in a category of continuations as defined in~\cite{SelingerControl}. We fix a signature of $\lambda\mu$-calculus and a category of continuations $\CatRC$. To each type $\LmSortA$ of $\lambda\mu$-calculus we associate an object $\CatInterpSortNeg{\LmSortA}\in\CatObj{\CatC}$ of continuations of type $\LmSortA$, and an object $\CatInterpSort{\LmSortA}=\CatExp{\CatR}{\CatInterpSortNeg{\LmSortA}}\in\CatObj{\CatRC}$ of computations of type $\LmSortA$. The objects $\CatInterpSortNeg{\LmSortBase}\in\CatObj{\CatC}$ where $\LmSortBase$ is a base type of the signature are parameters of the interpretation, and we define inductively:
\begin{align*}
\CatInterpSortNeg{\LmSortA\to\LmSortB}&=\CatInterpSort{\LmSortA}\CatTimes\CatInterpSortNeg{\LmSortB}\in\CatObj{\CatC}&\CatInterpSortNeg{\LmSortA\LmSortTimes\LmSortB}&=\CatInterpSortNeg{\LmSortA}\CatPlus\CatInterpSortNeg{\LmSortB}\in\CatObj{\CatC}\\
\CatInterpSortNeg{\LmSortBot}&=\CatTerm&\CatInterpSort{\LmSortA}&=\CatExp{\CatR}{\CatInterpSortNeg{\LmSortA}}\in\CatObj{\CatRC}
\end{align*}
We have in particular $\CatInterpSort{\LmSortA\LmSortTimes\LmSortB}=\CatExp{\CatR}{\CatInterpSortNeg{\LmSortA}\CatPlus\CatInterpSortNeg{\LmSortB}}=\CatExp{\CatR}{\CatInterpSortNeg{\LmSortA}}\CatTimes\CatExp{\CatR}{\CatInterpSortNeg{\LmSortB}}=\CatInterpSort{\LmSortA}\CatTimes\CatInterpSort{\LmSortB}$ where $\CatTimes$ is the cartesian product in $\CatRC$ defined above, $\CatInterpSort{\LmSortA\to\LmSortB}=\CatExp{\CatR}{\CatInterpSort{\LmSortA}\CatTimes\CatInterpSortNeg{\LmSortB}}=\CatExp{\CatR}{\CatExp{\CatR}{\CatInterpSortNeg{\LmSortA}}\CatTimes\CatInterpSortNeg{\LmSortB}}=\CatExp{\left(\CatExp{\CatR}{\CatInterpSortNeg{\LmSortB}}\right)}{\CatExp{\CatR}{\CatInterpSortNeg{\LmSortA}}}=\CatExp{\CatInterpSort{\LmSortB}}{\CatInterpSort{\LmSortA}}$, using the definition of the exponential in $\CatRC$ given above, and $\CatInterpSort{\LmSortBot}=\CatExp{\CatR}{\CatTerm}=\CatR$.\par
Once we have an interpretation of types in $\CatRC$, we define the interpretation of typed $\lambda\mu$-terms such that a term:
$$\Sequent{\LmTerm{\LmVarA_1}{\LmSortA_1},\ldots,\LmTerm{\LmVarA_n}{\LmSortA_n}}{\LmTerm{\LmTermA}{\LmSortC}}{\LmTerm{\LmMVarA_1}{\LmSortB_1},\ldots,\LmTerm{\LmMVarA_m}{\LmSortB_m}}$$
is interpreted as a morphism in $\CatRC$:
$$\CatInterpTerm{\Sequent{\LmTerm{\LmVarA_1}{\LmSortA_1},\ldots,\LmTerm{\LmVarA_n}{\LmSortA_n}}{\LmTerm{\LmTermA}{\LmSortC}}{\LmTerm{\LmMVarA_1}{\LmSortB_1},\ldots,\LmTerm{\LmMVarA_m}{\LmSortB_m}}}:\CatInterpSort{\LmSortA_1}\CatTimes\ldots\CatTimes\CatInterpSort{\LmSortA_n}\to\CatInterpSort{\LmSortC}\CatPar\CatInterpSort{\LmSortB_1}\CatPar\ldots\CatPar\CatInterpSort{\LmSortB_m}$$
where $\CatInterpSort{\LmSortA_1}\CatTimes\ldots\CatTimes\CatInterpSort{\LmSortA_n}$ associates to the left and $\CatInterpSort{\LmSortC}\CatPar\CatInterpSort{\LmSortB_1}\CatPar\ldots\CatPar\CatInterpSort{\LmSortB_m}$ associates to the right. In order to do that, we suppose given for each constant $\LmTerm{\LmConstA}{\LmSortA}\in\LmConsts$ of the signature a morphism in $\CatRC$:
$$\CatInterpTerm{\LmTerm{\LmConstA}{\LmSortA}}:\CatTerm\to\CatInterpSort{\LmSortA}$$
which is again a parameter of the interpretation. These parameters are summarized in the following definition:
\begin{defi}[Interpretation]
Given a signature and a category of continuations $\CatRC$, an interpretation of $\lambda\mu$-calculus is given by an object $\CatInterpSortNeg{\LmSortBase}\in\CatC$ for each base type $\LmSortBase$ of the signature and a morphism $\CatInterpTerm{\LmTerm{\LmConstA}{\LmSortA}}:\CatTerm\to\CatInterpSort{\LmSortA}$ in $\CatRC$ for each constant $\LmTerm{\LmConstA}{\LmSortA}\in\LmConsts$ of the signature.
\end{defi}
We now have all necessary material to interpret every typed $\lambda\mu$-term as a morphism in $\CatRC$. The interpretation of typed $\lambda\mu$-terms is almost identical to the interpretation of $\lambda$-calculus in a cartesian closed category (since as shown in the previous section, $\CatRC$ is cartesian closed). The first difference is that we must be able to carry over the $\mu$-context, so we want to build from $\CatRCHomA:\CatInterpSort{\LmSortA}\to\CatInterpSort{\LmSortB}$ a morphism $\CatRCHomA\CatPar\CatInterpSort{\LmSortC}:\CatInterpSort{\LmSortA}\CatPar\CatInterpSort{\LmSortC}\to\CatInterpSort{\LmSortB}\CatPar\CatInterpSort{\LmSortC}$. The second difference is that in order to interpret the introduction rules for $\mu\LmMVarA.\LmTermA$ and $\left[\LmMVarA\right]\,\LmTermA$, we also need to have canonical morphisms from $\CatInterpSort{\LmSortBot}\CatPar\CatInterpSort{\vec{\LmSortB}}\CatPar\CatInterpSort{\LmSortA}$ to $\CatInterpSort{\LmSortA}\CatPar\CatInterpSort{\vec{\LmSortB}}$ and from $\CatInterpSort{\LmSortA}\CatPar\CatInterpSort{\vec{\LmSortB}}\CatPar\CatInterpSort{\LmSortA}$ to $\CatInterpSort{\LmSortBot}\CatPar\CatInterpSort{\vec{\LmSortB}}\CatPar\CatInterpSort{\LmSortA}$. These requirements are axiomatized in~\cite{SelingerControl}, to which we refer for the full definition of the interpretation, and the proof that the axioms of call-by-name $\lambda\mu$-calculus are sound under this interpretation.\par
A model of a $\lambda\mu$ theory is then a sound interpretation:
\begin{defi}[Model of a $\lambda\mu$ theory]
\label{LambdaMuModel}
A model of a $\lambda\mu$ theory is a category of continuations together with an interpretation for the $\lambda\mu$ signature such that every equation $\LmTermA=\LmTermB$ of the theory is true in the model: $\CatInterpTerm{\LmTermA}=\CatInterpTerm{\LmTermB}$.
\end{defi}
If the $\lambda\mu$ theory is generated from a given set of equations, then any interpretation satisfying these equations is a model of the $\lambda\mu$ theory.\par
Since call-by-name $\lambda\mu$-calculus is the internal language of categories of continuations (as shown in~\cite{SelingerControl}), we can apply $\lambda\mu$-calculus constructs on morphisms of $\CatRC$ through the use of $\lambda\mu$-terms with parameters in $\CatRC$. Therefore, we will also drop the interpretation brackets for terms. For example, if $\CatRCHomA:\prod_{j\in J}\CatExp{\CatR}{\CatObjA_j}\to\CatExp{\CatR}{\CatObjB}\CatPar\left(\bigparr_{k\in K}\CatExp{\CatR}{\CatObjD_k}\right)$ and $\CatRCHomB:\prod_{j\in J}\CatExp{\CatR}{\CatObjA_j}\to\CatExp{\CatR}{\CatExp{\CatR}{\CatObjB}\CatTimes\CatObjC}\CatPar\left(\bigparr_{k\in K}\CatExp{\CatR}{\CatObjD_k}\right)$, then $\CatRCHomB\,\CatRCHomA:\prod_{j\in J}\CatExp{\CatR}{\CatObjA_j}\to\CatExp{\CatR}{\CatObjC}\CatPar\left(\bigparr_{k\in K}\CatExp{\CatR}{\CatObjD_k}\right)$, where formally $\CatRCHomB\,\CatRCHomA$ is the term with parameters $\left(\LmVarB\,\LmVarA\right)\LogSubst{\CatRCHomA/\LmVarA,\CatRCHomB/\LmVarB}$.
\subsubsection{Connection with the call-by-name CPS translation of \texorpdfstring{$\lambda\mu$}{lambda-mu}-calculus}
\label{cps}
Another interesting thing about the interpretation of $\lambda\mu$-calculus into a category of continuations is the exact correspondence with the interpretation of its call-by-name CPS translation in the underlying cartesian ``$\CatR$-closed'' category, as stressed in~\cite{StreicherContinuation}. The target of such a translation is a simply-typed $\lambda$-calculus $\Lam$ with product and sum types, a particular base type $\LamTypeR$, and the function types being restricted to $\LamTypeA\LamTypeTo\LamTypeR$. This particular $\lambda$-calculus can be interpreted in $\CatC$ by interpreting the product type as the product in $\CatC$, the sum type as the coproduct, and using the fact that function types are of the form $\LamTypeA\LamTypeTo\LamTypeR$, so having all exponentials $\CatExp{\CatR}{\CatObjA}$ in $\CatC$ is enough.\par
To be more precise, $\Lam$ has one base type $\tilde{\LmSortBase}$ for each base type $\LmSortBase$ of $\lambda\mu$-calculus and another particular base type $\LamTypeR$. From these we build the types:
$$\LamTypeA,\LamTypeB\GramDef\LamCPS{\LmSortBase}\BarSep\LamTypeA\LamTypeTo\LamTypeR\BarSep\LamTypeA\LamTypeTimes\LamTypeB\BarSep\LamTypeUnit\BarSep\LamTypeA\LamTypePlus\LamTypeB$$
The arrow types are syntactically restricted to be of the form $\LamTypeA\LamTypeTo\LamTypeR$. We map every type $\LmSortA$ of $\lambda\mu$-calculus to a type $\LamCPS{\LmSortA}$ of $\Lam$ (each base type $\LmSortBase$ being obviously mapped to $\LamCPS{\LmSortBase}$) as follows:
$$\LamCPS{\LmSortA\LmSortTo\LmSortB}=\left(\LamCPS{\LmSortA}\LamTypeTo\LamTypeR\right)\LamTypeTimes\LamCPS{\LmSortB}\qquad\LamCPS{\LmSortA\LmSortTimes\LmSortB}=\LamCPS{\LmSortA}\LamTypePlus\LamCPS{\LmSortB}\qquad\LamCPS{\LmSortBot}=\LamTypeUnit$$
$\Lam$ also has one constant $\LamCPS{\LmConstA}:\LamCPS{\LmSortA}\LamTypeTo\LamTypeR$ for each constant $\LmTerm{\LmConstA}{\LmSortA}$ of the source language. We also suppose given for each $\lambda$-variable $\LmTerm{\LmVarA}{\LmSortA}$ of the source language a variable $\LamCPS{\LmVarA}:\LamCPS{\LmSortA}\LamTypeTo\LamTypeR$ in the target language, and for each $\mu$-variable $\LmTerm{\LmMVarA}{\LmSortB}$ of the source language a variable $\LamCPS{\LmMVarA}:\LamCPS{\LmSortB}$ in the target language. A typed $\lambda\mu$-term:
$$\Sequent{\LmTerm{\LmVarA_1}{\LmSortA_1},\ldots,\LmTerm{\LmVarA_n}{\LmSortA_n}}{\LmTerm{\LmTermA}{\LmSortA}}{\LmTerm{\LmMVarA_1}{\LmSortB_1},\ldots,\LmTerm{\LmMVarA_m}{\LmSortB_m}}$$
will then be translated to a typed $\lambda$-term:
\begin{equation}
\label{LamTerm}
\LamCPS{\LmVarA_1}:\LamCPS{\LmSortA_1}\LamTypeTo\LamTypeR,\ldots,\LamCPS{\LmVarA_n}:\LamCPS{\LmSortA_n}\LamTypeTo\LamTypeR,\LamCPS{\LmMVarA_1}:\LamCPS{\LmSortB_1},\ldots,\LamCPS{\LmMVarA_m}:\LamCPS{\LmSortB_m}\vdash\LamCPS{\LmTermA}:\LamCPS{\LmSortA}\LamTypeTo\LamTypeR
\end{equation}
Before defining the translation, we give the typing rules of $\Lam$ in Figure~\ref{LamRules}.
\begin{figure}
\begin{gather*}
\LmRuleAx{\LmTerm{\vec{\LamVarA}}{\vec{\LamTypeA}}}{}{\LamVarA}{\LamTypeA}
\qquad
\LmRuleAxSig{\LmTerm{\vec{\LamVarA}}{\vec{\LamTypeA}}}{}{\LamCPS{\LmConstA}}{\left(\LamCPS{\LmSortA}\LamTypeTo\LamTypeR\right)}
\\[5pt]
\LmRuleImpIntro{\LmTerm{\vec{\LamVarA}}{\vec{\LamTypeA}}}{}{\LamVarA}{\LamTypeA}{\LamTermA}{\LamTypeR}
\qquad
\LmRuleImpElim{\LmTerm{\vec{\LamVarA}}{\vec{\LamTypeA}}}{}{\LamTermA}{\LamTypeR}{\LamTermB}{\LamTypeA}
\\[5pt]
\LamRuleTop{\Gamma}
\qquad
\LmRuleAndIntro{\LmTerm{\vec{\LamVarA}}{\vec{\LamTypeA}}}{}{\LamTermA}{\LamTypeA}{\LamTermB}{\LamTypeB}
\qquad
\LmRuleAndElim{\LmTerm{\vec{\LamVarA}}{\vec{\LmSortA}}}{}{\LamTermA}{\LamTypeA}\\[5pt]
\LamRuleOrIntro{\Gamma}{\LamTermA}{\LamTypeA}
\qquad
\LamRuleOrElim{\Gamma}{\LamVarA}{\LamTermA}{\LamTermB}{\LamTypeA}{\LamTypeB}
\end{gather*}
\caption{Typing rules of $\Lam$}
\label{LamRules}
\end{figure}
The typing rules for the arrow type are restricted to the case $\LamTypeA\LamTypeTo\LamTypeR$. The translation of a variable $\LmVarA$ or a constant $\LmConstA$ is $\LamCPS{\LmVarA}$ or $\LamCPS{\LmConstA}$ as defined above, and the remaining part is given in Figure~\ref{LamTrans}, where $\LamVarA$ is always a fresh variable.
\begin{figure}
\begin{align*}
\LamCPS{\lambda\LmVarA.\LmTermA}&=\lambda\LamVarA.\left(\lambda\LamCPS{\LmVarA}.\LamCPS{\LmTermA}\right)\left(\LmProj_1\,\LamVarA\right)\left(\LmProj_2\,\LamVarA\right)&\LamCPS{\LmTermA\,\LmTermB}&=\lambda\LamVarA.\LamCPS{\LmTermA}\left\langle\LamCPS{\LmTermB},\LamVarA\right\rangle\\
\LamCPS{\LmPair{\LmTermA}{\LmTermB}}&=\lambda\LamVarA.\LamCase{\LamVarA}{\LamVarB}{\LamCPS{\LmTermA}\,\LamVarB}{\LamCPS{\LmTermB}\,\LamVarB}&\LamCPS{\LmProj_i\,\LmTermA}&=\lambda\LamVarA.\LamCPS{\LmTermA}\,\LamIn_i\,\LamVarA\\
\LamCPS{\mu\LmMVarA.\LmTermA}&=\lambda\LamCPS{\LmMVarA}.\LamCPS{\LmTermA}\,\LamUnit&\LamCPS{\left[\LmMVarA\right]\,\LmTermA}&=\lambda\LamVarA.\LamCPS{\LmTermA}\,\LamCPS{\LmMVarA}
\end{align*}
\caption{Translation of $\lambda\mu$-calculus in $\Lam$}
\label{LamTrans}
\end{figure}
We define the interpretation of $\Lam$ in $\CatC$ by first giving an object $\CatInterpSortNeg{\LamTypeA}$ of $\CatC$ for each type $\LamTypeA$:
$$\CatInterpSortNeg{\LamCPS{\LmSortBase}}=\CatInterpSortNeg{\LmSortBase}\;\quad\CatInterpSortNeg{\LamTypeA\LamTypeTo\LamTypeR}=\CatExp{\CatR}{\CatInterpSortNeg{\LamTypeA}}\;\quad\CatInterpSortNeg{\LamTypeA\LamTypeTimes\LamTypeB}=\CatInterpSortNeg{\LamTypeA}\LamTypeTimes\CatInterpSortNeg{\LamTypeB}\;\quad\CatInterpSortNeg{\LamTypeUnit}=\CatTerm\;\quad\CatInterpSortNeg{\LamTypeA\LamTypePlus\LamTypeB}=\CatInterpSortNeg{\LamTypeA}\CatPlus\CatInterpSortNeg{\LamTypeB}$$
The notation $\CatInterpSortNeg{\,\underline{\ \ }\,}$ may seem misleading, but it is on purpose, since one can easily see that if $\LmSortA$ is a type of $\lambda\mu$-calculus, then $\CatInterpSortNeg{\LamCPS{\LmSortA}}=\CatInterpSortNeg{\LmSortA}$, where on the left the type $\LamCPS{\LmSortA}$ of $\Lam$ is interpreted in $\CatC$, and on the right the type $\LmSortA$ of $\lambda\mu$-calculus is interpreted in $\CatRC$. Since the only function types of our $\lambda$-calculus are of the form $\LamTypeA\LamTypeTo\LamTypeR$ and since $\CatC$ has all exponentials $\CatExp{\CatR}{\CatObjA}$, we can interpret $\LamCPS{\LmTermA}$ in $\CatC$ the same way we would interpret simply typed $\lambda$-calculus in a cartesian closed category. The term $\LamCPS{\LmTermA}$ of (\ref{LamTerm}) is interpreted as:
$$\LamCPS{\LmTermA}:\CatExp{\CatR}{\CatInterpSortNeg{\LamCPS{\LmSortA_1}}}\CatTimes\ldots\CatTimes\CatExp{\CatR}{\CatInterpSortNeg{\LamCPS{\LmSortA_n}}}\CatTimes\CatInterpSortNeg{\LamCPS{\LmSortB_1}}\CatTimes\ldots\CatTimes\CatInterpSortNeg{\LamCPS{\LmSortB_m}}\to\CatExp{\CatR}{\CatInterpSortNeg{\LamCPS{\LmSortA}}}$$
which is, by the above observation that $\CatInterpSortNeg{\LamCPS{\LmSortA}}=\CatInterpSortNeg{\LmSortA}$:
$$\LamCPS{\LmTermA}:\CatExp{\CatR}{\CatInterpSortNeg{\LmSortA_1}}\CatTimes\ldots\CatTimes\CatExp{\CatR}{\CatInterpSortNeg{\LmSortA_n}}\CatTimes\CatInterpSortNeg{\LmSortB_1}\CatTimes\ldots\CatTimes\CatInterpSortNeg{\LmSortB_m}\to\CatExp{\CatR}{\CatInterpSortNeg{\LmSortA}}$$
Now, by currying we obtain:
$$\CatLambda\left(\LamCPS{\LmTermA}\right):\CatExp{\CatR}{\CatInterpSortNeg{\LmSortA_1}}\CatTimes\ldots\CatTimes\CatExp{\CatR}{\CatInterpSortNeg{\LmSortA_n}}\to\CatExp{\CatR}{\CatInterpSortNeg{\LmSortA}\CatTimes\CatInterpSortNeg{\LmSortB_1}\CatTimes\ldots\CatTimes\CatInterpSortNeg{\LmSortB_m}}$$
and we have the following result:
$$\CatLambda\left(\LamCPS{\LmTermA}\right)=\LmTermA$$
where the brackets on the left represent the interpretation of $\Lam$ in the cartesian ``$\CatR$-closed'' category $\CatC$, and the brackets on the right represent the interpretation of $\lambda\mu$-calculus in the category of continuations $\CatRC$.\par
On the equational side, the equations of $\Lam$ are given in Figure~\ref{LamAxioms}, where the two terms are of the same type:
\begin{figure}
\begin{alignat*}{2}
\left(\beta_\LamTypeTo^\lambda\right)\quad&\left(\lambda\LamVarA.\LamTermA\right)\LamTermB=\LamTermA\LamSubst{\LamTermB/\LamVarA}
\qquad&
\left(\eta_\LamTypeTo^\lambda\right)\quad&\lambda\LamVarA.\LamTermA\,\LamVarA=\LamTermA\quad\left(\LamVarA\notin\FV{\LamTermA}\right)
\\
\left(\beta_\LamTypeTimes^\lambda\right)\quad&\LmProj_i\LmPair{\LamTermA_1}{\LamTermA_2}=\LamTermA_i
\qquad&
\left(\eta_\LamTypeTimes^\lambda\right)\quad&\LmPair{\LmProj_1\,\LamTermA}{\LmProj_2\,\LamTermA}=\LamTermA
\\
\left(\beta_\LamTypePlus^\lambda\right)\quad&\LamCaseBlock{\left(\LamIn_i\,\LamTermA\right)}{\LamVarA}{\LamTermB_1}{\LamTermB_2}=\LamTermB_i\LamSubst{\LamTermA/\LamVarA}
\qquad&
\left(\eta_\LamTypePlus^\lambda\right)\quad&\LamCaseBlock{\LamTermA}{\LamVarA}{\LamIn_1\,\LamVarA}{\LamIn_2\,\LamVarA}=\LamTermA
\\
\quad&
\qquad&
\left(\eta_\LamTypeUnit^\lambda\right)\quad&\LamUnit=\LamTermA
\end{alignat*}
\caption{Axioms of $\Lam$}
\label{LamAxioms}
\end{figure}
Since these equations are typed, $\LmTermA$ is of type $\LamTypeUnit$ in $\left(\eta_1^\lambda\right)$. If $\LmTermA$ and $\LmTermB$ are $\lambda\mu$-terms of the same type, then $\LmTermA=\LmTermB$ holds using the equations $\left(\beta_\LmSortTo\right)$, $\left(\eta_\LmSortTo\right)$, $\left(\zeta_\LmSortTo\right)$, $\left(\beta_\LmSortTimes\right)$, $\left(\eta_\LmSortTimes\right)$, $\left(\zeta_\LmSortTimes\right)$, $\left(\beta_\LmSortBot\right)$, $\left(\eta_\LmSortBot\right)$ and $\left(\zeta_\LmSortBot\right)$ of section~\ref{lmtheo} if and only if $\LamCPS{\LmTermA}=\LamCPS{\LmTermB}$ holds using the equations $\left(\beta_\LamTypeTo^\lambda\right)$, $\left(\eta_\LamTypeTo^\lambda\right)$, $\left(\beta_\LamTypeTimes^\lambda\right)$, $\left(\eta_\LamTypeTimes^\lambda\right)$, $\left(\beta_\LamTypePlus^\lambda\right)$, $\left(\eta_\LamTypePlus^\lambda\right)$ and $\left(\eta_\LamTypeUnit^\lambda\right)$.\par
Just as we did for $\lambda\mu$-calculus and categories of continuations, we use terms of $\Lam$ with parameters in $\CatC$. For example, if $\CatCHomA:\prod_{j\in J}\CatObjA_i\to\CatObjB$ and $\CatCHomB:\prod_{j\in J}\CatObjA_i\to\CatExp{\CatR}{\CatObjB}$, then $\CatCHomB\,\CatCHomA:\prod_{j\in J}\CatObjA_i\to\CatR$, where formally $\CatCHomB\,\CatCHomA$ is the term with parameters $\left(\LmVarB\,\LmVarA\right)\LogSubst{\CatCHomA/\LmVarA,\CatCHomB/\LmVarB}$.\par
\subsubsection{Interactions between \texorpdfstring{$\CatC$}{C} and \texorpdfstring{$\CatRC$}{RC}}
\label{interaction}
We will also extend the $\lambda\mu$-terms with parameters of section~\ref{CatInterp} by allowing the substitution of terms of $\Lam$ with parameters in $\CatC$ (defined in section~\ref{cps}) for $\mu$-variables of $\lambda\mu$-terms with parameters in $\CatRC$. For example, if $\CatRCHomA:\prod_{j\in J}\CatExp{\CatR}{\CatObjA_j}\to\CatExp{\CatR}{\CatObjB}\CatPar\left(\bigparr_{k\in K}\CatExp{\CatR}{\CatObjC_k}\right)$ in $\CatRC$ and $\CatCHomA:\left(\prod_{j\in J}\CatExp{\CatR}{\CatObjA_j}\right)\CatTimes\left(\prod_{k\in K}\CatObjC_k\right)\to\CatObjB$ in $\CatC$, then $\left[\CatCHomA\right]\,\CatRCHomA:\prod_{j\in J}\CatExp{\CatR}{\CatObjA_j}\to\CatExp{\CatR}{\CatTerm}\CatPar\left(\bigparr_{k\in K}\CatExp{\CatR}{\CatObjC_k}\right)$ in $\CatRC$. As expected, we can prove the equations of Figure~\ref{LmLamEq}
\begin{figure}
\begin{align*}
\left[\LmPair{\LamCPS{\CatRCHomA}}{\CatCHomA}\right]\,\CatRCHomB&=\left[\CatCHomA\right]\,\CatRCHomB\,\CatRCHomA&\text{if }&\left\{\begin{gathered}\CatRCHomA:\prod_{j\in J}\CatExp{\CatR}{\CatObjA_j}\to\CatExp{\CatR}{\CatObjB}\CatPar\left(\bigparr_{k\in K}\CatExp{\CatR}{\CatObjD_k}\right)\\\CatRCHomB:\prod_{j\in J}\CatExp{\CatR}{\CatObjA_j}\to\CatExp{\CatR}{\CatExp{\CatR}{\CatObjB}\CatTimes\CatObjC}\CatPar\left(\bigparr_{k\in K}\CatExp{\CatR}{\CatObjD_k}\right)\\\CatCHomA:\left(\prod_{j\in J}\CatExp{\CatR}{\CatObjA_j}\right)\CatTimes\left(\prod_{k\in K}\CatObjD_k\right)\to\CatObjC\end{gathered}\right.\\
\left[\LamUnit\right]\,\CatRCHomA&=\CatRCHomA&\text{if }&\CatRCHomA:\prod_{j\in J}\CatExp{\CatR}{\CatObjA_j}\to\CatExp{\CatR}{\CatTerm}\CatPar\left(\bigparr_{k\in K}\CatExp{\CatR}{\CatObjD_k}\right)\\
\left[\LamIn_i\,\CatCHomA\right]\,\CatRCHomA&=\left[\CatCHomA\right]\,\LmProj_i\,\CatRCHomA&\text{if }&\left\{\begin{gathered}\CatRCHomA:\prod_{j\in J}\CatExp{\CatR}{\CatObjA_j}\to\left(\CatExp{\CatR}{\CatObjB_1}\CatTimes\CatExp{\CatR}{\CatObjB_2}\right)\CatPar\left(\bigparr_{k\in K}\CatExp{\CatR}{\CatObjD_k}\right)\\\CatCHomA:\left(\prod_{j\in J}\CatExp{\CatR}{\CatObjA_j}\right)\CatTimes\left(\prod_{k\in K}\CatObjD_k\right)\to\CatObjB_i\end{gathered}\right.
\end{align*}
\caption{Interactions between \texorpdfstring{$\CatC$}{C} and \texorpdfstring{$\CatRC$}{RC}}
\label{LmLamEq}
\end{figure}
This possibility of having terms of $\Lam$ inside the brackets of $\lambda\mu$-terms is closely related to the $\overline{\lambda}\mu\widetilde{\mu}$-calculus of~\cite{HerbelinPhD,CurienHerbelinDuality}, extended to handle products (using sums of $\Lam$).
\section{A realizability model for classical analysis}
\label{modelOrthogonal}
In this section, we define a realizability model which is based on the duality between realizers and counter-realizers. First, we explain the relativization needed to interpret the recursion scheme and the axiom of choice, then we define the realizability relation and we prove its adequacy for first-order logic, Peano arithmetic and the axiom of choice. Finally, we use our orthogonality-based model to extract computational content from classical proofs.
\subsection{The relativization predicate}
\label{RelPred}
In the following we will consider two kinds of quantifications: the uniform and the relativized ones. From the point of view of provability in the two theories $\PAom$ and $\CAom$ considered in this article, this will not change anything. Indeed, we will define for $\PAom$ and $\CAom$ (which have uniform quantifications only) their relativized versions $\LogRelForm{\PAom}$ and $\LogRelForm{\CAom}$ (which have both uniform and relativized quantifications). Then we will describe a mapping from $\PAom$ (resp. $\CAom$) to $\LogRelForm{\PAom}$ (resp.$\LogRelForm{\CAom}$), and we will perform our realizability interpretation in the relativized theories.\par
Relativization is a technique that was already used in Krivine's models~\cite{KrivineStorage,OlivaStreicher,MiquelWitness}, and its utility will appear more clearly in the realizability interpretation. A realizer of $\forall\LogVarA\,\LogFormA$ (uniform quantification) is an element which must be a realizer of $\LogFormA\LogSubst{\ModElemA/\LogVarA}$ for every instance $\ModElemA$ of $\LogVarA$, whereas a realizer of $\LogForallRel\LogVarA\,\LogFormA$ (relativized quantification) is a function that turns any instance $\ModElemA$ of $\LogVarA$ into a realizer of $\LogFormA\LogSubst{\ModElemA/\LogVarA}$. The uniform quantifiers suffice to realize the rules of first-order logic and Leibniz equality, but when it comes to Peano arithmetic, and more particularly to the axiom scheme of induction, the recursor of G\"odel's system T needs to know on which particular natural number to recurse. A simple and intuitive example arises when we use the induction scheme to perform case-analysis: we prove on one hand $\LogFormA\LogSubst{\CALogZ}$, and on the other hand $\LogFormA\LogSubst{n}$ for $n\neq\CALogZ$, so by case-analysis we get $\LogForallRel\LogVarA\,\LogFormA$ (note the relativized quantification). Then the corresponding program must read the natural number, test if it is zero, and then branch to the corresponding program. One solution would have been to relativize all quantifiers, however relativizing only when it is necessary gives a better understanding of the realizability interpretation and simpler extracted programs.\par
We explain now how we map proofs of a theory into proofs of a relativized version of the theory. Fix a theory $\LogAxioms$ on a signature $\Sigma$. Let $\LogRelForm{\Sigma}$ be the signature obtained by adding to $\Sigma$ a predicate symbol for each base sort $\SortBase$ of $\Sigma$:
$$\LogRel{\LogSortedTerm{.}{\SortBase}}$$
We lift it to every sort with the following syntactic sugar:
$$\LogRel{\LogSortedTerm{\LogTermA}{\SortA\SortTo\SortB}}\Def\forall\LogSortedTerm{\LogVarA}{\SortA}\,\LogRel{\LogSortedTerm{\LogVarA}{\SortA}}\LogImp\LogRel{\LogSortedTerm{\LogTermA\,\LogVarA}{\SortB}}\qquad\text{(so in particular }\LmInterpForm{\LogRel{\LogSortedTerm{.}{\SortA\SortTo\SortB}}}=\LmInterpForm{\LogRel{\LogSortedTerm{.}{\SortA}}}\LmSortTo\LmInterpForm{\LogRel{\LogSortedTerm{.}{\SortB}}}\text{)}$$
First, since $\Sigma\subseteq\LogRelForm{\Sigma}$, $\LogAxioms$ can be considered as a theory on $\LogRelForm{\Sigma}$. We also define syntactic sugar for relativized quantifications:
$$\LogForallRel\LogSortedTerm{\LogVarA}{\SortA}\,\LogFormA\Def\forall\LogSortedTerm{\LogVarA}{\SortA}\left(\LogRel{\LogSortedTerm{\LogVarA}{\SortA}}\LogImp\LogFormA\right)\qquad\LogExistsRel\LogSortedTerm{\LogVarA}{\SortA}\,\LogFormA\Def\neg\LogForallRel\LogSortedTerm{\LogVarA}{\SortA}\,\neg\LogFormA$$
and we define inductively the relativization $\LogRelForm{\LogFormA}$ (on the signature $\LogRelForm{\Sigma}$) of a formula $\LogFormA$ (on the signature $\Sigma$):
$$\LogRelForm{\LogPredA\left(\LogSortedTerm{\LogTermA_1}{\SortA_1},\ldots,\LogSortedTerm{\LogTermA_n}{\SortA_n}\right)}\Def\LogPredA\left(\LogSortedTerm{\LogTermA_1}{\SortA_1},\ldots,\LogSortedTerm{\LogTermA_n}{\SortA_n}\right)\qquad\LogRelForm{\LogBot}\Def\LogBot\\$$
$$\LogRelForm{\left(\LogFormA\LogImp\LogFormB\right)}\Def\LogRelForm{\LogFormA}\LogImp\LogRelForm{\LogFormB}\qquad\LogRelForm{\left(\LogFormA\LogAnd\LogFormB\right)}\Def\LogRelForm{\LogFormA}\LogAnd\LogRelForm{\LogFormB}\qquad\LogRelForm{\left(\forall\LogSortedTerm{\LogVarA}{\SortA}\,\LogFormA\right)}\Def\LogForallRel\LogSortedTerm{\LogVarA}{\SortA}\,\LogRelForm{\LogFormA}$$
This translation is extended to contexts the obvious way, so we write $\LogRelForm{\Gamma}$ and $\LogRelForm{\Delta}$.\par
Before going through the relativization of a theory, we first parameterize the syntax of our logic by splitting the set of basic predicates between positive and negative predicates. The negative predicates are those $\LogPredA$ for which $\neg\neg\LogPredA\LogImp\LogPredA$ is valid under the realizability interpretation, and this notion is fundamental when it comes to relativized theories. Indeed, while the inequality predicate of $\PAom$ is negative, the relativization predicate is positive and cannot be defined from a negative one.\par
Formally, we extend the definition of the signature of a logical system by distinguishing the negative predicate symbols (written $\LogNeg{\LogPredA}$) from the positive ones (written $\LogPos{\LogPredA}$). We then extend this to all formulas of the logic by defining the following subsets of negative and positive formulas:
\begin{align*}
\LogNeg{\LogFormA},\LogNeg{\LogFormB}&\GramDef\LogNeg{\LogPredA}\left(\LogSortedTerm{\LogTermA_1}{\SortA_1},\ldots,\LogSortedTerm{\LogTermA_n}{\SortA_n}\right)\BarSep\LogBot\BarSep\LogFormA\LogImp\LogNeg{\LogFormB}\BarSep\LogNeg{\LogFormA}\LogAnd\LogNeg{\LogFormB}\BarSep\forall\LogSortedTerm{\LogVarA}{\SortA}\LogNeg{\LogFormA}\\
\LogPos{\LogFormA},\LogPos{\LogFormB}&\GramDef\LogPos{\LogPredA}\left(\LogSortedTerm{\LogTermA_1}{\SortA_1},\ldots,\LogSortedTerm{\LogTermA_n}{\SortA_n}\right)\BarSep\LogFormA\LogImp\LogPos{\LogFormB}\BarSep\LogPos{\LogFormA}\LogAnd\LogFormB\BarSep\LogFormA\LogAnd\LogPos{\LogFormB}\BarSep\forall\LogSortedTerm{\LogVarA}{\SortA}\LogPos{\LogFormA}
\end{align*}
where the absence of polarity means that it can be either. In usual interpretations of classical proofs every formula has both a positive and a negative interpretation, which are orthogonal to each other. This is is reflected on the computational side by the duality between terms and contexts. The relativization predicate, however, only has a positive interpretation. In order for our logical rules to be valid under the realizability interpretation, we restrict our sequents by asking the right-hand context to contain only negative formulas. Formally, in a sequent:
$$\Sequent{\Gamma}{\LogFormA}{\Delta}$$
the formulas appearing in $\Delta$ must all be negative, which will be emphasized from now on by writing $\LogNeg{\Delta}$. There are no other requirements, so $\LogFormA$ and formulas of $\Gamma$ can be either positive or negative. This restriction on the syntax of the sequents roughly amounts to forbid the structural rules on the right for positive formulas. In particular $\Sequent{}{\LogBot\LogImp\LogFormA}{}$ and $\Sequent{}{\neg\neg\LogFormA\LogImp\LogFormA}{}$ are provable only if $\LogFormA$ is a negative formula. This restriction will appear as necessary in the proof of the adequacy lemma for first-order logic~\ref{adequacyLemma}.\par
One could think at first sight that these polarities are related to the ones of LC~\cite{GirardLC}, however there is an important mismatch in the case of implication, where in LC $\LogFormA\LogImp\LogFormB$ is negative iff $\LogFormA$ is positive or $\LogFormB$ is positive, while in our setting it has the same polarity as $\LogFormB$. After a discussion with Olivier Laurent, it appeared that the polarities defined here are more related to the system LU~\cite{GirardLU}, where our positive formulas are LU formulas with polarity $0$, our negative formulas are LU formulas with polarity $-1$, our $\LogImp$ is LU's $\supset$, our $\LogAnd$ is LU's $\&$ and our $\forall\LogVarA$ is LU's $\bigwedge\LogVarA$. A sequent $\Sequent{\Gamma}{\LogFormA}{\Delta}$ in our setting then corresponds to a sequent $;\Gamma\vdash\Delta;\LogFormA$ in LU.\par
For equational theories, the inequality predicate is defined to be negative. Since the inequality predicate is the only predicate of $\Sigma_\PAom=\Sigma_\CAom$, all the formulas written on this signature are negative, and therefore the proofs in $\PAom$ and $\CAom$ automatically respect the negativeness condition.\par
Since the negative predicates do not carry computational content, we also fix their interpretation to be the empty type of $\lambda\mu$-calculus:
$$\LmInterpForm{\LogNeg{\LogPredA}}\Def\LmSortBot$$
We now describe the process of relativization of a theory. In this process, the new predicate symbol $\LogRel{.}$ is defined to be positive, while the other predicates of $\Sigma$ keep their polarity. First, an important fact is that the relativization of formulas preserves polarity:
\begin{lem}
\label{RelNeg}
If $\LogNeg{\LogFormA}$ (resp. $\LogPos{\LogFormA}$) is a negative (resp. positive) formula over the signature $\Sigma$, then $\LogRelForm{\LogNeg{\LogFormA}}$ (resp. $\LogRelForm{\LogPos{\LogFormA}}$) is a negative (resp. positive) formula over the signature $\LogRelForm{\Sigma}$ (obtained by adding a positive relativization predicate $\LogRel{\LogSortedTerm{.}{\SortBase}}$ for each base sort $\SortBase$).
\end{lem}
\proof
By induction on $\LogNeg{\LogFormA}$ (resp. $\LogPos{\LogFormA}$).
\qed
A relativized version of a theory $\LogAxioms$ on $\Sigma$ is then a set of axioms $\LogRelForm{\LogAxioms}$ on the signature $\LogRelForm{\Sigma}$ such that if $\LogFormA$ is provable in $\LogAxioms$, then $\LogRelForm{\LogFormA}$ is provable in $\LogRelForm{\LogAxioms}$. Be careful however that $\LogRelForm{\LogAxioms}$ is in general different from $\SetSuch{\LogRelForm{\LogFormA}}{\LogFormA\in\LogAxioms}$. The following lemma gives a sufficient condition for $\LogRelForm{\LogAxioms}$ to be a relativized version of $\LogAxioms$:
\begin{lem}
\label{RelTheo}
If $\LogRelForm{\LogAxioms}$ is a theory on $\LogRelForm{\Sigma}$ such that:
\begin{itemize}
\item for any constant individual $\LogSortedTerm{\LogConstA}{\SortA}$ of $\Sigma$, $\LogRelForm{\LogAxioms}\Derives\LogRel{\LogSortedTerm{\LogConstA}{\SortA}}$
\item for each $\LogFormA\in\LogAxioms$, $\LogRelForm{\LogAxioms}\Derives\LogRelForm{\LogFormA}$
\item for each sort $\SortA$ of $\Sigma$, there is a closed term $\LogSortedTerm{\LogTermA}{\SortA}$
\end{itemize}
then for any closed formula $\LogFormA$ on $\Sigma$:
$$\LogAxioms\Derives\LogFormA\qquad\Longrightarrow\qquad\LogRelForm{\LogAxioms}\Derives\LogRelForm{\LogFormA}$$
\end{lem}
\proof
We translate inductively every proof in the theory $\LogAxioms$ (on the signature $\Sigma$) of a sequent:
$$\Sequent{\Gamma}{\LogFormA}{\LogNeg{\Delta}}$$
to a proof in $\LogRelForm{\LogAxioms}$ (on the signature $\LogRelForm{\Sigma}$) of a sequent:
$$\Sequent{\LogRel{\LogSortedTerm{\vec{\LogVarA}}{\vec{\SortA}}},\LogRelForm{\Gamma}}{\LogRelForm{\LogFormA}}{\LogRelForm{\LogNeg{\Delta}}}$$
where $\FV{\Gamma,\LogFormA,\Delta}\subseteq\LogSortedTerm{\vec{\LogVarA}}{\vec{\SortA}}$. The translation of a proof $\Sequent{}{\LogFormA}{}$ in $\LogAxioms$ for $\LogFormA$ a closed formula is then a proof in $\LogRelForm{\LogAxioms}$ of a sequent $\Sequent{\LogRel{\LogSortedTerm{\vec{\LogVarA}}{\vec{\SortA}}}}{\LogRelForm{\LogFormA}}{}$. Indeed, some dummy variables may appear during the translation, without appearing free in $\LogFormA$ (this is due to our system not satisfying the subformula property). In order to eliminate these, we use the last hypothesis: from $\Sequent{\LogRel{\LogSortedTerm{\vec{\LogVarA}}{\vec{\SortA}}}}{\LogRelForm{\LogFormA}}{}$ where $\LogFormA$ (and therefore $\LogRelForm{\LogFormA}$) is closed we can derive $\Sequent{}{\LogForallRel\LogSortedTerm{\vec{\LogVarA}}{\vec{\SortA}}\LogRelForm{\LogFormA}}{}$, and then $\Sequent{}{\LogRel{\LogSortedTerm{\vec{\LogTermA}}{\vec{\SortA}}}\LogImp\LogRelForm{\LogFormA}}{}$ where $\LogSortedTerm{\vec{\LogTermA}}{\vec{\SortA}}$ are closed individuals obtained from the last hypothesis. Finally, using lemma~\ref{relindiv} below, we can combine the proof of $\Sequent{}{\LogRel{\LogSortedTerm{\vec{\LogTermA}}{\vec{\SortA}}}\LogImp\LogRelForm{\LogFormA}}{}$ with proofs of $\LogRel{\LogSortedTerm{\vec{\LogTermA}}{\vec{\SortA}}}$ and get a proof of $\Sequent{}{\LogRelForm{\LogFormA}}{}$.\par
The translation of the axiom rule follows from the hypotheses of the lemma, the identity rule is translated as follows, with $\LogSortedTerm{\vec{\LogVarA}}{\vec{\SortA}}=\FV{\Gamma,\LogFormA,\Delta}$:
$$\LogRuleAx{\Gamma}{\LogNeg{\Delta}}{\LogFormA}\;\rightsquigarrow\;\LogRuleAx{\LogRel{\LogSortedTerm{\vec{\LogVarA}}{\vec{\SortA}}},\LogRelForm{\Gamma}}{\LogRelForm{\LogNeg{\Delta}}}{\LogRelForm{\LogFormA}}$$
The rules for logical connectives are translated trivially as in Figure~\ref{RelatRules},
\begin{figure}
\begin{align*}
\LogRuleBotIntro{\Gamma}{\LogNeg{\Delta}}{\LogNeg{\LogFormA}}&\;\;\rightsquigarrow\;\;\LogRuleBotIntro{\LogRel{\LogSortedTerm{\vec{\LogVarA}}{\vec{\SortA}}},\LogRelForm{\Gamma}}{\LogRelForm{\LogNeg{\Delta}}}{\LogRelForm{\LogNeg{\LogFormA}}}\\[5pt]
\LogRuleBotElim{\Gamma}{\LogNeg{\Delta}}{\LogNeg{\LogFormA}}&\;\;\rightsquigarrow\;\;\LogRuleBotElim{\LogRel{\LogSortedTerm{\vec{\LogVarA}}{\vec{\SortA}}},\LogRelForm{\Gamma}}{\LogRelForm{\LogNeg{\Delta}}}{\LogRelForm{\LogNeg{\LogFormA}}}\\[5pt]
\LogRuleImpIntro{\Gamma}{\LogNeg{\Delta}}{\LogFormA}{\LogFormB}&\;\;\rightsquigarrow\;\;\LogRuleImpIntro{\LogRel{\LogSortedTerm{\vec{\LogVarA}}{\vec{\SortA}}},\LogRelForm{\Gamma}}{\LogRelForm{\LogNeg{\Delta}}}{\LogRelForm{\LogFormA}}{\LogRelForm{\LogFormB}}\\[5pt]
\LogRuleImpElim{\Gamma}{\LogNeg{\Delta}}{\LogFormA}{\LogFormB}&\;\;\rightsquigarrow\;\;\LogRuleImpElim{\LogRel{\LogSortedTerm{\vec{\LogVarA}}{\vec{\SortA}}},\LogRelForm{\Gamma}}{\LogRelForm{\LogNeg{\Delta}}}{\LogRelForm{\LogFormA}}{\LogRelForm{\LogFormB}}\\[5pt]
\LogRuleAndIntro{\Gamma}{\LogNeg{\Delta}}{\LogFormA}{\LogFormB}&\;\;\rightsquigarrow\;\;\LogRuleAndIntro{\LogRel{\LogSortedTerm{\vec{\LogVarA}}{\vec{\SortA}}},\LogRelForm{\Gamma}}{\LogRelForm{\LogNeg{\Delta}}}{\LogRelForm{\LogFormA}}{\LogRelForm{\LogFormB}}\\[5pt]
\LogRuleAndElim{\Gamma}{\LogNeg{\Delta}}{\LogFormA}&\;\;\rightsquigarrow\;\;\LogRuleAndElim{\LogRel{\LogSortedTerm{\vec{\LogVarA}}{\vec{\SortA}}},\LogRelForm{\Gamma}}{\LogRelForm{\LogNeg{\Delta}}}{\LogRelForm{\LogFormA}}
\end{align*}
\caption{Relativization of the rules for logical connectives}
\label{RelatRules}
\end{figure}
where in the cases of introduction of implication and conjunction, the two premises can get the same relativized variables on the left by applying the (admissible) left weakening rule when necessary. The translation of the introduction of universal quantification is given by:
$$\LogRuleForallIntro{\Gamma}{\LogNeg{\Delta}}{\LogSortedTerm{\LogVarB}{\SortB}}{\LogFormA}\;\;\rightsquigarrow\;\;\AXM{\Sequent{\LogRel{\LogSortedTerm{\LogVarB}{\SortB}},\LogRel{\LogSortedTerm{\vec{\LogVarA}}{\vec{\SortA}}},\LogRelForm{\Gamma}}{\LogRelForm{\LogFormA}}{\LogRelForm{\LogNeg{\Delta}}}}\UIM{\Sequent{\LogRel{\LogSortedTerm{\vec{\LogVarA}}{\vec{\SortA}}},\LogRelForm{\Gamma}}{\LogRel{\LogSortedTerm{\LogVarB}{\SortB}}\LogImp\LogRelForm{\LogFormA}}{\LogRelForm{\LogNeg{\Delta}}}}\RLM{\left(\LogSortedTerm{\LogVarB}{\SortB}\notin\FV{\LogRel{\LogSortedTerm{\vec{\LogVarA}}{\vec{\SortA}}},\LogRelForm{\Gamma},\LogRelForm{\LogNeg{\Delta}}}\right)}\UIM{\Sequent{\LogRel{\LogSortedTerm{\vec{\LogVarA}}{\vec{\SortA}}},\LogRelForm{\Gamma}}{\forall\LogSortedTerm{\LogVarB}{\SortB}\left(\LogRel{\LogSortedTerm{\LogVarB}{\SortB}}\LogImp\LogRelForm{\LogFormA}\right)}{\LogRelForm{\LogNeg{\Delta}}}}\DP$$
and preserves the fact that all the free variables of a sequent are relativized in the context. Finally, in order to translate the elimination of the universal quantification we must prove that we can lift relativization to all constructs on individuals. This can be obtained using the hypothesis of the lemma requiring that for every individual constant $\LogSortedTerm{\LogConstA}{\SortA}$ of $\Sigma$, $\LogRelForm{\LogAxioms}\Derives\LogRel{\LogSortedTerm{\LogConstA}{\SortA}}$. Indeed, we have the following lemma:
\begin{lem}
\label{relindiv}
Suppose that for every individual constant $\LogSortedTerm{\LogConstA}{\SortA}$ of $\Sigma$, $\LogRelForm{\LogAxioms}\Derives\LogRel{\LogSortedTerm{\LogConstA}{\SortA}}$. Let $\LogSortedTerm{\LogTermA}{\SortA}$ be an individual of the logic with free variables $\LogSortedTerm{\vec{\LogVarA}}{\vec{\SortC}}$. We have:
$$\LogRelForm{\LogAxioms}\Derives\LogForallRel\LogSortedTerm{\vec{\LogVarA}}{\vec{\SortC}}\,\LogRel{\LogSortedTerm{\LogTermA}{\SortA}}$$
\end{lem}
\proof
We prove it by induction on $\LogSortedTerm{\LogTermA}{\SortA}$. If $\LogSortedTerm{\LogTermA}{\SortA}$ is some $\LogSortedTerm{\LogConstA}{\SortA}$, then this is an assumption of the lemma. If $\LogSortedTerm{\LogTermA}{\SortA}$ is a variable $\LogSortedTerm{\LogVarA}{\SortA}$, then $\LogForallRel\LogSortedTerm{\LogVarA}{\SortA}\,\LogRel{\LogSortedTerm{\LogVarA}{\SortA}}\equiv\forall\LogSortedTerm{\LogVarA}{\SortA}\left(\LogRel{\LogSortedTerm{\LogVarA}{\SortA}}\LogImp\LogRel{\LogSortedTerm{\LogVarA}{\SortA}}\right)$, which is trivially derivable, and if $\LogSortedTerm{\LogTermA}{\SortA}$ is $\LogSortedTerm{\LogTermB}{\SortB\SortTo\SortA}\,\LogSortedTerm{\LogTermC}{\SortB}$, then we get $\Sequent{\LogRel{\LogSortedTerm{\vec{\LogVarA}}{\vec{\SortC}}}}{\forall\LogSortedTerm{\LogVarB}{\SortB}\left(\LogRel{\LogSortedTerm{\LogVarB}{\SortB}}\LogImp\LogRel{\LogSortedTerm{\left(\LogTermB\,\LogVarB\right)}{\SortA}}\right)}{}$ and $\Sequent{\LogRel{\LogSortedTerm{\vec{\LogVarA}}{\vec{\SortC}}}}{\LogRel{\LogSortedTerm{\LogTermC}{\SortB}}}{}$ from the induction hypotheses, so we obtain $\Sequent{\LogRel{\LogSortedTerm{\vec{\LogVarA}}{\vec{\SortC}}}}{\LogRel{\LogSortedTerm{\left(\LogTermB\,\LogTermC\right)}{\SortA}}}{}$ and $\Sequent{}{\LogForallRel\LogSortedTerm{\vec{\LogVarA}}{\vec{\SortC}}\,\LogRel{\LogSortedTerm{\left(\LogTermB\,\LogTermC\right)}{\SortA}}}{}$.
\qed
Now we can describe the translation of the elimination rule of the universal quantifier:
$$
\LogRuleForallElim{\Gamma}{\LogNeg{\Delta}}{\LogSortedTerm{\LogVarB}{\SortB}}{\LogSortedTerm{\LogTermA}{\SortA}}{\LogFormA}\;\;\rightsquigarrow\;\;
\AXM{\Sequent{\LogRel{\LogSortedTerm{\vec{\LogVarA}}{\vec{\SortA}}},\LogRelForm{\Gamma}}{\LogForallRel\LogSortedTerm{\LogVarB}{\SortB}\,\LogRelForm{\LogFormA}}{\LogRelForm{\LogNeg{\Delta}}}}
\UIM{\Sequent{\LogRel{\LogSortedTerm{\vec{\LogVarA}}{\vec{\SortA}}},\LogRelForm{\Gamma}}{\LogRel{\LogSortedTerm{\LogTermA}{\SortA}}\LogImp\LogRelForm{\LogFormA\LogSubst{\LogTermA/\LogVarB}}}{\LogRelForm{\LogNeg{\Delta}}}}
\AXM{\vdots}
\UIM{\Sequent{\LogRel{\LogSortedTerm{\vec{\LogVarA}}{\vec{\SortA}}},\LogRelForm{\Gamma}}{\LogRel{\LogSortedTerm{\LogTermA}{\SortA}}}{\LogRelForm{\LogNeg{\Delta}}}}
\BIM{\Sequent{\LogRel{\LogSortedTerm{\vec{\LogVarA}}{\vec{\SortA}}},\LogRelForm{\Gamma}}{\LogRelForm{\LogFormA\LogSubst{\LogTermA/\LogVarB}}}{\LogRelForm{\LogNeg{\Delta}}}}
\DP
$$
where we can suppose without loss of generality that $\FV{\LogSortedTerm{\LogTermA}{\SortA}}\subseteq\LogSortedTerm{\vec{\LogVarA}}{\vec{\SortA}}$ (using the admissible left weakening rule if necessary), and $\Sequent{\LogRel{\LogSortedTerm{\vec{\LogVarA}}{\vec{\SortA}}},\LogRelForm{\Gamma}}{\LogRel{\LogSortedTerm{\LogTermA}{\SortA}}}{\LogRelForm{\Delta}}$ is easily derivable using $\LogForallRel\LogSortedTerm{\vec{\LogVarC}}{\vec{\SortC}}\,\LogRel{\LogSortedTerm{\LogTermA}{\SortA}}$ (with $\LogSortedTerm{\vec{\LogVarC}}{\vec{\SortC}}=\FV{\LogSortedTerm{\LogTermA}{\SortA}}\subseteq\LogSortedTerm{\vec{\LogVarA}}{\vec{\SortA}}$) from lemma~\ref{relindiv}.\par
Finally, the negativeness of the formulas in the right-hand context is preserved through the relativization, thanks to lemma~\ref{RelNeg}.
\qed
\subsubsection{Relativized \texorpdfstring{$\PAom$}{PAomega}: \texorpdfstring{$\LogRelForm{\PAom}$}{PAomegar}}
We now define the relativized version $\LogRelForm{\PAom}$ of $\PAom$, to which we apply lemma~\ref{RelTheo}. First, $\Sigma_\LogRelForm{\PAom}$ is $\LogRelForm{\Sigma_\PAom}$, that is $\Sigma_\PAom$ augmented with the positive predicate symbol $\LogRel{\LogSortedTerm{\LogTermA}{\CASort}}$. The axioms of $\LogRelForm{\PAom}$ are those of $\PAom$ (unrelativized) where the induction scheme is replaced with:
$$\CAindName\quad\CAindNoType{\LogFormA}$$
plus the axioms $\CARelZ$ and $\CARelS$. Notice that $\CAindName$ is different from $\LogRelForm{\CAindUnrelName}$ (the parameters $\vec{\LogVarB}$ are not relativized). In order to use lemma~\ref{RelTheo} we need the following lemmas:
\begin{lem}
$$\LogRelForm{\PAom}\Derives\CARelsNoType\qquad\LogRelForm{\PAom}\Derives\CARelkNoType\qquad\LogRelForm{\PAom}\Derives\CARelrecNoType$$
\end{lem}
\proof
The formulas $\CARelsNoType$ and $\CARelkNoType$ are provable using the axioms of equality and respectively $\CAdefsName$ and $\CAdefkName$. Indeed, $\CARelsNoType$ is the following formula:
$$\forall\LogSortedTerm{\LogVarA}{\SortA\SortTo\SortB\SortTo\SortC}\left(\LogRel{\LogVarA}\LogImp\forall\LogSortedTerm{\LogVarB}{\SortA\SortTo\SortB}\left(\LogRel{\LogVarB}\LogImp\forall\LogSortedTerm{\LogVarC}{\sigma}\left(\LogRel{\LogVarC}\LogImp\LogRel{\CALogs\,\LogVarA\,\LogVarB\,\LogVarC}\right)\right)\right)$$
which is equivalent in first-order logic to:
$$\forall\LogSortedTerm{\LogVarA}{\SortA\SortTo\SortB\SortTo\SortC}\forall\LogSortedTerm{\LogVarB}{\SortA\SortTo\SortB}\forall\LogSortedTerm{\LogVarC}{\SortA}\left(\LogRel{\LogVarA}\LogImp\LogRel{\LogVarB}\LogImp\LogRel{\LogVarC}\LogImp\LogRel{\CALogs\,\LogVarA\,\LogVarB\,\LogVarC}\right)$$
using $\CAdefsName$ and $\CALeibName$ this is equivalent to:
$$\forall\LogSortedTerm{\LogVarA}{\SortA\SortTo\SortB\SortTo\SortC}\forall\LogSortedTerm{\LogVarB}{\SortA\SortTo\SortB}\forall\LogSortedTerm{\LogVarC}{\SortA}\left(\LogRel{\LogVarA}\LogImp\LogRel{\LogVarB}\LogImp\LogRel{\LogVarC}\LogImp\LogRel{\LogVarA\,\LogVarC\left(\LogVarB\,\LogVarC\right)}\right)$$
and since $\LogRel{\LogVarA}$ and $\LogRel{\LogVarB}$ are the following formulas:
$$\forall\LogSortedTerm{\LogVarA_1}{\SortA}\left(\LogRel{\LogVarA_1}\LogImp\forall\LogSortedTerm{\LogVarA_2}{\SortB}\left(\LogRel{\LogVarA_2}\LogImp\LogRel{\LogVarA\,\LogVarA_1\,\LogVarA_2}\right)\right)\qquad\forall\LogSortedTerm{\LogVarB_1}{\SortA}\left(\LogRel{\LogVarB_1}\LogImp\LogRel{\LogVarB\,\LogVarB_1}\right)$$
we can instantiate these with $\LogVarA_1\equiv\LogVarC$, $\LogVarA_2\equiv\LogVarB\,\LogVarC$ and $\LogVarB_1\equiv\LogVarC$ to obtain a proof of $\CARelsNoType$. The case of $\CARelkNoType$ is similar.\par
Similarly, $\CARelrecNoType$ is provable using the axioms of equality, $\CAdefrecZName$, $\CAdefrecSName$ and $\CAindName$. $\CARelrecNoType$ is equivalent in first-order logic to:
$$\forall\LogSortedTerm{\LogVarA}{\SortA}\forall\LogSortedTerm{\LogVarB}{\CASort\SortTo\SortA\SortTo\SortA}\left(\LogRel{\LogVarA}\LogImp\LogForallRel\LogSortedTerm{\LogVarB_1}{\CASort}\forall\LogSortedTerm{\LogVarB_2}{\SortA}\left(\LogRel{\LogVarB_2}\LogImp\LogRel{\LogVarB\,\LogVarB_1\,\LogVarB_2}\right)\LogImp\LogForallRel\LogSortedTerm{\LogVarC}{\CASort}\LogRel{\CALogrec\,\LogVarA\,\LogVarB\,\LogVarC}\right)$$
if we instantiate $\LogVarB_2$ with $\CALogrec\,\LogVarA\,\LogVarB\,\LogVarB_1$ it is sufficient to prove:
$$\forall\LogSortedTerm{\LogVarA}{\SortA}\forall\LogSortedTerm{\LogVarB}{\CASort\SortTo\SortA\SortTo\SortA}\left(\LogRel{\LogVarA}\LogImp\LogForallRel\LogSortedTerm{\LogVarB_1}{\CASort}\left(\LogRel{\CALogrec\,\LogVarA\,\LogVarB\,\LogVarB_1}\LogImp\LogRel{\LogVarB\,\LogVarB_1\left(\CALogrec\,\LogVarA\,\LogVarB\,\LogVarB_1\right)}\right)\LogImp\LogForallRel\LogSortedTerm{\LogVarC}{\CASort}\LogRel{\CALogrec\,\LogVarA\,\LogVarB\,\LogVarC}\right)$$
which is equivalent, using $\CAdefrecZName$, $\CAdefrecSName$ and $\CALeibName$, to:
$$\forall\LogSortedTerm{\LogVarA}{\SortA}\forall\LogSortedTerm{\LogVarB}{\CASort\SortTo\SortA\SortTo\SortA}\left(\LogRel{\CALogrec\,\LogVarA\,\LogVarB\,\CALogZ}\LogImp\LogForallRel\LogSortedTerm{\LogVarB_1}{\CASort}\left(\LogRel{\CALogrec\,\LogVarA\,\LogVarB\,\LogVarB_1}\LogImp\LogRel{\CALogrec\,\LogVarA\,\LogVarB\left(\CALogS\,\LogVarB_1\right)}\right)\LogImp\LogForallRel\LogSortedTerm{\LogVarC}{\CASort}\LogRel{\CALogrec\,\LogVarA\,\LogVarB\,\LogVarC}\right)$$
which is an instance of $\CAindName$ with the formula $\LogRel{\CALogrec\,\LogVarA\,\LogVarB\,\LogVarC}$.\par
Moreover, these proofs of $\CARelsNoType$, $\CARelkNoType$ and $\CARelrecNoType$ respect the condition of having only negative formulas in the right-hand context, since this context is empty (which means that the proofs are valid in minimal logic).
\qed
\begin{lem}
For any $\LogFormA\in\PAom$, $\LogRelForm{\PAom}\Derives\LogRelForm{\LogFormA}$.
\end{lem}
\proof
For $\CAdefsName$, $\CAdefkName$, $\CAdefrecZName$, $\CAdefrecSName$ and $\CASnZName$ it follows from the fact that the formula $\forall\LogVarA\,\LogFormA\LogImp\LogForallRel\LogVarA\,\LogFormA$ is derivable in first-order logic. For $\CAindUnrelName$ it comes from this and the fact that the following formula:
$$\forall\vec{\LogVarB}\left(\LogRelForm{\LogFormA}\LogSubst{\CALogZ/\LogVarA}\LogImp\LogForallRel\LogVarA\left(\LogRelForm{\LogFormA}\LogImp\LogRelForm{\LogFormA}\LogSubst{\CALogS\,\LogVarA/\LogVarA}\right)\LogImp\LogForallRel\LogVarA\,\LogRelForm{\LogFormA}\right)$$
is an instance of $\CAindName$. Here again, the right-hand context is empty so the proofs are correct.
\qed
\begin{lem}
For every sort $\SortA$ on $\Sigma_\PAom$, there is a closed individual $\LogSortedTerm{\LogTermA}{\SortA}$.
\end{lem}
\proof
$\LogSortedTerm{\CALogZ}{\SortA}$ from section~\ref{PeanoTheory} is such a term.
\qed
Using these three lemmas, it follows from lemma~\ref{RelTheo} that for any closed formula $\LogFormA$ on the signature $\Sigma_\PAom$:
$$\PAom\Derives\LogFormA\qquad\Longrightarrow\qquad\LogRelForm{\PAom}\Derives\LogRelForm{\LogFormA}$$
\subsubsection{Interpreting \texorpdfstring{$\LogRelForm{\PAom}$}{PAomegar} in system T + \texorpdfstring{$\CALmom$}{omega}}
\label{LmInterpPA}
System T was introduced by G\"odel in~\cite{GodelDialectica} in order to give a consistency proof of Heyting arithmetic (and therefore of Peano arithmetic by double-negation translation). This system can be equivalently formulated as a system of primitive recursive functionals, which is an extension of primitive recursive functions to higher types. It is strictly more powerful than primitive recursion, since for example the Ackermann's function is expressible in system T.\par
System T has one base type $\CALmnSort$ for natural numbers, product and function types, constants for $0$ and successor, and a recursion operator of type $\CALmrecSort{\LmSortA}$ for any type $\LmSortA$. Restricting the type of the recursor to $\LmSortA=\CALmnSort$ gives back the usual primitive recursive functions.\par
Since $\mu$PCF contains constants for every natural number, successor, predecessor and general recursion, it is easy to encode system T in it. Indeed, if we define:
$$\CALmrec\;\Def\;\lambda\LmVarA\LmVarB.\CALmfix\left(\lambda\LmVarC\LmVarD.\CALmifz\,\LmVarD\,\LmVarA\left(\LmVarB\left(\CALmpred\,\LmVarD\right)\left(\LmVarC\left(\CALmpred\,\LmVarD\right)\right)\right)\right)$$
Then it is easy to derive:
$$\LmTerm{\CALmrec}{\CALmrecSort{\LmSortA}}$$
and in order to prove that it implements G\"odel's recursor we must prove that it satisfies the corresponding equations:
\begin{lem}
Let $\LmTerm{\LmTermA}{\LmSortA}$ and $\LmTerm{\LmTermB}{\CALmnSort\LmSortTo\LmSortA\LmSortTo\LmSortA}$. We have:
$$\CALmrec\,\LmTermA\,\LmTermB\,\CALmn{0}=\LmTermA$$
and for any $n\in\mathbb{N}$:
$$\CALmrec\,\LmTermA\,\LmTermB\,\CALmn{n+1}=\LmTermB\,\CALmn{n}\left(\CALmrec\,\LmTermA\,\LmTermB\,\CALmn{n}\right)$$
\end{lem}
\proof
This follows easily from the definition of $\CALmrec$ and the equations of $\mu$PCF for $\CALmpred$, $\CALmifz$ and $\CALmfix$.
\qed
Therefore in the following we will consider system T as a subsystem of $\mu$PCF. We will also use the constant $\CALmom$ to interpret the fact that $\CALogZ$ is not a successor. In order to interpret $\LogRelForm{\PAom}$ in system T + $\CALmom$, we first fix the interpretation of the relativization predicate:
$$\LmInterpForm{\LogRel{\LogSortedTerm{.}{\SortBase}}}\Def\CALmnSort$$
The inequality predicate being a negative one, it is interpreted as $\LmSortBot$. Finally, we provide a term $\LmInterpAxiom{\LogFormA}$ of type $\LmInterpForm{\LogFormA}$ for each $\LogFormA\in\LogRelForm{\PAom}$ in Figure~\ref{AxiomTrans}, where $\LogFormB$ is a formula with free variables among $\LogSortedTerm{\LogVarA}{\CASort},\LogSortedTerm{\vec{\LogVarB}}{\vec{\SortA}}$.
\begin{figure}
\begin{align*}
\LmInterpAxiom{\CAReflName}&\Def\LmTerm{\lambda\LmVarA.\LmVarA}{\CAReflSort}&\;\LmInterpAxiom{\CAdefkName}&\Def\LmTerm{\CAdefkTerm}{\CAdefkSort}\\
\LmInterpAxiom{\CALeibName}&\Def\LmTerm{\lambda\LmVarA.\LmVarA}{\CALeibSort{\LogFormA}}&\;\LmInterpAxiom{\CAdefsName}&\Def\LmTerm{\CAdefsTerm}{\CAdefsSort}\\
\LmInterpAxiom{\CASnZName}&\Def\LmTerm{\CASnZTerm}{\CASnZSort}&\;\LmInterpAxiom{\CAdefrecZName}&\Def\LmTerm{\CAdefrecZTerm}{\CAdefrecZSort}\\
\LmInterpAxiom{\CAindName}&\Def\LmTerm{\CAindTerm}{\CAindSort{\LogFormB}}&\;\LmInterpAxiom{\CAdefrecSName}&\Def\LmTerm{\CAdefrecSTerm}{\CAdefrecSSort}
\end{align*}
\begin{align*}
\LmInterpAxiom{\CARelkNoType}&\Def\LmTerm{\CARelkTerm}{\CARelkSort{\LmSortA}{\LmSortB}}&\;\LmInterpAxiom{\CARelsNoType}&\Def\LmTerm{\CARelsTerm}{\CARelsSort{\LmSortA}{\LmSortB}{\LmSortC}}
\end{align*}
\begin{align*}
\LmInterpAxiom{\CARelZNoType}&\Def\LmTerm{\CARelZTerm}{\CARelZSort}&\;\LmInterpAxiom{\CARelSNoType}&\Def\LmTerm{\CARelSTerm}{\CARelSSort}&\;\LmInterpAxiom{\CARelrecNoType}&\Def\LmTerm{\CARelrecTerm}{\CARelrecSort{\LmSortA}}
\end{align*}
\caption{Interpretation of the axioms of $\LogRelForm{\PAom}$}
\label{AxiomTrans}
\end{figure}
\subsubsection{Relativized axiom of choice}
\label{RelCAom}
As we did for $\PAom$, we define the relativized version $\LogRelForm{\CAom}$ of $\CAom$. First, the signature is the same as $\Sigma_\LogRelForm{\PAom}$: it is $\Sigma_\CAom=\Sigma_\PAom$ augmented with a positive predicate symbol $\LogRel{\LogSortedTerm{.}{\CASort}}$. In this section, $\LogFormA$ denotes a formula over $\Sigma_\LogRelForm{\CAom}=\Sigma_\LogRelForm{\PAom}$ with free variables among $\LogSortedTerm{\LogVarA}{\CASort},\LogSortedTerm{\LogVarB}{\SortA},\LogSortedTerm{\LogVarC}{\SortA},\LogSortedTerm{\vec{\LogVarD}}{\vec{\SortB}}$. For clarity, we write $\LogFormA\LogSubst{\LogTermA,\LogTermB,\LogTermC}$ instead of $\LogFormA\LogSubst{\LogTermA/\LogVarA,\LogTermB/\LogVarB,\LogTermC/\LogVarC}$. The axioms of $\LogRelForm{\CAom}$ are those of $\LogRelForm{\PAom}$ plus the following version of dependent choice:
$$\CADCName\;\CADC{\SortA}{\LogFormA}{\SortB}$$
where $\LogFormA\LogSubst{\LogVarA,\LogVarB,\LogVarC}$ is of the shape $\LogRel{\LogVarC}\LogAnd\LogFormC$. The formula $\LogFormA$ in $\CADCName$ sould be understood as an abbreviation, and the axiom is schematic in $\LogFormC$. This version is quite different from $\LogRelForm{\CADCUnrelName}$ which is:
$$\LogForallRel\LogSortedTerm{\vec{\LogVarD}}{\vec{\SortB}}\left(\LogForallRel\LogSortedTerm{\LogVarA}{\CASort}\,\LogForallRel\LogSortedTerm{\LogVarB}{\SortA}\,\LogExistsRel\LogSortedTerm{\LogVarC}{\SortA}\,\LogRelForm{\LogFormB}\LogSubst{\LogVarA,\LogVarB,\LogVarC}\;\;\LogImp\;\;\LogExistsRel\LogSortedTerm{\LogVarE}{\CASort\SortTo\SortA}\,\LogForallRel\LogSortedTerm{\LogVarA}{\CASort}\,\LogRelForm{\LogFormB}\LogSubst{\LogVarA,\LogVarE\,\LogVarA,\LogVarE\left(\CALogS\,\LogVarA\right)}\right)$$
where $\LogFormB$ is a formula over $\Sigma_\CAom=\Sigma_\PAom$ with free variables among $\LogSortedTerm{\LogVarA}{\CASort},\LogSortedTerm{\LogVarB}{\SortA},\LogSortedTerm{\LogVarC}{\SortA},\LogSortedTerm{\vec{\LogVarD}}{\vec{\SortB}}$, for which we use again the notation $\LogRelForm{\LogFormB}\LogSubst{\LogTermA,\LogTermB,\LogTermC}$. First, some quantifications have been unrelativized, but another important difference is that we replaced $\LogBot$ with $\forall\LogVarA'\,\LogFormA\LogSubst{\LogVarA',\LogVarB,\LogVarB}$. This change is in the spirit of~\cite{EscardoOlivaPeirce} and will allow an easier realizability interpretation in section~\ref{ChoiceAdequacy}. In order to use lemma~\ref{RelTheo}, we need to prove that $\LogRelForm{\CADCUnrelName}$ is derivable in $\LogRelForm{\CAom}$. We will actually prove that the instance of $\CADCName$ with:
$$\LogFormA\LogSubst{\LogVarA,\LogVarB,\LogVarC}\equiv\LogRel{\LogVarC}\LogAnd\LogRel{\LogVarB}\LogAnd\LogRelForm{\LogFormB}\LogSubst{\LogVarA,\LogVarB,\LogVarC}$$
implies $\LogRelForm{\CADCUnrelName}$ in first-order logic. This is indeed an instance of $\CADCName$ since $\LogFormA$ is of the shape $\LogRel{\LogVarC}\LogAnd\LogFormC$. Since $\forall\vec{\LogVarD}\,\LogFormC\LogImp\LogForallRel\vec{\LogVarD}\,\LogFormC$ is derivable, it is sufficient to prove the following lemma:
\begin{lem}
The following formula over the signature $\Sigma_\LogRelForm{\PAom}=\Sigma_\LogRelForm{\CAom}$:
\begin{multline*}
\LogForallRel\LogVarA\,\LogForallRel\LogVarB\left(\forall\LogVarC\,\neg\LogFormA\LogSubst{\LogVarA,\LogVarB,\LogVarC}\LogImp\forall\LogVarA'\,\LogFormA\LogSubst{\LogVarA',\LogVarB,\LogVarB}\right)\;\;\LogImp\;\;\forall\LogVarE\,\neg\LogForallRel\LogVarA\,\LogFormA\LogSubst{\LogVarA,\LogVarE\,\LogVarA,\LogVarE\left(\CALogS\,\LogVarA\right)}\;\;\LogImp\;\;\LogBot\\
\LogImp\\
\LogForallRel\LogVarA\,\LogForallRel\LogVarB\,\LogExistsRel\LogVarC\,\LogRelForm{\LogFormB}\LogSubst{\LogVarA,\LogVarB,\LogVarC}\;\;\LogImp\;\;\LogForallRel\LogVarE\,\neg\LogForallRel\LogVarA\,\LogRelForm{\LogFormB}\LogSubst{\LogVarA,\LogVarE\,\LogVarA,\LogVarE\left(\CALogS\,\LogVarA\right)}\;\;\LogImp\;\;\LogBot
\end{multline*}
where $\LogFormA\LogSubst{\LogVarA,\LogVarB,\LogVarC}\equiv\LogRel{\LogVarC}\LogAnd\LogRel{\LogVarB}\LogAnd\LogRelForm{\LogFormB}\LogSubst{\LogVarA,\LogVarB,\LogVarC}$, is provable in the logical system with relativization defined in section~\ref{RelPred}.
\end{lem}
\proof
We do this by proving the following two sequents:
\begin{align*}
\Sequent{\LogForallRel\LogVarA\,\LogForallRel\LogVarB\,\LogExistsRel\LogVarC\,\LogRelForm{\LogFormB}\LogSubst{\LogVarA,\LogVarB,\LogVarC}}{&\LogForallRel\LogVarA\,\LogForallRel\LogVarB\left(\forall\LogVarC\,\neg\LogFormA\LogSubst{\LogVarA,\LogVarB,\LogVarC}\LogImp\forall\LogVarA'\,\LogFormA\LogSubst{\LogVarA',\LogVarB,\LogVarB}\right)}{}\\
\Sequent{\LogForallRel\LogVarE\,\neg\LogForallRel\LogVarA\,\LogRelForm{\LogFormB}\LogSubst{\LogVarA,\LogVarE\,\LogVarA,\LogVarE\left(\CALogS\,\LogVarA\right)}}{&\forall\LogVarE\,\neg\LogForallRel\LogVarA\,\LogFormA\LogSubst{\LogVarA,\LogVarE\,\LogVarA,\LogVarE\left(\CALogS\,\LogVarA\right)}}{}
\end{align*}
\begin{itemize}
\item For the first one, we suppose $\LogForallRel\LogVarA\,\LogForallRel\LogVarB\,\LogExistsRel\LogVarC\,\LogRelForm{\LogFormB}\LogSubst{\LogVarA,\LogVarB,\LogVarC}$, $\LogRel{\LogVarA}$, $\LogRel{\LogVarB}$ and $\forall\LogVarC\,\neg\LogFormA\LogSubst{\LogVarA,\LogVarB,\LogVarC}$, and we want to prove $\LogRel{\LogVarB}\LogAnd\LogRel{\LogVarB}\LogAnd\LogRelForm{\LogFormB}\LogSubst{\LogVarA',\LogVarB,\LogVarB}$. $\LogRel{\LogVarB}$ is an hypothesis and we deduce $\LogRelForm{\LogFormB}\LogSubst{\LogVarA',\LogVarB,\LogVarB}$ from $\LogBot$ (which is valid since $\LogRelForm{\LogFormB}\LogSubst{\LogVarA',\LogVarB,\LogVarB}$ is a negative formula by lemma~\ref{RelNeg} and therefore $\LogBot\LogImp\LogRelForm{\LogFormB}\LogSubst{\LogVarA',\LogVarB,\LogVarB}$ is derivable) by applying the hypothesis $\LogForallRel\LogVarA\,\LogForallRel\LogVarB\left(\LogForallRel\LogVarC\,\neg\LogRelForm{\LogFormB}\LogSubst{\LogVarA,\LogVarB,\LogVarC}\LogImp\LogBot\right)$ with $\LogRel{\LogVarA}$ and $\LogRel{\LogVarB}$, so the only thing left to prove is $\LogForallRel\LogVarC\neg\LogRelForm{\LogFormB}\LogSubst{\LogVarA,\LogVarB,\LogVarC}$. We derive it from $\forall\LogVarC\neg\LogFormA\LogSubst{\LogVarA,\LogVarB,\LogVarC}$ and $\LogRel{\LogVarB}$, which amounts to proving the following sequent:
$$\Sequent{\forall\LogVarC\left(\left(\LogRel{\LogVarC}\LogAnd\LogRel{\LogVarB}\LogAnd\LogRelForm{\LogFormB}\LogSubst{\LogVarA,\LogVarB,\LogVarC}\right)\LogImp\LogBot\right),\LogRel{\LogVarB}}{\forall\LogVarC\left(\LogRel{\LogVarC}\LogImp\LogRelForm{\LogFormB}\LogSubst{\LogVarA,\LogVarB,\LogVarC}\LogImp\LogBot\right)}{}$$
which is immediate
\item The second sequent can be rewritten as:
$$\Sequent{\forall\LogVarE\left(\LogRel{\LogVarE}\LogImp\LogForallRel\LogVarA\,\LogRelForm{\LogFormB}\LogSubst{\LogVarA,\LogVarE\,\LogVarA,\LogVarE\left(\CALogS\,\LogVarA\right)}\LogImp\LogBot\right)}{\forall\LogVarE\left(\LogForallRel\LogVarA\,\LogFormA\LogSubst{\LogVarA,\LogVarE\,\LogVarA,\LogVarE\left(\CALogS\,\LogVarA\right)}\LogImp\LogBot\right)}{}$$
Therefore, it is sufficient to prove:
$$\Sequent{\LogForallRel\LogVarA\,\LogFormA\LogSubst{\LogVarA,\LogVarE\,\LogVarA,\LogVarE\left(\CALogS\,\LogVarA\right)}}{\LogRel{\LogVarE}\LogAnd\LogForallRel\LogVarA\,\LogRelForm{\LogFormB}\LogSubst{\LogVarA,\LogVarE\,\LogVarA,\LogVarE\left(\CALogS\,\LogVarA\right)}}{}$$
which is after unfolding some definitions:
$$\Sequent{\LogForallRel\LogVarA\left(\LogRel{\LogVarE\left(\CALogS\,\LogVarA\right)}\LogAnd\LogRel{\LogVarE\,\LogVarA}\LogAnd\LogRelForm{\LogFormB}\LogSubst{\LogVarA,\LogVarE\,\LogVarA,\LogVarE\left(\CALogS\,\LogVarA\right)}\right)}{\LogForallRel\LogVarA\,\LogRel{\LogVarE\,\LogVarA}\LogAnd\LogForallRel\LogVarA\,\LogRelForm{\LogFormB}\LogSubst{\LogVarA,\LogVarE\,\LogVarA,\LogVarE\left(\CALogS\,\LogVarA\right)}}{}$$
and this sequent is indeed provable in our logical system.\qed\end{itemize}
Thanks to this lemma, we can now apply lemma~\ref{RelTheo} to get for any $\LogFormA$ on the signature $\Sigma_\CAom$:
$$\CAom\Derives\LogFormA\qquad\Longrightarrow\qquad\LogRelForm{\CAom}\Derives\LogRelForm{\LogFormA}$$
\subsubsection{Interpreting the axiom of choice with bar recursion}
\label{barrecursion}
Bar recursion is an operator which can be seen as recursion on well-founded trees. It was first introduced by Spector in~\cite{Spector} to extend G\"odel's Dialectica interpretation to Heyting arithmetic augmented with the axiom of countable choice. This operator was studied in~\cite{KohlenbachThesis}, and a more uniform operator which is very similar to bar recursion was introduced in~\cite{BerardiBezemCoquand} and used in a realizability setting. A version in which the well-foundedness of trees is implicit was proposed in~\cite{BergerOlivaChoice} under the name of modified bar recursion, and we use this version here. For a comparison between these different forms of bar recursion and other similar principles we refer the reader to~\cite{PowellThesis,PowellEquivalence}.\par
In this section, we first encode lists and list operators in $\mu$PCF. Then we define the modified bar recursion operator of~\cite{BergerOlivaChoice} that we will use to provide our computational interpretation of the axiom of dependent choice.
\paragraph{Encoding lists in $\mu$PCF}
In order to define bar recursion, we first need to encode lists and operations on lists in $\mu$PCF. We choose to represent a list by a natural number (the size of the list) together with a (partial) function on natural numbers. We define the type of lists and notations for the size of a list, the empty list, and the access to a particular element:
$$\CASortList{\LmSortA}\Def\CALmnSort\LmSortTimes\left(\CALmnSort\LmSortTo\LmSortA\right)
\qquad
\CALmlen{\LmTermA}\Def\LmProj_1\,\LmTermA
\qquad
\CALmnil\Def\LmPair{\CALmn{0}}{\lambda\LmVarA.\CALmom}
\qquad
\CALmind{\LmTermA}{\LmTermB}\Def\LmProj_2\,\LmTermA\,\LmTermB$$
If $\LmTerm{\LmTermA}{\CASortList{\LmSortA}}$ and $\LmTerm{\LmTermB}{\CALmnSort}$, then it is easy to prove:
$$\LmTerm{\CALmlen{\LmTermA}}{\CALmnSort}
\qquad
\LmTerm{\CALmnil}{\CASortList{\LmSortA}}
\qquad
\CALmlen{\CALmnil}=\CALmn{0}
\qquad
\LmTerm{\CALmind{\LmTermA}{\LmTermB}}{\LmSortA}
\qquad
\CALmind{\CALmnil}{\LmTermB}=\CALmom$$
In order to define extensions of lists, we need subtraction on natural numbers $\LmTerm{\CALmsub}{\CALmsubSort}$ and tests of equality $\LmTerm{\CALmife}{\CALmifeSort{\LmSortA}}$ and strict ordering $\LmTerm{\CALmifl}{\CALmiflSort{\LmSortA}}$, which we can define in $\mu$PCF. These operators satisfy the following equations for $m,n\in\mathbb{N}$:
$$\CALmsub\,\CALmn{m}\,\CALmn{n}=\left\{\begin{aligned}&\CALmn{m-n}\text{ if }n\leq m\\&\CALmn{0}\text{ otherwise}\end{aligned}\right.
\;\;
\CALmife\,\CALmn{m}\,\CALmn{n}\,\LmTermA\,\LmTermB=\left\{\begin{aligned}&\LmTermA\text{ if }m=n\\&\LmTermB\text{ otherwise}\end{aligned}\right.
\;\;
\CALmifl\,\CALmn{m}\,\CALmn{n}\,\LmTermA\,\LmTermB=\left\{\begin{aligned}&\LmTermA\text{ if }m<n\\&\LmTermB\text{ otherwise}\end{aligned}\right.$$
We are now able to define the extension of a list by a single element:
$$\LmTermA\CALmextend\LmTermB\Def\LmPair{\CALmsucc\CALmlen{\LmTermA}}{\lambda\LmVarA.\CALmife\,\LmVarA\CALmlen{\LmTermA}\LmTermB\left(\CALmind{\LmTermA}{\LmVarA}\right)}$$
as expected, if $\LmTerm{\LmTermA}\CASortList{\LmSortA}$ and $\LmTerm{\LmTermB}{\LmSortA}$ then $\LmTerm{\LmTermA\CALmextend\LmTermB}{\CASortList{\LmSortA}}$, $\CALmlen{\LmTermA\CALmextend\LmTermB}=\CALmsucc\,\CALmlen{\LmTermA}$ and $\CALmind{\left(\LmTermA\CALmextend\LmTermB\right)}{\CALmlen{\LmTermA}}=\LmTermB$. Finally, we define infinite extension of a list with a constant element:
$$\LmTermA\CALmconcat\LmTermB\Def\lambda\LmVarA.\CALmifl\,\LmVarA\CALmlen{\LmTermA}\left(\CALmind{\LmTermA}{\LmVarA}\right)\LmTermB$$
We can derive from $\LmTerm{\LmTermA}\CASortList{\LmSortA}$ and $\LmTerm{\LmTermB}{\LmSortA}$ that $\LmTerm{\LmTermA\CALmconcat\LmTermB}{\CALmnSort\LmSortTo\LmSortA}$ and:
$$\CALmind{\left(\LmTermA_0\CALmextend\LmTermA_1\CALmextend\ldots\CALmextend\LmTermA_{n-1}\CALmconcat\LmTermB\right)}{\CALmn{m}}=\begin{cases}\LmTermA_m\text{ if }m<n\\\LmTermB\text{ otherwise}\end{cases}$$
\paragraph{The bar recursion operator}
\label{barrec}
We have now all necessary material to define formally the bar recursion operator:
$$\CALmbarrec\Def\lambda\LmVarD\LmVarE.\CALmfix\left(\lambda\LmVarC\LmVarB.\LmVarE\left(\LmVarB\CALmconcat\left(\LmVarD\,\LmVarB\left(\lambda\LmVarA.\LmVarC\left(\LmVarB\CALmextend\LmVarA\right)\right)\right)\right)\right)$$
Bar recursor can be typed as expected:
$$\LmTerm{\CALmbarrec}{\CALmbarrecSort{\LmSortA}{\LmSortB}}$$
And it verifies indeed the equation:
$$\CALmbarrec\,\LmTermA\,\LmTermB\,\LmTermC=\LmTermB\left(\LmTermC\CALmconcat\LmTermA\,\LmTermC\left(\lambda\LmVarA.\CALmbarrec\,\LmTermA\,\LmTermB\left(\LmTermC\CALmextend\LmVarA\right)\right)\right)$$
\paragraph{Interpreting dependent choice using bar recursion}
We use here the bar-recursion operator to provide the term $\LmInterpAxiom{\CADCName}$ interpreting the axiom of dependent choice. Remember that $\CADCName$ is:
$$\CADC{\SortA}{\LogFormA}{\SortB}$$
where $\LogFormA$ is a formula over $\Sigma_\LogRelForm{\CAom}=\Sigma_\LogRelForm{\PAom}$ with free variables among $\LogSortedTerm{\LogVarA}{\CASort},\LogSortedTerm{\LogVarB}{\SortA},\LogSortedTerm{\LogVarC}{\SortA},\LogSortedTerm{\vec{\LogVarD}}{\vec{\SortB}}$ and which is of the shape $\LogRel{\LogVarC}\LogAnd\LogFormC$. As in the previous section, we write $\LogFormA\LogSubst{\LogTermA,\LogTermB,\LogTermC}$ instead of $\LogFormA\LogSubst{\LogTermA/\LogVarA,\LogTermB/\LogVarB,\LogTermC/\LogVarC}$. The type of $\LmInterpAxiom{\CADCName}$, $\LmInterpForm{\CADCName}$ is:
$$\LmTerm{\LmInterpAxiom{\CADCName}}{\left(\CALmnSort\LmSortTo\LmSortA\LmSortTo\left(\LmInterpForm{\LogFormA}\LmSortTo\LmSortBot\right)\LmSortTo\LmInterpForm{\LogFormA}\right)\LmSortTo\left(\left(\CALmnSort\LmSortTo\LmInterpForm{\LogFormA}\right)\LmSortTo\LmSortBot\right)\LmSortTo\LmSortBot}$$
In order to define $\LmInterpAxiom{\CADCName}$ we make an informal reasoning. Suppose $\LmTerm{\LmTermA}{\CASort\LmSortTo\LmSortA\LmSortTo\left(\LmInterpForm{\LogFormA}\LmSortTo\LmSortBot\right)\LmSortTo\LmInterpForm{\LogFormA}}$ is a witness of:
$$\LogForallRel\LogSortedTerm{\LogVarA}{\CASort}\LogForallRel\LogSortedTerm{\LogVarB}{\SortA}\left(\forall\LogSortedTerm{z}{\SortA}\neg\LogFormA\LogSubst{\LogVarA,\LogVarB,\LogVarC}\LogImp\forall\LogSortedTerm{\LogVarA'}{\CASort}\LogFormA\LogSubst{\LogVarA',\LogVarB,\LogVarB}\right)$$
and $\LmTerm{\LmTermB}{\left(\CALmnSort\LmSortTo\LmInterpForm{\LogFormA}\right)\LmSortTo\LmSortBot}$ is a witness of:
$$\forall\LogSortedTerm{\LogVarE}{\CASort\to\SortA}\neg\LogForallRel\LogSortedTerm{\LogVarA}{\CASort}\LogFormA\LogSubst{\LogVarA,\LogVarE\,\LogVarA,\LogVarE\left(\CALogS\,\LogVarA\right)}$$
We want to build from this, using $\CALmbarrec$, a witness of $\LogBot$. We will use the following instance of $\CALmbarrec$:
$$\LmTerm{\CALmbarrec}{\CALmbarrecSort{\LmInterpForm{\LogFormA}}{\LmSortBot}}$$
The idea now is that $\CALmbarrec$ will build a sequence of witnesses of $\LogFormA\LogSubst{\LogVarA,\LogVarE\,\LogVarA,\LogVarE\left(\CALogS\,\LogVarA\right)}$. The first argument represents the recursive step. If we have an element $\LmTerm{\LmVarF}{\CASortList{\LogFormA}}$ which represents the sequence of witnesses already computed, then $\LmTermA$ computes the next element of $\LmVarF$, given its length and last element. Here we have two cases, the first one is when $\CALmlen{\LmVarF}=\CALmn{0}$, so we must initialize the sequence with an element of type $\LmInterpForm{\LogRel{\LogSortedTerm{.}{\SortA}}}$ that we can choose arbitrarily. Since $\SortA$ is a sort of the logic, we can write it as $\SortA_1\SortTo\ldots\SortTo\SortA_n\SortTo\SortBase$ so:
$$\LmInterpForm{\LogRel{\LogSortedTerm{.}{\SortA}}}=\LmInterpForm{\LogRel{\LogSortedTerm{.}{\SortA_1}}}\LmSortTo\ldots\LmSortTo\LmInterpForm{\LogRel{\LogSortedTerm{.}{\SortA_n}}}\LmSortTo\CALmnSort$$
and we define this arbitrary element to be $\lambda\LmVarA_1\ldots\LmVarA_n.\CALmn{0}$. In the following, we will simply write this term as $\CALmn{0}$, leaving the type implicit. In the second case, the last element of $\LmVarF$ is $\CALmind{\LmVarF}{\CALmpred\,\CALmlen{\LmVarF}}$. Therefore, we provide $\LmTermA$ with $\CALmlen{\LmVarF}$ and $\CALmifz\,\CALmlen{\LmVarF}\,\CALmn{0}\left(\LmProj_1\,\CALmind{\LmVarF}{\CALmpred\,\CALmlen{\LmVarF}}\right)$. The $\LmProj_1$ is because the last element of $\LmVarF$ is a witness of $\LogFormA\LogSubst{n-2,\LogVarE\,\left(n-2\right),\LogVarE\left(n-1\right)}$ where $n$ is the length of $\LmVarF$, so since $\LogFormA$ is of shape $\LogRel{\LogVarC}\LogAnd\LogFormC$, $\LmProj_1\,\CALmind{\LmVarF}{\CALmpred\,\CALmlen{\LmVarF}}$ is a witness of $\LogRel{\LogVarE\left(n-1\right)}$. The first argument is then:
$$\lambda\LmVarF.\LmTermA\,\CALmlen{\LmVarF}\left(\CALmifz\,\CALmlen{\LmVarF}\,\CALmn{0}\left(\LmProj_1\,\CALmind{\LmVarF}{\CALmpred\,\CALmlen{\LmVarF}}\right)\right)$$
The second argument represents the behavior if we have an infinite sequence of witnesses $\LmTerm{\LmVarF}{\CALmnSort\LmSortTo\LogFormB}$. In that case we simply provide $\LmTermB$ with this argument $\LmVarF$, so the third argument is just $\LmTermB$. Finally, the last argument of $\CALmbarrec$ is the initial sequence of witnesses, that is the empty sequence $\CALmnil$. We have then:
$$\LmTerm{\CALmbarrec\left(\lambda\LmVarF.\LmTermA\,\CALmlen{\LmVarF}\left(\CALmifz\,\CALmlen{\LmVarF}\,\CALmn{0}\left(\LmProj_1\,\CALmind{\LmVarF}{\CALmpred\,\CALmlen{\LmVarF}}\right)\right)\right)\LmTermB\,\CALmnil}{\LmSortBot}$$
The interpretation of $\CADCName$ is therefore defined as:
$$\LmInterpAxiom{\CADCName}\Def\CADCTerm$$
\subsection{The realizability relation}
\subsubsection{Negative translation and orthogonality}
\label{NegTrans}
The first realizability models for classical logic were obtained by combining G\"odel's negative translation with intuitionistic realizability~\cite{BerardiBezemCoquand,BergerOlivaChoice,KohlenbachProofTheory}. G\"odel's negative translation from Peano arithmetic $\PA^\omega_=$ (equivalent to $\PAom$) to $\HA^\omega_=$ (see section~\ref{usualtheories}) maps a formula $\LogFormA$ to $\LogFormA^\neg$ by prefixing inductively all the positive connectives and atomic predicates of $\LogFormA$ with a double negation. It holds that if $\PA^\omega_=\Derives\LogFormA$, then $\HA^\omega_=\Derives\LogFormA^\neg$, and the proof of this relies on the fact that for every axiom $\LogFormA$ of $\PA^\omega_=$, $\HA^\omega_=\Derives\LogFormA\LogImp\LogFormA^\neg$. Therefore, a realizability model for $\PA^\omega_=$ can be obtained from a realizability model for $\HA^\omega_=$ using G\"odel's negative translation. Concerning the extraction of witnesses, if $\PA^\omega_=\Derives\exists\LogSortedTerm{\LogVarA}{\CASort}\left(\LogTermA=\CALogZ\right)$, then $\HA^\omega_=\Derives\neg\neg\exists\LogSortedTerm{\LogVarA}{\CASort}\,\neg\neg\left(\LogTermA=\CALogZ\right)$ from which we easily get $\HA^\omega_=\Derives\neg\neg\exists\LogSortedTerm{\LogVarA}{\CASort}\left(\LogTermA=\CALogZ\right)$. While in usual intuitionistic realizability the formula $\LogBot$ has no realizer so the model is sound, Friedman's trick is to allow $\LogBot$ to have realizers. If we then take the realizers of $\LogBot$ to be the same as those of $\exists\LogSortedTerm{\LogVarA}{\CASort}\left(\LogTermA=\CALogZ\right)$, then combining the proof of $\neg\neg\exists\LogSortedTerm{\LogVarA}{\CASort}\left(\LogTermA=\CALogZ\right)$ with the identity gives a realizer of $\exists\LogSortedTerm{\LogVarA}{\CASort}\left(\LogTermA=\CALogZ\right)$, and therefore the witness.\par
In Krivine's~\cite{KrivinePanoramas} classical realizability models this double step (negative translation + intuitionistic realizability) is avoided through the use of orthogonality in system F. In these models there is a set of terms $\Lambda$ and a set of stacks $\Pi$. Each formula has a set of realizers (the truth value, subset of $\Lambda$) and a set of counter-realizers (the falsity value, subset of $\Pi$). The falsity values are primitive, an orthogonality relation is defined between $\Lambda$ and $\Pi$, and the truth values are defined as the orthogonals of the falsity values, so they are orthogonally closed.\par
Here we work in a typed setting so we must choose the types of realizers and counter-realizers so they can interact. Given a formula $\LogFormA$, the set of realizers of $\LogFormA$ would normally be a set of morphisms in a category of continuations $\CatRC$ from the terminal object $\CatTerm$ to the interpretation of $\LogFormA$: $\CatInterpSort{\LmInterpForm{\LogFormA}}=\CatExp{\CatR}{\CatInterpSortNeg{\LmInterpForm{\LogFormA}}}$, and under the duality between terms and contexts of $\lambda\mu$-calculus, a natural choice for the counter-realizers of $\LogFormA$ is a set of morphisms in $\CatC$ from $\CatTerm$ to $\CatInterpSortNeg{\LmInterpForm{\LogFormA}}$. Then we can combine a potential realizer of $\LogFormA$ with a potential counter-realizer using the evaluation morphism $\CatEval:\CatExp{\CatR}{\CatInterpSortNeg{\LmInterpForm{\LogFormA}}}\CatTimes\CatInterpSortNeg{\LmInterpForm{\LogFormA}}\to\CatR$ so we obtain a morphism from $\CatTerm$ to $\CatR$. The potential realizer and counter-realizer are orthogonal to each other or not, depending on the result. Therefore, in order to define a non-trivial orthogonality relation, there must be at least two morphisms from $\CatTerm$ to $\CatR$, and if we want to perform extraction this homset has to be isomorphic to the set of values we want to extract. In realizability for arithmetic, this is usually done by choosing an object $\CatR$ which is the same as the interpretation of natural numbers, however this choice has some drawbacks. Indeed, some computational models can be naturally seen as categories of continuations for a given $\CatR$, and this $\CatR$ may not be isomorphic to the object of natural numbers. Take for example the model of Hyland-Ong games~\cite{HO}. We know that by relaxing the well-bracketing condition we obtain a fully abstract model of $\mu$PCF~\cite{LairdControl}. It is therefore not a surprise that the same game model is a category of continuations $\CatRC$, and it turns out that in this category of continuations, the object $\CatR$ is the one-move arena. The key point for this is that the arena of natural numbers is the exponential of the one-move arena by the countable product of one-move arenas, details can be found in~\cite{BlotThesis}, section 4.4. But since there is only one strategy on the one-move arena (the empty strategy), we cannot easily define a non-trivial orthogonality relation.\par
Therefore we choose here a different approach and rely on Friedman's trick directly in the definition of the realizability relation: our orthogonality relation relies on an artificially added output channel. Formally we add a $\mu$-variable in the process of interpreting logic in $\lambda\mu$-calculus. A proof:
$$\AXM{\LogProofA}\UIM{\Sequent{\LogFormA_1,\ldots,\LogFormA_n}{\LogFormA}{\LogNeg{\LogFormB_1},\ldots,\LogNeg{\LogFormB_m}}}\DP$$
is now translated to a $\lambda\mu$-term:
$$\Sequent{\LmTerm{\LmVarA_1}{\LmInterpForm{\LogFormA_1}},\ldots,\LmTerm{\LmVarA_n}{\LmInterpForm{\LogFormA_n}}}{\LmTerm{\LmInterpProof{\LogProofA}}{\LmInterpForm{\LogFormA}}}{\LmTerm{\LmMVarA_1}{\LmInterpForm{\LogFormB_1}},\ldots,\LmTerm{\LmMVarA_n}{\LmInterpForm{\LogFormB_m}},\LmTerm{\kappa}{\LmSortExtract}}$$
where $\LmSortExtract$ is some fixed base type, by applying the (admissible) rule of right weakening of $\lambda\mu$-calculus. The $\mu$-variable $\kappa$ is, intuitively, a continuation variable which can be used by a realizer or a counter-realizer to stop computation and give an answer. Apart from its use in the definition of the realizability relation, this feature will also be used in the proof of the extraction result, for which $\LmSortExtract$ will be instantiated with the type of natural numbers. After translating the proof $\LogProofA$ to a $\lambda\mu$-term $\LmInterpProof{\LogProofA}$, we interpret it in $\CatRC$ as a morphism:
$$\LmInterpProof{\LogProofA}\in\CatRC\left(\CatInterpSort{\LmInterpForm{\LogFormA_1}}\CatTimes\ldots\CatTimes\CatInterpSort{\LmInterpForm{\LogFormA_n}},\CatInterpSort{\LmInterpForm{\LogFormA}}\CatPar\CatInterpSort{\LmInterpForm{\LogFormB_1}}\CatPar\ldots\CatPar\CatInterpSort{\LmInterpForm{\LogFormB_m}}\CatPar\CatInterpSort{\LmSortExtract}\right)$$
In the particular case of empty contexts, $\LmInterpProof{\LogProofA}$ is a morphism in $\CatRC\left(\CatTerm,\CatInterpSort{\LmInterpForm{\LogFormA}}\CatPar\CatInterpSort{\LmSortExtract}\right)$, and we therefore choose the potential realizers of a closed formula $\LogFormA$ to be such morphisms. Similarly we choose the potential counter-realizers of $\LogFormA$ to be morphisms in $\CatC\left(\CatInterpSortNeg{\LmSortExtract},\CatInterpSortNeg{\LmInterpForm{\LogFormA}}\right)$. As explained in section~\ref{CatInterp} and by taking the convention that the potential realizers have a free $\mu$-variable $\kappa$ of type $\LmSortExtract$, we use the syntax of $\lambda\mu$-calculus (and possibly $\left[\kappa\right]$ and $\mu\kappa$) to manipulate these. We also substitute morphisms of $\CatC$ for $\mu$-variables, as in section~\ref{interaction}, so if $\CatRCHomA\in\CatRC\left(\CatTerm,\CatInterpSort{\LmInterpForm{\LogFormA}}\CatPar\CatInterpSort{\LmSortExtract}\right)$ is a potential realizer, and if $\CatCHomA\in\CatC\left(\CatInterpSortNeg{\LmSortExtract},\CatInterpSortNeg{\LmInterpForm{\LogFormA}}\right)$ is a potential counter-realizer, then $\left[\CatCHomA\right]\,\CatRCHomA\in\CatRC\left(\CatTerm,\CatInterpSort{\LmSortBot}\CatPar\CatInterpSort{\LmSortExtract}\right)$, since $\CatInterpSort{\LmInterpForm{\LogFormA}}=\CatExp{\CatR}{\CatInterpSortNeg{\LmInterpForm{\LogFormA}}}$ and $\CatInterpSort{\LmSortBot}=\CatR$, and therefore $\mu\kappa.\left[\CatCHomA\right]\,\CatRCHomA\in\CatRC\left(\CatTerm,\CatInterpSort{\LmSortExtract}\right)$. Since the object $\CatInterpSort{\LmSortExtract}=\CatExp{\CatR}{\CatInterpSortNeg{\LmSortExtract}}$ can (and will) be chosen larger than $\CatInterpSort{\LmSortBot}=\CatR$, we can now define when $\CatCHomA$ is orthogonal to $\CatRCHomA$ depending on $\mu\kappa.\left[\CatCHomA\right]\,\CatRCHomA$. Typically, in the model of unbracketed games, $\CatR$ is the one-move arena and $\CatInterpSortNeg{\LmSortExtract}$ is the countable product of the one-move arena, so $\CatInterpSort{\LmSortExtract}=\CatExp{\CatR}{\CatInterpSortNeg{\LmSortExtract}}$ is indeed the usual arena of natural numbers.\par
The choice of having a separate $\mu$-variable instead of choosing a big enough object $\CatR$ can also provide a simpler interpretation of proofs in $\CatRC$. We give a comparison of our interpretation of the identity proof of $\LogBot\LogImp\LogBot$ in arithmetic with the usual interpretation in figure~\ref{Comparison}, where we take $\CatRC$ to be the unbracketed games model.
\begin{figure}
$$\begin{array}{|c@{\hspace{20pt}}|@{\hspace{20pt}}c@{\hspace{40pt}}c|}
\hline
&
\text{usual interpretation:}
&
\text{our interpretation:}
\\[10pt]
\hline
\text{arena:}
&
\vcenter{\xymatrix@!0@R=30pt@C=15pt
{
&&&&&&q\\
&&q'\ar@{-}[urrrr]&&0\ar@{-}[urr]&1\ar@{-}[ur]&\ldots&n\ar@{-}[ul]&\ldots\\
0'\ar@{-}[urr]&1'\ar@{-}[ur]&\ldots&n'\ar@{-}[ul]&\ldots
}}
&
\vcenter{\xymatrix@!0@R=30pt@C=15pt
{
&&&&q\\
q'\ar@{-}[urrrr]&&0_1\ar@{-}[urr]&1_1\ar@{-}[ur]&\ldots&n_1\ar@{-}[ul]&\ldots\\
}}\\[40pt]
\text{views of the strategy:}
&
\SetSuch{\epsilon;qq'}{}\cup\SetSuch{qq'mm'}{m\in\mathbb{N}}
&
\SetSuch{\epsilon;qq'}{}
\\
\hline
\end{array}$$
\caption{Comparison of our interpretation with the usual one}
\label{Comparison}
\end{figure}
We can see that with the usual interpretation where $\CatR$ is the object of natural numbers, this proof is interpreted as a strategy from natural numbers to natural numbers. However, in our interpretation the same proof is interpreted as a strategy on the arena $\CatExp{\CatInterpSort{\LmSortBot}}{\CatInterpSort{\LmSortBot}}\CatPar\CatInterpSort{\LmSortExtract}$ where $\CatInterpSort{\LmSortBot}$ is the one-move arena and $\CatInterpSort{\LmSortExtract}$ is the arena of natural numbers (indeed, the $\CatPar$ operation on arenas is the merge of roots). This simplification doesn't happen in every model however. For example, in Scott domains, the natural choice of taking the singleton set for $\CatR$ leads to a degenerated model, since $\CatExp{\CatR}{\CatObjA}$ is then isomorphic to $\CatR$ for any domain $\CatObjA$. In order to get a non-degenerated model, we would need to choose for $\CatR$ a bigger domain, and therefore we would lose the benefit of our simpler interpretation.\par
In~\cite{BlotRibaBarRec} the goals of orthogonally-defined realizability and extraction were achieved similarly by taking realizers of a formula $\LogFormA$ to be interpretations of closed terms of type $\CatInterpSortNeg{\LmSortExtract}\to\CatInterpSort{\LmInterpForm{\LogFormA}}$, but also interpreting the atomic formulas and $\LogBot$ with the base type $\LmSortExtract$ instead of the empty type $\LmSortBot$, which led to unnecessary complex types for the realizers.\par
To summarize, in order to have the benefit of the simpler interpretation, our categorical model $\CatRC$ needs to have an object $\CatR$ which is significantly simpler than $\CatInterpSort{\LmSortExtract}=\CatExp{\CatR}{\CatInterpSortNeg{\LmSortExtract}}$, and $\CatInterpSort{\LmSortExtract}$ should be large enough to interpret natural numbers. This rules out Scott domains, but unbracketed games models do have this property. It would be interesting to find other examples of such models, the main candidates being Laird's bistable biorders~\cite{LairdBistable} and Berry and Curien's sequential algorithms~\cite{BerryCurienSequential}, which are both fully abstract models of $\mu$PCF.
\subsubsection{Truth values, falsity values}
\label{OrthoReal}
We fix a first-order signature $\Sigma$ and a $\Sigma$-structure $\ModM$. We also fix a corresponding $\lambda\mu$ signature and a type $\LmInterpForm{\LogPredA}$ in this signature for each positive predicate $\LogPredA$ of $\Sigma$, as in section~\ref{LmInterp}. We interpret $\lambda\mu$-calculus in a category of continuations $\CatRC$ as in section~\ref{CatInterp} and we use the syntax of $\lambda\mu$-calculus to describe strategies of $\CatRC$, so we omit the interpretation brackets. All the $\lambda\mu$-terms that we write from now on are to be understood as morphisms in $\CatRC$.\par
In order to build our realizability relation by orthogonality, and later on to perform extraction on $\Pi^0_2$ formulas using Friedman's trick, our model is parameterized with a set
$$\RealBot\subseteq\CatRC\left(\CatTerm,\CatInterpSort{\LmSortExtract}\right)$$
which is, intuitively, the set of ``correct'' values that can be output through the variable $\kappa$.\par
We now define the set of realizers of a formula, that we call its truth value. We fix for each positive predicate $\LogPredA$ of $\Sigma$ and each $\ModElemA_1,\ldots,\ModElemA_n\in\ModMInterp{\SortA_1}\times\ldots\times\ModMInterp{\SortA_n}$ a set of morphisms $\RealVal{\LogPredA\left(\ModElemA_1,\ldots,\ModElemA_n\right)}\subseteq\CatRC\left(\CatTerm,\CatInterpSort{\LmInterpForm{\LogPredA}}\CatPar\CatInterpSort{\LmSortExtract}\right)$. The extension of this to every closed formula $\LogFormA$ on $\Sigma$ with parameters in $\ModM$ is given in Figure~\ref{TruthValues},
\begin{figure}
\begin{align*}
\RealVal{\LogFormA\LogImp\LogFormB}&=\SetSuch{\CatRCHomA}{\forall\CatRCHomB\in\RealVal{\LogFormA},\CatRCHomA\,\CatRCHomB\in\RealVal{\LogFormB}}&
\RealVal{\forall\LogSortedTerm{\LogVarA}{\SortA}\,\LogFormA}&=\bigcap_{\ModElemA\in\ModMInterp{\SortA}}\RealVal{\LogFormA\LogSubst{\ModElemA/\LogVarA}}\\
\RealVal{\LogFormA\LogAnd\LogFormB}&=\SetSuch{\CatRCHomA}{\LmProj_1\,\CatRCHomA\in\RealVal{\LogFormA}\wedge\LmProj_2\,\CatRCHomA\in\RealVal{\LogFormB}}&
\RealVal{\LogBot}&=\SetSuch{\CatRCHomA}{\mu\kappa.\CatRCHomA\in\RealBot}
\end{align*}
\begin{align*}
\RealVal{\LogNeg{\LogPredA}\left(\ModElemA_1,\ldots,\ModElemA_n\right)}&=\left\{\begin{aligned}&\CatRC\left(\CatTerm,\CatInterpSort{\LmSortBot}\CatPar\CatInterpSort{\LmSortExtract}\right)&&\text{ if }\ModM\Models\LogNeg{\LogPredA}\left(\ModElemA_1,\ldots,\ModElemA_n\right)\\&\SetSuch{\CatRCHomA}{\mu\kappa.\CatRCHomA\in\RealBot}&&\text{ otherwise}\end{aligned}\right.
\end{align*}
\caption{Truth values}
\label{TruthValues}
\end{figure}
so $\RealVal{\LogFormA}\subseteq\CatRC\left(\CatTerm,\CatInterpSort{\LmInterpForm{\LogFormA}}\CatPar\CatInterpSort{\LmSortExtract}\right)$. Remark that contrary to~\cite{KrivinePanoramas}, we do not define the truth values as the orthogonals of the falsity values. In our logical system, only some predicates are negative, and therefore only some formulas are negative. In the realizability interpretation, only the negative formulas are given a falsity value and their truth values will be proved to be orthogonal to their falsity values. For the other formulas the truth value is primitive and may not be bi-orthogonally closed.\par
For every closed negative formula $\LogNeg{\LogFormA}$ with parameters in $\ModM$ the falsity value of $\LogNeg{\LogFormA}$ is given in Figure~\ref{FalsityValues},
\begin{figure}
\begin{align*}
\RealValNeg{\LogFormA\LogImp\LogNeg{\LogFormB}}&=\SetSuch{\LmPair{\LamCPS{\CatRCHomA}}{\CatCHomA}}{\CatRCHomA\in\RealVal{\LogFormA}\wedge\CatCHomA\in\RealValNeg{\LogNeg{\LogFormB}}}&
\RealValNeg{\forall\LogSortedTerm{\LogVarA}{\SortA}\,\LogNeg{\LogFormA}}&=\bigcup_{\ModElemA\in\ModMInterp{\SortA}}\RealValNeg{\LogNeg{\LogFormA}\LogSubst{\ModElemA/\LogVarA}}\\
\RealValNeg{\LogNeg{\LogFormA}\LogAnd\LogNeg{\LogFormB}}&=\SetSuch{\LamIn_1\,\CatCHomA}{\CatCHomA\in\RealValNeg{\LogNeg{\LogFormA}}}\cup\SetSuch{\LamIn_2\,\CatCHomA}{\CatCHomA\in\RealValNeg{\LogNeg{\LogFormB}}}&
\RealValNeg{\LogBot}&=\SetSuch{\LamUnit}{}
\end{align*}
\begin{align*}
\RealValNeg{\LogNeg{\LogPredA}\left(\ModElemA_1,\ldots,\ModElemA_n\right)}&=\left\{\begin{aligned}&\emptyset&&\text{ if }\ModM\Models\LogNeg{\LogPredA}\left(\ModElemA_1,\ldots,\ModElemA_n\right)\\&\SetSuch{\LamUnit}{}&&\text{ otherwise}\end{aligned}\right.
\end{align*}
\caption{Falsity values}
\label{FalsityValues}
\end{figure}
so $\RealValNeg{\LogNeg{\LogFormA}}\subseteq\CatC\left(\CatInterpSortNeg{\LmSortExtract},\CatInterpSortNeg{\LmInterpForm{\LogFormA}}\right)$. As explained in section~\ref{cps}, we use here the syntax of $\Lam$ to manipulate morphisms in $\CatC$. We define now an orthogonality relation between $\CatRC\left(\CatTerm,\CatInterpSort{\LmInterpForm{\LogFormA}}\CatPar\CatInterpSort{\LmSortExtract}\right)$ and $\CatC\left(\CatInterpSortNeg{\LmSortExtract},\CatInterpSortNeg{\LmInterpForm{\LogFormA}}\right)$: if $\CatRCHomA\in\CatRC\left(\CatTerm,\CatInterpSort{\LmInterpForm{\LogFormA}}\CatPar\CatInterpSort{\LmSortExtract}\right)$ and $\CatCHomA\in\CatC\left(\CatInterpSortNeg{\LmSortExtract},\CatInterpSortNeg{\LmInterpForm{\LogFormA}}\right)$, then $\mu\kappa.\left[\CatCHomA\right]\,\CatRCHomA\in\CatRC\left(\CatTerm,\CatInterpSort{\LmSortExtract}\right)$, so we define:
$$\CatRCHomA\ \bot\ \CatCHomA\quad\Def\quad\mu\kappa.\left[\CatCHomA\right]\,\CatRCHomA\in\RealBot$$
The following lemma states that the truth values are indeed the orthogonals of the falsity values defined above for negative formulas:
\begin{lem}
\label{orthogonal}
For every closed negative formula $\LogNeg{\LogFormA}$ with parameters in $\ModM$:
$$\RealVal{\LogNeg{\LogFormA}}=\SetSuch{\CatRCHomA}{\forall\CatCHomA\in\RealValNeg{\LogNeg{\LogFormA}},\ \CatRCHomA\ \bot\ \CatCHomA}$$
\end{lem}
\proof
We prove this result by induction on the structure of the formula:
\begin{itemize}
\item$\LogBot$: if $\CatRCHomA\in\CatRC\left(\CatTerm,\CatInterpSort{\LmInterpForm{\LogBot}}\CatPar\CatInterpSort{\LmSortExtract}\right)$ then:
$$\CatRCHomA\in\RealVal{\LogBot}\Leftrightarrow\mu\kappa.\CatRCHomA\in\RealBot\Leftrightarrow\mu\kappa.\left[\LamUnit\right]\,\CatRCHomA\in\RealBot\Leftrightarrow\CatRCHomA\ \bot\ \LamUnit$$
from which we conclude since $\RealValNeg{\LogBot}=\SetSuch{\LamUnit}{}$.
\item$\LogNeg{\LogPredA}\left(\ModElemA_1,\ldots,\ModElemA_n\right)$: if $\ModM\Models\LogNeg{\LogPredA}\left(\ModElemA_1,\ldots,\ModElemA_n\right)$ then the result is immediate, and otherwise the proof is the same as for $\LogBot$.
\item$\LogNeg{\LogFormA}\LogAnd\LogNeg{\LogFormB}$: if $\CatRCHomA\in\CatRC\left(\CatTerm,\left(\CatInterpSort{\LmInterpForm{\LogFormA}}\CatTimes\CatInterpSort{\LmInterpForm{\LogFormB}}\right)\CatPar\CatInterpSort{\LmSortExtract}\right)$, $\CatCHomA_1\in\CatC\left(\CatInterpSortNeg{\LmSortExtract},\CatInterpSortNeg{\LmInterpForm{\LogFormA}}\right)$, $\CatCHomA_2\in\CatC\left(\CatInterpSortNeg{\LmSortExtract},\CatInterpSortNeg{\LmInterpForm{\LogFormB}}\right)$, then $\mu\kappa.\left[\LamIn_i\,\CatCHomA_i\right]\,\CatRCHomA=\mu\kappa.\left[\CatCHomA_i\right]\,\LmProj_i\,\CatRCHomA$, therefore:
$$\LmProj_i\,\CatRCHomA\ \bot\ \CatCHomA_i\Leftrightarrow\mu\kappa.\left[\CatCHomA_i\right]\,\LmProj_i\,\CatRCHomA\in\RealBot\Leftrightarrow\mu\kappa.\left[\LamIn_i\,\CatCHomA_i\right]\,\CatRCHomA\in\RealBot\Leftrightarrow\CatRCHomA\ \bot\ \LamIn_i\,\CatCHomA_i$$
and finally:
\begin{align*}
\sigma\in\RealVal{\LogNeg{\LogFormA}\LogAnd\LogNeg{\LogFormB}}&\Leftrightarrow\left\{\begin{gathered}\LmProj_1\,\CatRCHomA\in\RealVal{\LogNeg{\LogFormA}}\\\LmProj_2\,\CatRCHomA\in\RealVal{\LogNeg{\LogFormB}}\end{gathered}\right.\\
&\Leftrightarrow\left\{\begin{gathered}\forall\CatCHomA_1\in\RealValNeg{\LogNeg{\LogFormA}},\LmProj_1\,\CatRCHomA\ \bot\ \CatCHomA_1\\\forall\CatCHomA_2\in\RealValNeg{\LogNeg{\LogFormB}},\LmProj_2\,\CatRCHomA\ \bot\ \CatCHomA_2\end{gathered}\right.\text{ by induction hypothesis}\\
&\Leftrightarrow\left\{\begin{gathered}\forall\CatCHomA_1\in\RealValNeg{\LogNeg{\LogFormA}},\CatRCHomA\ \bot\ \LamIn_1\,\CatCHomA_1\\\forall\CatCHomA_2\in\RealValNeg{\LogNeg{\LogFormB}},\CatRCHomA\ \bot\ \LamIn_2\,\CatCHomA_2\end{gathered}\right.\\
&\Leftrightarrow\forall\CatCHomA\in\RealValNeg{\LogNeg{\LogFormA}\LogAnd\LogNeg{\LogFormB}},\CatRCHomA\ \bot\ \CatCHomA
\end{align*}
\item$\LogFormA\LogImp\LogNeg{\LogFormB}$: if $\CatRCHomA\in\CatRC\left(\CatTerm,\CatExp{\CatInterpSort{\LmInterpForm{\LogFormB}}}{\CatInterpSort{\LmInterpForm{\LogFormA}}}\CatPar\CatInterpSort{\LmSortExtract}\right)$, $\CatRCHomB\in\CatRC\left(\CatTerm,\CatInterpSort{\LmInterpForm{\LogFormA}}\CatPar\CatInterpSort{\LmSortExtract}\right)$ and $\CatCHomA\in\CatC\left(\CatInterpSortNeg{\LmSortExtract},\CatInterpSortNeg{\LmInterpForm{\LogFormB}}\right)$, then $\mu\kappa.\left[\LmPair{\LamCPS{\CatRCHomB}}{\CatCHomA}\right]\,\CatRCHomA=\left[\CatCHomA\right]\,\CatRCHomA\,\CatRCHomB$, therefore:
$$\CatRCHomA\,\CatRCHomB\ \bot\ \CatCHomA\Leftrightarrow\mu\kappa.\left[\CatCHomA\right]\,\CatRCHomA\,\CatRCHomB\in\RealBot\Leftrightarrow\mu\kappa.\left[\LmPair{\LamCPS{\CatRCHomB}}{\CatCHomA}\right]\,\CatRCHomA\in\RealBot\Leftrightarrow\CatRCHomA\ \bot\ \LmPair{\LamCPS{\CatRCHomB}}{\CatCHomA}$$
and finally:
\begin{align*}
\CatRCHomA\in\RealVal{\LogFormA\LogImp\LogNeg{\LogFormB}}&\Leftrightarrow\forall\CatRCHomB\in\RealVal{\LogFormA},\CatRCHomA\,\CatRCHomB\in\RealVal{\LogNeg{\LogFormB}}\\
&\Leftrightarrow\forall\CatRCHomB\in\RealVal{\LogFormA},\forall\CatCHomA\in\RealValNeg{\LogNeg{\LogFormB}},\CatRCHomA\,\CatRCHomB\ \bot\ \CatCHomA\qquad\text{by induction hypothesis}\\
&\Leftrightarrow\forall\CatRCHomB\in\RealVal{\LogFormA},\forall a\in\RealValNeg{\LogNeg{\LogFormB}},\CatRCHomA\ \bot\ \LmPair{\LamCPS{\CatRCHomB}}{a}\\
&\Leftrightarrow\forall\CatCHomA'\in\RealValNeg{\LogFormA\LogImp\LogNeg{\LogFormB}},\CatRCHomB\ \bot\ \CatCHomA'
\end{align*}
\item$\forall\LogSortedTerm{\LogVarA}{\SortA}\,\LogNeg{\LogFormA}$: if $\CatRCHomA\in\CatRC\left(\CatTerm,\CatInterpSort{\LmInterpForm{\LogFormA}}\CatPar\CatInterpSort{\LmSortExtract}\right)$, then:
\begin{align*}
\CatRCHomA\in\RealVal{\forall\LogSortedTerm{\LogVarA}{\SortA}\,\LogNeg{\LogFormA}}&\Leftrightarrow\forall\ModElemA\in\ModMInterp{\SortA},\CatRCHomA\in\RealVal{\LogNeg{\LogFormA}\LogSubst{\ModElemA/\LogVarA}}\\
&\Leftrightarrow\forall\ModElemA\in\ModMInterp{\SortA},\forall\CatCHomA\in\RealValNeg{\LogNeg{\LogFormA}\LogSubst{\ModElemA/\LogVarA}},\CatRCHomA\ \bot\ \CatCHomA\\
&\Leftrightarrow\forall\CatCHomA\in\CatC\left(\CatInterpSortNeg{\LmSortExtract},\CatInterpSortNeg{\LmInterpForm{\LogFormA}}\right),\left(\exists\ModElemA\in\ModMInterp{\SortA},\CatCHomA\in\RealValNeg{\LogNeg{\LogFormA}\LogSubst{\ModElemA/\LogVarA}}\right)\Rightarrow\CatRCHomA\ \bot\ \CatCHomA\\
&\Leftrightarrow\forall\CatCHomA\in\RealValNeg{\forall\LogSortedTerm{\LogVarA}{\SortA}\,\LogNeg{\LogFormA}},\CatRCHomA\ \bot\ \CatCHomA\rlap{\hbox to204 pt{\hfill\qEd}}
\end{align*}
\end{itemize}
\subsection{Adequacy}
\subsubsection{Adequacy for first-order logic}
We will now state the adequacy lemma, which states the soundness of our realizability interpretation with respect to first-order classical logic. It is interesting to remark that the only cases which depends on the orthogonality relation are those of introduction and elimination of the $\LogBot$ formula. It is not much of a surprise, since these rules are the ones that make our proof system classical. Most of the other cases are straightforward, though some care must be taken for the $\forall$ rules.\par
In order to prove the adequacy lemma, we suppose that the interpretations of the terms associated to the axioms are realizers of these axioms: for every $\LogFormA\in\LogAxioms$, $\LmInterpAxiom{\LogFormA}\in\RealVal{\LogFormA}$. The adequacy lemma is then as follows:
\begin{lem}
\label{adequacyLemma}
Suppose $\LogProofA$ is a proof of $\Sequent{\Gamma}{\LogFormA}{\LogNeg{\Delta}}$ with $\FV{\Gamma,\LogFormA,\Delta}\subseteq\LogSortedTerm{\vec{\LogVarB}}{\vec{\SortA}}$, so:
$$\Sequent{\LmTerm{\vec{\LmVarA}}{\LmInterpForm{\Gamma}}}{\LmTerm{\LmInterpProof{\LogProofA}}{\LmInterpForm{\LogFormA}}}{\LmTerm{\vec{\LmMVarA}}{\LmInterpForm{\Delta}},\LmTerm{\kappa}{\LmSortExtract}}$$
then for any $\vec{\ModElemA}\in\ModMInterp{\vec{\SortA}},\vec{\CatRCHomA}\in\RealVal{\Gamma\LogSubst{\vec{\ModElemA}/\vec{\LogVarB}}},\vec{\CatCHomA}\in\RealValNeg{\LogNeg{\Delta}\LogSubst{\vec{\ModElemA}/\vec{\LogVarB}}}$, we have:
$$\LmInterpProof{\LogProofA}\LogSubst{\vec{\CatRCHomA}/\vec{\LmVarA},\vec{\CatCHomA}/\vec{\LmMVarA}}\in\RealVal{\LogFormA\LogSubst{\vec{\ModElemA}/\vec{\LogVarB}}}$$
In particular if $\LogFormA$ is a closed formula, $\LmInterpProof{\LogProofA}\in\RealVal{\LogFormA}$
\end{lem}
\proof
By induction on the proof tree:
\begin{itemize}
\item If $\LogProofA$ is the identity rule then $\LmInterpProof{\LogProofA}$ is:
$$\LmRuleAx{\LmTerm{\vec{\LmVarA}}{\LmInterpForm{\Gamma}}}{\LmTerm{\vec{\LmMVarA}}{\LmInterpForm{\Delta}},\LmTerm{\kappa}{\LmSortExtract}}{\LmVarC}{\LmInterpForm{\LogFormA}}$$
Let $\vec{\ModElemA}\in\ModMInterp{\vec{\SortA}}$, $\vec{\CatRCHomA}\in\RealVal{\Gamma\LogSubst{\vec{\ModElemA}/\vec{\LogVarB}}}$, $\CatRCHomB\in\RealVal{\LogFormA\LogSubst{\vec{\ModElemA}/\vec{\LogVarB}}}$ and $\vec{\CatCHomA}\in\RealValNeg{\LogNeg{\Delta}\LogSubst{\vec{\ModElemA}/\vec{\LogVarB}}}$. Then we have:
$$\LmInterpProof{\LogProofA}\LogSubst{\vec{\CatRCHomA}/\vec{\LmVarA},\CatRCHomB/\LmVarC,\vec{\CatCHomA}/\vec{\LmMVarA}}=\CatRCHomB\in\RealVal{\LogFormA\LogSubst{\vec{\ModElemA}/\vec{\LogVarB}}}$$
\item If $\LogProofA$ is the axiom rule then $\LmInterpProof{\LogProofA}$ is:
$$\Sequent{\LmTerm{\vec{\LmVarA}}{\LmInterpForm{\Gamma}}}{\LmTerm{\LmInterpAxiom{\LogFormA}}{\LmInterpForm{\LogFormA}}}{\LmTerm{\vec{\LmMVarA}}{\LmInterpForm{\Delta}},\LmTerm{\kappa}{\LmSortExtract}}$$
Let $\vec{\ModElemA}\in\ModMInterp{\vec{\SortA}}$, $\vec{\CatRCHomA}\in\RealVal{\Gamma\LogSubst{\vec{\ModElemA}/\vec{\LogVarB}}}$ and $\vec{\CatCHomA}\in\RealValNeg{\LogNeg{\Delta}\LogSubst{\vec{\ModElemA}/\vec{\LogVarB}}}$. Then we have:
$$\LmInterpProof{\LogProofA}\LogSubst{\vec{\CatRCHomA}/\vec{\LmVarA},\vec{\CatCHomA}/\vec{\LmMVarA}}=\LmInterpAxiom{\LogFormA}\in\RealVal{\LogFormA}=\RealVal{\LogFormA\LogSubst{\vec{\ModElemA}/\vec{\LogVarB}}}$$
since $\LogFormA$ is closed and $\LmInterpAxiom{\LogFormA}\in\RealVal{\LogFormA}$ by assumption.
\item If $\LogProofA$ ends with an introduction of $\LogImp$ then $\LmInterpProof{\LogProofA}$ is:
$$\LmRuleImpIntro{\LmTerm{\vec{\LmVarA}}{\LmInterpForm{\Gamma}}}{\LmTerm{\vec{\LmMVarA}}{\LmInterpForm{\Delta}},\LmTerm{\kappa}{\LmSortExtract}}{\LmVarA}{\LmInterpForm{\LogFormA}}{\LmInterpProof{\LogProofB}}{\LmInterpForm{\LogFormB}}$$
Let $\vec{\ModElemA}\in\ModMInterp{\vec{\SortA}}$, $\vec{\CatRCHomA}\in\RealVal{\Gamma\LogSubst{\vec{\ModElemA}/\vec{\LogVarB}}}$ and $\vec{\CatCHomA}\in\RealValNeg{\LogNeg{\Delta}\LogSubst{\vec{\ModElemA}/\vec{\LogVarB}}}$. Then for any $\CatRCHomB\in\RealVal{\LogFormA\LogSubst{\vec{\ModElemA}/\vec{\LogVarB}}}$ we have:
$$\left(\LmInterpProof{\LogProofA}\LogSubst{\vec{\CatRCHomA}/\vec{\LmVarA},\vec{\CatCHomA}/\vec{\LmMVarA}}\right)\CatRCHomB=\LmInterpProof{\LogProofB}\LogSubst{\vec{\CatRCHomA}/\vec{\LmVarA},\CatRCHomB/\LmVarB,\vec{\CatCHomA}/\vec{\LmMVarA}}\in\RealVal{\LogFormB\LogSubst{\vec{\ModElemA}/\vec{\LogVarB}}}$$
by induction hypothesis, since $\CatRCHomB\in\RealVal{\LogFormA\LogSubst{\vec{\ModElemA}/\vec{\LogVarB}}}$. Therefore:
$$\LmInterpProof{\LogProofA}\LogSubst{\vec{\CatRCHomA}/\vec{\LmVarA},\vec{\CatCHomA}/\vec{\LmMVarA}}\in\RealVal{\left(\LogFormA\LogImp\LogFormB\right)\LogSubst{\vec{\ModElemA}/\vec{\LogVarB}}}$$
\item If $\LogProofA$ ends with an elimination of $\LogImp$ then $\LmInterpProof{\LogProofA}$ is:
$$\LmRuleImpElim{\LmTerm{\vec{\LmVarA}}{\LmInterpForm{\Gamma}}}{\LmTerm{\vec{\LmMVarA}}{\LmInterpForm{\Delta}},\LmTerm{\kappa}{\LmSortExtract}}{\LmInterpProof{\LogProofB}}{\LmInterpForm{\LogFormB}}{\LmInterpProof{\LogProofC}}{\LmInterpForm{\LogFormA}}$$
Let $\vec{\ModElemA}\in\ModMInterp{\vec{\SortA}}$, $\vec{\CatRCHomA}\in\RealVal{\Gamma\LogSubst{\vec{\ModElemA}/\vec{\LogVarB}}}$ and $\vec{\CatCHomA}\in\RealValNeg{\LogNeg{\Delta}\LogSubst{\vec{\ModElemA}/\vec{\LogVarB}}}$. Then we have:
$$\LmInterpProof{\LogProofA}\LogSubst{\vec{\CatRCHomA}/\vec{\LmVarA},\vec{\CatCHomA}/\vec{\LmMVarA}}=\left(\LmInterpProof{\LogProofB}\LogSubst{\vec{\CatRCHomA}/\vec{\LmVarA},\vec{\CatCHomA}/\vec{\LmMVarA}}\right)\left(\LmInterpProof{\LogProofC}\LogSubst{\vec{\CatRCHomA}/\vec{\LmVarA},\vec{\CatCHomA}/\vec{\LmMVarA}}\right)\in\RealVal{\LogFormB\LogSubst{\vec{\ModElemA}/\vec{\LogVarB}}}$$
since by induction hypothesis:
$$\LmInterpProof{\LogProofB}\LogSubst{\vec{\CatRCHomA}/\vec{\LmVarA},\vec{\CatCHomA}/\vec{\LmMVarA}}\in\RealVal{\left(\LogFormA\LogImp\LogFormB\right)\LogSubst{\vec{\ModElemA}/\vec{\LogVarB}}}\text{ and }\LmInterpProof{\LogProofC}\LogSubst{\vec{\CatRCHomA}/\vec{\LmVarA},\vec{\CatCHomA}/\vec{\LmMVarA}}\in\RealVal{\LogFormA\LogSubst{\vec{\ModElemA}/\vec{\LogVarB}}}$$
\item If $\LogProofA$ ends with an introduction of $\LogAnd$ then $\LmInterpProof{\LogProofA}$ is:
$$\LmRuleAndIntro{\LmTerm{\vec{\LmVarA}}{\LmInterpForm{\Gamma}}}{\LmTerm{\vec{\LmMVarA}}{\LmInterpForm{\Delta}},\LmTerm{\kappa}{\LmSortExtract}}{\LmInterpProof{\LogProofB}}{\LmInterpForm{\LogFormA}}{\LmInterpProof{\LogProofC}}{\LmInterpForm{\LogFormB}}$$
Let $\vec{\ModElemA}\in\ModMInterp{\vec{\SortA}}$, $\vec{\CatRCHomA}\in\RealVal{\Gamma\LogSubst{\vec{\ModElemA}/\vec{\LogVarB}}}$ and $\vec{\CatCHomA}\in\RealValNeg{\LogNeg{\Delta}\LogSubst{\vec{\ModElemA}/\vec{\LogVarB}}}$. Then we have:
$$\LmInterpProof{\LogProofA}\LogSubst{\vec{\CatRCHomA}/\vec{\LmVarA},\vec{\CatCHomA}/\vec{\LmMVarA}}=\LmPair{\LmInterpProof{\LogProofB}\LogSubst{\vec{\CatRCHomA}/\vec{\LmVarA},\vec{\CatCHomA}/\vec{\LmMVarA}}}{\LmInterpProof{\LogProofC}\LogSubst{\vec{\CatRCHomA}/\vec{\LmVarA},\vec{\CatCHomA}/\vec{\LmMVarA}}}\in\RealVal{\left(\LogFormA\LogAnd\LogFormB\right)\LogSubst{\vec{\ModElemA}/\vec{\LogVarB}}}$$
since by induction hypothesis:
$$\LmInterpProof{\LogProofB}\LogSubst{\vec{\CatRCHomA}/\vec{\LmVarA},\vec{\CatCHomA}/\vec{\LmMVarA}}\in\RealVal{\LogFormA\LogSubst{\vec{\ModElemA}/\vec{\LogVarB}}}\text{ and }\LmInterpProof{\LogProofC}\LogSubst{\vec{\CatRCHomA}/\vec{\LmVarA},\vec{\CatCHomA}/\vec{\LmMVarA}}\in\RealVal{\LogFormB\LogSubst{\vec{\ModElemA}/\vec{\LogVarB}}}$$
\item If $\LogProofA$ ends with an elimination of $\LogAnd$ then $\LmInterpProof{\LogProofA}$ is:
$$\LmRuleAndElim{\LmTerm{\vec{\LmVarA}}{\LmInterpForm{\Gamma}}}{\LmTerm{\vec{\LmMVarA}}{\LmInterpForm{\Delta}},\LmTerm{\kappa}{\LmSortExtract}}{\LmInterpProof{\LogProofB}}{\LmInterpForm{\LogFormA}}$$
Let $\vec{\ModElemA}\in\ModMInterp{\vec{\SortA}}$, $\vec{\CatRCHomA}\in\RealVal{\Gamma\LogSubst{\vec{\ModElemA}/\vec{\LogVarB}}}$ and $\vec{\CatCHomA}\in\RealValNeg{\LogNeg{\Delta}\LogSubst{\vec{\ModElemA}/\vec{\LogVarB}}}$. Then we have:
$$\LmInterpProof{\LogProofA}\LogSubst{\vec{\CatRCHomA}/\vec{\LmVarA},\vec{\CatCHomA}/\vec{\LmMVarA}}=\LmProj_i\left(\LmInterpProof{\LogProofB}\LogSubst{\vec{\CatRCHomA}/\vec{\LmVarA},\vec{\CatCHomA}/\vec{\LmMVarA}}\right)\in\RealVal{\LogFormA_i\LogSubst{\vec{\ModElemA}/\vec{\LogVarB}}}$$
since by induction hypothesis:
$$\LmInterpProof{\LogProofB}\LogSubst{\vec{\CatRCHomA}/\vec{\LmVarA},\vec{\CatCHomA}/\vec{\LmMVarA}}\in\RealVal{\left(\LogFormA_1\LogAnd\LogFormA_2\right)\LogSubst{\vec{\ModElemA}/\vec{\LogVarB}}}$$
\item If $\LogProofA$ ends with an introduction of $\forall$ then $\LmInterpProof{\LogProofA}$ is:
$$\Sequent{\LmTerm{\vec{\LmVarA}}{\LmInterpForm{\Gamma}}}{\LmTerm{\LmInterpProof{\LogProofB}}{\LmInterpForm{\left(\forall\LogSortedTerm{\LogVarC}{\SortB}\LogFormA\right)}}}{\LmTerm{\vec{\LmMVarA}}{\LmInterpForm{\Delta}},\LmTerm{\kappa}{\LmSortExtract}}\qquad=\qquad\Sequent{\LmTerm{\vec{\LmVarA}}{\LmInterpForm{\Gamma}}}{\LmTerm{\LmInterpProof{\LogProofB}}{\LmInterpForm{\LogFormA}}}{\LmTerm{\vec{\LmMVarA}}{\LmInterpForm{\Delta}},\LmTerm{\kappa}{\LmSortExtract}}$$
Let $\vec{\ModElemA}\in\ModMInterp{\vec{\SortA}}$, $\vec{\CatRCHomA}\in\RealVal{\Gamma\LogSubst{\vec{\ModElemA}/\vec{\LogVarB}}}$ and $\vec{\CatCHomA}\in\RealValNeg{\LogNeg{\Delta}\LogSubst{\vec{\ModElemA}/\vec{\LogVarB}}}$. Then for any $\ModElemB\in\ModMInterp{\SortB}$, $\vec{\CatRCHomA}\in\RealVal{\Gamma\LogSubst{\vec{\ModElemA}/\vec{\LogVarB},\ModElemB/\LogVarC}}$ and $\vec{\CatCHomA}\in\RealValNeg{\LogNeg{\Delta}\LogSubst{\vec{\ModElemA}/\vec{\LogVarB},\ModElemB/\LogVarC}}$ (since $\LogSortedTerm{\LogVarC}{\SortB}\notin\FV{\Gamma,\Delta}$), so by induction hypothesis:
$$\LmInterpProof{\LogProofB}\LogSubst{\vec{\CatRCHomA}/\vec{\LmVarA},\vec{\CatCHomA}/\vec{\LmMVarA}}\in\RealVal{\LogFormA\LogSubst{\vec{\ModElemA}/\vec{\LogVarB},\ModElemB/\LogVarC}}$$
Therefore $\LmInterpProof{\LogProofA}\LogSubst{\vec{\CatRCHomA}/\vec{\LmVarA},\vec{\CatCHomA}/\vec{\LmMVarA}}\in\RealVal{\left(\forall\LogSortedTerm{\LogVarC}{\SortB}\,\LogFormA\right)\LogSubst{\vec{\ModElemA}/\vec{\LogVarB}}}$.
\item If $\LogProofA$ ends with an elimination of $\forall$ then $\LmInterpProof{\LogProofA}$ is:
$$\Sequent{\LmTerm{\vec{\LmVarA}}{\LmInterpForm{\Gamma}}}{\LmTerm{\LmInterpProof{\LogProofB}}{\LmInterpForm{\left(\LogFormA\LogSubst{\LogSortedTerm{\LogTermA}{\SortB}/\LogSortedTerm{\LogVarC}{\SortB}}\right)}}}{\LmTerm{\vec{\LmMVarA}}{\LmInterpForm{\Delta}},\LmTerm{\kappa}{\LmSortExtract}}\qquad=\qquad\Sequent{\LmTerm{\vec{\LmVarA}}{\LmInterpForm{\Gamma}}}{\LmTerm{\LmInterpProof{\LogProofB}}{\LmInterpForm{\left(\forall\LogSortedTerm{\LogVarC}{\SortB}\LogFormA\right)}}}{\LmTerm{\vec{\LmMVarA}}{\LmInterpForm{\Delta}},\LmTerm{\kappa}{\LmSortExtract}}$$
Let $\vec{\ModElemA}\in\ModMInterp{\vec{\SortA}}$, $\vec{\CatRCHomA}\in\RealVal{\Gamma\LogSubst{\vec{\ModElemA}/\vec{\LogVarB}}}$ and $\vec{\CatCHomA}\in\RealValNeg{\LogNeg{\Delta}\LogSubst{\vec{\ModElemA}/\vec{\LogVarB}}}$. By induction hypothesis:
$$\LmInterpProof{\LogProofB}\LogSubst{\vec{\CatRCHomA}/\vec{\LmVarA},\vec{\CatCHomA}/\vec{\LmMVarA}}\in\bigcap_{\ModElemB\in\ModMInterp{\SortB}}\RealVal{\LogFormA\LogSubst{\vec{\ModElemA}/\vec{\LogVarB},\ModElemB/\LogVarC}}$$
so taking $\ModElemB=\ModMInterp{\left(\LogTermA\LogSubst{\vec{\ModElemA}/\vec{\LogVarB}}\right)}\in\ModMInterp{\SortB}$ we get:
$$\LmInterpProof{\LogProofB}\LogSubst{\vec{\CatRCHomA}/\vec{\LmVarA},\vec{\CatCHomA}/\vec{\LmMVarA}}\in\RealVal{\LogFormA\LogSubst{\vec{\ModElemA}/\vec{\LogVarB},\ModMInterp{\left(\LogTermA\LogSubst{\vec{\ModElemA}/\vec{\LogVarB}}\right)}/\LogVarC}}$$
and since $\LmInterpProof{\LogProofA}=\LmInterpProof{\LogProofB}$ and $\LogFormA\LogSubst{\vec{\ModElemA}/\vec{\LogVarB},\ModMInterp{\left(\LogTermA\LogSubst{\vec{\ModElemA}/\vec{\LogVarB}}\right)}/\LogVarC}=\LogFormA\LogSubst{\LogTermA/\LogVarC}\LogSubst{\vec{\ModElemA}/\vec{\LogVarB}}$ we get:
$$\LmInterpProof{\LogProofA}\LogSubst{\vec{\CatRCHomA}/\vec{\LmVarA},\vec{\CatCHomA}/\vec{\LmMVarA}}\in\RealVal{\LogFormA\LogSubst{\LogTermA/\LogVarC}\LogSubst{\vec{\ModElemA}/\vec{\LogVarB}}}$$
\item If $\LogProofA$ ends with an introduction of $\LogBot$ then $\LmInterpProof{\LogProofA}$ is:
$$\LmRuleBotIntro{\LmTerm{\vec{\LmVarA}}{\LmInterpForm{\Gamma}}}{\LmTerm{\vec{\LmMVarA}}{\LmInterpForm{\Delta}},\LmTerm{\kappa}{\LmSortExtract}}{\LmMVarB}{\LmInterpProof{\LogProofB}}{\LmInterpForm{\LogFormA}}$$
Let $\vec{\ModElemA}\in\ModMInterp{\vec{\SortA}}$, $\vec{\CatRCHomA}\in\RealVal{\Gamma\LogSubst{\vec{\ModElemA}/\vec{\LogVarB}}}$, $\vec{\CatCHomA}\in\RealValNeg{\LogNeg{\Delta}\LogSubst{\vec{\ModElemA}/\vec{\LogVarB}}}$ and $\CatCHomB\in\RealValNeg{\LogNeg{\LogFormA}\LogSubst{\vec{\ModElemA}/\vec{\LogVarB}}}$. Then we have:
\begin{align*}
\LmInterpProof{\LogProofA}\LogSubst{\vec{\CatRCHomA}/\vec{\LmVarA},\vec{\CatCHomA}/\vec{\LmMVarA},\CatCHomB/\LmMVarB}&=\left(\LmInterpForm{\left[\LmMVarB\right]\,\LogProofB}\right)\LogSubst{\vec{\CatRCHomA}/\vec{\LmVarA},\vec{\CatCHomA}/\vec{\LmMVarA},\CatCHomB/\LmMVarB}\\
&=\left[\CatCHomB\right]\left(\LmInterpProof{\LogProofB}\LogSubst{\vec{\CatRCHomA}/\vec{\LmVarA},\vec{\CatCHomA}/\vec{\LmMVarA},\CatCHomB/\LmMVarB}\right)\in\RealVal{\LogBot}=\RealVal{\LogBot\LogSubst{\vec{\ModElemA}/\vec{\LogVarB}}}
\end{align*}
since by induction hypothesis:
$$\LmInterpProof{\LogProofB}\LogSubst{\vec{\CatRCHomA}/\vec{\LmVarA},\vec{\CatCHomA}/\vec{\LmMVarA},\CatCHomB/\LmMVarB}\in\RealVal{\LogNeg{\LogFormA}\LogSubst{\vec{\ModElemA}/\vec{\LogVarB}}}$$
so $\LmInterpProof{\LogProofB}\LogSubst{\vec{\CatRCHomA}/\vec{\LmVarA},\vec{\CatCHomA}/\vec{\LmMVarA},\CatCHomB/\LmMVarB}\ \bot\ \CatCHomB$ by lemma~\ref{orthogonal}, and $\mu\kappa.\left[\CatCHomB\right]\left(\LmInterpProof{\LogProofB}\LogSubst{\vec{\CatRCHomA}/\vec{\LmVarA},\vec{\CatCHomA}/\vec{\LmMVarA},\CatCHomB/\LmMVarB}\right)\in\RealBot$.
\item If $\LogProofA$ ends with an elimination of $\LogBot$ then $\LmInterpProof{\LogProofA}$ is:
$$\LmRuleBotElim{\LmTerm{\vec{\LmVarA}}{\LmInterpForm{\Gamma}}}{\LmTerm{\vec{\LmMVarA}}{\LmInterpForm{\Delta}},\LmTerm{\kappa}{\LmSortExtract}}{\LmMVarB}{\LmInterpProof{\LogProofB}}{\LmInterpForm{\LogFormA}}$$
Let $\vec{\ModElemA}\in\ModMInterp{\vec{\SortA}}$, $\vec{\CatRCHomA}\in\RealVal{\Gamma\LogSubst{\vec{\ModElemA}/\vec{\LogVarB}}}$ and $\vec{\CatCHomA}\in\RealValNeg{\LogNeg{\Delta}\LogSubst{\vec{\ModElemA}/\vec{\LogVarB}}}$. Then for any $\CatCHomB\in\RealValNeg{\LogNeg{\LogFormA}\LogSubst{\vec{\ModElemA}/\vec{\LogVarB}}}$ we have:
\begin{align*}\mu\kappa.\left[\CatCHomB\right]\left(\LmInterpProof{\LogProofA}\LogSubst{\vec{\CatRCHomA}/\vec{\LmVarA},\vec{\CatCHomA}/\vec{\LmMVarA}}\right)&=\mu\kappa.\left(\left[\CatCHomB\right]\,\mu\LmMVarB.\LmInterpProof{\LogProofB}\LogSubst{\vec{\CatRCHomA}/\vec{\LmVarA},\vec{\CatCHomA}/\vec{\LmMVarA}}\right)\\
&=\mu\kappa.\left(\LmInterpProof{\LogProofB}\LogSubst{\vec{\CatRCHomA}/\vec{\LmVarA},\vec{\CatCHomA}/\vec{\LmMVarA},\CatCHomB/\LmMVarB}\right)\in\RealBot\end{align*}
since by induction hypothesis, $\LmInterpProof{\LogProofB}\LogSubst{\vec{\CatRCHomA}/\vec{\LmVarA},\vec{\CatCHomA}/\vec{\LmMVarA},\CatCHomB/\LmMVarB}\in\RealVal{\LogBot\LogSubst{\vec{\ModElemA}/\vec{\LogVarB}}}=\RealVal{\LogBot}$. Therefore:
$$\LmInterpProof{\LogProofA}\LogSubst{\vec{\CatRCHomA}/\vec{\LmVarA},\vec{\CatCHomA}/\vec{\LmMVarA}}\ \bot\ \CatCHomB$$ and so by lemma~\ref{orthogonal}:
$$\LmInterpProof{\LogProofA}\LogSubst{\vec{\CatRCHomA}/\vec{\LmVarA},\vec{\CatCHomA}/\vec{\LmMVarA}}\in\RealVal{\LogNeg{\LogFormA}\LogSubst{\vec{\ModElemA}/\vec{\LogVarB}}}\eqno{\qEd}$$

\end{itemize}
\subsubsection{Adequacy for Peano arithmetic}
\label{OrthoRealPeano}
We fix now the theory to be $\LogRelForm{\PAom}$ on $\Sigma_\LogRelForm{\PAom}$ (the inequality predicate being negative, and the relativization predicate being positive), and the structure $\ModM$ to be a model of $\PAom$, the inequality predicate being interpreted in $\ModM$ as in section~\ref{models}. For simplicity we write $n$ for $\ModMInterp{\left(\CALogS^n\,\CALogZ\right)}$ where $n\in\mathbb{N}$. The $\lambda\mu$ signature is that of $\mu$PCF (see section~\ref{muPCF}) and we interpret the two predicates as in sections~\ref{RelPred} and~\ref{LmInterpPA}:
$$\LmInterpForm{\neq_\SortA}\quad\Def\quad\LmSortBot\qquad\qquad\LmInterpForm{\LogRel{\LogSortedTerm{.}{\CASort}}}\quad\Def\quad\CALmnSort$$
and the axioms as in section~\ref{LmInterpPA}. We suppose that the category of continuations $\CatRC$ is a model of $\mu$PCF (see definition~\ref{LambdaMuModel}). Since we want to extract algorithms on natural numbers, we fix $\LmSortExtract=\CALmnSort$.\par
The realizability value for the relativization predicate is:
$$\RealVal{\LogRel{\LmTerm{\ModElemA}{\CASort}}}\Def\begin{cases}\begin{aligned}&\SetSuch{\CALmn{n}}{}&&\text{ if }\ModElemA=n\text{ for some }n\in\mathbb{N}\\&\emptyset&&\text{otherwise}\end{aligned}\end{cases}$$
First, all equalities which are true in the model are trivially realized:
\begin{lem}
Let $\LogSortedTerm{\LogTermA}{\SortA}$ and $\LogSortedTerm{\LogTermB}{\SortA}$ be first-order terms with $\FV{\LogTermA,\LogTermB}=\LogSortedTerm{\vec{\LogVarA}}{\vec{\SortB}}$.
$$\text{If }\ModM\Models\forall\LogSortedTerm{\vec{\LogVarA}}{\vec{\SortB}}\left(\LogTermA=\LogTermB\right)\text{ then }\lambda\LmVarA.\LmVarA\in\RealVal{\forall\LogSortedTerm{\vec{\LogVarA}}{\vec{\SortB}}\left(\LogTermA=\LogTermB\right)}$$
\end{lem}
\proof
Let $\vec{\ModElemA}\in\ModMInterp{\vec{\SortB}}$. Since $\ModM\Models\forall\LogSortedTerm{\vec{\LogVarA}}{\vec{\SortB}}\left(\LogTermA=\LogTermB\right)$, we have $\ModMInterp{\left(\LogTermA\LogSubst{\vec{\ModElemA}/\vec{\LogVarA}}\right)}=\ModMInterp{\left(\LogTermB\LogSubst{\vec{\ModElemA}/\vec{\LogVarA}}\right)}$, so:
$$\RealVal{\LogTermA\LogSubst{\vec{\ModElemA}/\vec{\LogVarA}}\neq\LogTermB\LogSubst{\vec{\ModElemA}/\vec{\LogVarA}}}=\RealVal{\LogBot}$$
Therefore, for any $\CatRCHomA\in\RealVal{\LogTermA\LogSubst{\vec{\ModElemA}/\vec{\LogVarA}}\neq\LogTermB\LogSubst{\vec{\ModElemA}/\vec{\LogVarA}}}$:
$$\left(\lambda\LmVarA.\LmVarA\right)\CatRCHomA=\CatRCHomA\in\RealVal{\LogTermA\LogSubst{\vec{\ModElemA}/\vec{\LogVarA}}\neq\LogTermB\LogSubst{\vec{\ModElemA}/\vec{\LogVarA}}}=\RealVal{\LogBot}$$
and so:
$$\lambda\LmVarA.\LmVarA\in\RealVal{\forall\LogSortedTerm{\vec{\LogVarA}}{\vec{\SortB}}\left(\LogTermA\neq\LogTermB\LogImp\LogBot\right)}=\RealVal{\forall\LogSortedTerm{\vec{\LogVarA}}{\vec{\SortB}}\left(\LogTermA=\LogTermB\right)}\eqno{\qEd}$$
Therefore, since $\ModM$ is a model of $\PAom$ we have immediately the following results:
\begin{align*}
\LmInterpAxiom{\CAReflName}=\CAReflTerm&\in\RealVal{\CARefl{\SortA}}\\
\LmInterpAxiom{\CAdefsName}=\CAdefsTerm&\in\RealVal{\CAdefs{\SortA}{\SortB}{\SortC}}\\
\LmInterpAxiom{\CAdefkName}=\CAdefkTerm&\in\RealVal{\CAdefk{\SortA}{\SortB}}\\
\LmInterpAxiom{\CAdefrecZName}=\CAdefrecZTerm&\in\RealVal{\CAdefrecZ{\SortA}}\\
\LmInterpAxiom{\CAdefrecSName}=\CAdefrecSTerm&\in\RealVal{\CAdefrecS{\SortA}}
\end{align*}
The non-confusion axiom and Leibniz scheme are easy:
\begin{lem}
\begin{align*}
\LmInterpAxiom{\CASnZName}=\CASnZTerm&\in\RealVal{\CASnZ}\\
\LmInterpAxiom{\CALeibName}=\CALeibTerm&\in\RealVal{\CALeib{\SortA}{\LogFormA}{\SortB}}
\end{align*}
\end{lem}
\proof
Since $\RealVal{\LogRel{\ModElemA}}=\emptyset$ if $\ModElemA\notin\mathbb{N}$, it is sufficient to prove that for any $n\in\mathbb{N}$, $\CALmom\in\RealVal{n+1\neq0}$, which is true since $\ModM\Models\CASnZName$ so $\RealVal{n+1\neq0}=\CatRC\left(\CatTerm,\CatInterpSort{\LmSortBot}\CatPar\CatInterpSort{\CALmnSort}\right)$. Let now $\vec{\ModElemA}\in\ModMInterp{\vec{\SortB}}$, $\ModElemB,\ModElemC\in\ModMInterp{\SortA}$, $\CatRCHomA\in\RealVal{\neg\LogFormA\LogSubst{\ModElemB/\LogVarA,\vec{\ModElemA}/\vec{\LogVarC}}}$ and $\CatRCHomB\in\RealVal{\LogFormA\LogSubst{\ModElemC/\LogVarA,\vec{\ModElemA}/\vec{\LogVarC}}}$. If $\ModElemB\neq\ModElemC$, then $\RealVal{\ModElemB\neq\ModElemC}=\CatRC\left(\CatTerm,\CatInterpSort{\LmSortBot}\CatPar\CatInterpSort{\CALmnSort}\right)$ and therefore $\left(\CALeibTerm\right)\CatRCHomA\,\CatRCHomB\in\RealVal{\ModElemB\neq\ModElemC}$. Otherwise, $\ModElemB=\ModElemC$ so $\RealVal{\LogFormA\LogSubst{\ModElemB/\LogVarA,\vec{\ModElemA}/\vec{\LogVarC}}}=\RealVal{\LogFormA\LogSubst{\ModElemC/\LogVarA,\vec{\ModElemA}/\vec{\LogVarC}}}$ and by definition of $\RealVal{\LogFormA\LogImp\LogFormB}$ we get $\CatRCHomA\,\CatRCHomB\in\RealVal{\LogBot}$. Moreover, since $\ModElemB=\ModElemC$, $\RealVal{\ModElemB\neq\ModElemC}=\RealVal{\LogBot}$ and $\CatRCHomA\,\CatRCHomB\in\RealVal{\ModElemB\neq\ModElemC}$. Finally we get $\LmInterpAxiom{\CALeibName}\in\RealVal{\CALeib{\SortA}{\LogFormA}{\SortB}}$.
\qed
The adequacy for the induction axiom scheme is as follows:
\begin{lem}
$$\LmInterpAxiom{\CAindName}=\CAindTerm\in\RealVal{\CAind{\LogFormA}{\SortB}}$$
\end{lem}
\proof
Let $\vec{\ModElemA}\in\ModMInterp{\vec{\SortB}}$, $\CatRCHomA\in\RealVal{\LogFormA\LogSubst{\CALogZ/\LogVarA,\vec{\ModElemA}/\vec{\LogVarB}}}$, $\CatRCHomB\in\RealVal{\LogForallRel\LogSortedTerm{\LogVarA}{\CASort}\left(\LogFormA\LogSubst{\vec{\ModElemA}/\vec{\LogVarB}}\LogImp\LogFormA\LogSubst{\CALogS\,\LogVarA/\LogVarA,\vec{\ModElemA}/\vec{\LogVarB}}\right)}$. Since $\RealVal{\LogRel{\ModElemB}}=\emptyset$ if $\ModElemB\notin\mathbb{N}$, we can use induction to prove that for any $n\in\mathbb{N}$, $\CAindTerm\,\CatRCHomA\,\CatRCHomB\,\CALmn{n}\in\RealVal{\LogFormA\LogSubst{n/\LogVarA,\vec{\ModElemA}/\vec{\LogVarB}}}$:
\begin{itemize}
\item$n=0$: $\CAindTerm\,\CatRCHomA\,\CatRCHomB\,\CALmn{0}=\CatRCHomA\in\RealVal{\LogFormA\LogSubst{0/\LogVarA,\vec{\ModElemA}/\vec{\LogVarB}}}$
\item$n+1$: $\CAindTerm\,\CatRCHomA\,\CatRCHomB\,\CALmn{n+1}=\CAindTerm\,\CatRCHomA\,\CatRCHomB\,\CALmsucc\,\CALmn{n}=\CatRCHomB\,\CALmn{n}\left(\CAindTerm\,\CatRCHomA\,\CatRCHomB\,\CALmn{n}\right)$. Since $\CALmn{n}\in\RealVal{\LogRel{n}}$, we get:
$$\CatRCHomB\,\CALmn{n}\in\RealVal{\LogFormA\LogSubst{n/\LogVarA,\vec{\ModElemA}/\vec{\LogVarB}}\LogImp\LogFormA\LogSubst{n+1/\LogVarA,\vec{\ModElemA}/\vec{\LogVarB}}}$$
and since by induction hypothesis we have $\CAindTerm\,\CatRCHomA\,\CatRCHomB\,\CALmn{n}\in\RealVal{\LogFormA\LogSubst{n/\LogVarA,\vec{\ModElemA}/\vec{\LogVarB}}}$ we get:
$$\CatRCHomB\,\CALmn{n}\left(\CAindTerm\,\CatRCHomA\,\CatRCHomB\,\CALmn{n}\right)\in\RealVal{\LogFormA\LogSubst{n+1/\LogVarA,\vec{\ModElemA}/\vec{\LogVarB}}}\eqno{\qEd}$$
\end{itemize}
The last axioms are the relativization ones:
\begin{lem}
$$\LmInterpAxiom{\CARelZNoType}=\CARelZTerm\in\RealVal{\CARelZNoType}\qquad\LmInterpAxiom{\CARelSNoType}=\CARelSTerm\in\RealVal{\CARelSNoType}$$
\end{lem}
\proof
The first one is immediate, since $\RealVal{\LogRel{0}}=\SetSuch{\CALmn{0}}{}$. For the second one, remember that:
$$\LogRel{\CALogS}\equiv\LogForallRel\LogSortedTerm{\LogVarA}{\CASort}\LogRel{\CALogS\,\LogVarA}\equiv\forall\LogSortedTerm{\LogVarA}{\CASort}\left(\LogRel{\LogVarA}\LogImp\LogRel{\CALogS\,\LogVarA}\right)$$
Since $\RealVal{\LogRel{\ModElemA}}=\emptyset$ if $\ModElemA\notin\mathbb{N}$, it is sufficient to prove that for any $n\in\mathbb{N}$, $\CARelSTerm\in\RealVal{\LogRel{n}\LogImp\LogRel{n+1}}$. Since $\RealVal{\LogRel{n}}=\SetSuch{\CALmn{n}}{}$, it follows from $\CALmsucc\,\CALmn{n}=\CALmn{n+1}\in\RealVal{\LogRel{n+1}}$.
\qed
\subsubsection{Adequacy for the axiom of choice}
\label{ChoiceAdequacy}
When it comes to the axiom of countable choice $\mathrm{AC}$, the usual route of negative translation followed by intuitionistic realizability becomes much more difficult. Indeed, $\mathrm{AC}\LogImp\mathrm{AC}^\neg$ is not provable in intuitionistic logic, therefore the path described in section~\ref{NegTrans} cannot be followed as-is and an intuitionistic realizer of $\mathrm{AC}^\neg$ must be provided. In~\cite{BerardiBezemCoquand}, a variant of bar recursion was used to realize $\mathrm{AC}^\neg$, while in~\cite{BergerOlivaChoice} the principle of double negation shift $\forall\LogVarA\,\neg\neg\LogFormA\LogImp\neg\neg\forall\LogVarA\,\LogFormA$ was realized using bar recursion (see~\ref{barrecursion}). With this principle, it becomes possible to derive $\mathrm{AC}\LogImp\mathrm{AC}^\neg$ in intuitionistic logic, and since $\mathrm{AC}$ is realized in intuitionistic models by the identity, one obtains a realizer of $\mathrm{AC}^\neg$.\par
We follow here a different approach, since our realizability model is for classical logic, and we prove that bar recursion realizes the axiom of dependent choice $\CADCName$ in our classical model. The negative translation of proofs corresponds to the continuation-passing-style translation on terms and the semantics of $\lambda\mu$-calculus in a category of continuations $\CatRC$ corresponds to the semantics of its CPS-translation into $\Lam$ in the cartesian ``$\CatR$-closed'' category $\CatC$ as stated in section~\ref{cps}, so in order to compare more closely our model with the usual indirect realizability interpretation, one would need to define a realizability relation for intuitionistic logic directly in $\CatC$. Our choice of working in the category of continuations $\CatRC$ is mainly motivated by the existence of computational models with a natural structure of category of continuations (as the unbracketed Hyland-Ong games, see section~\ref{NegTrans}).\par
We suppose now that $\ModM$ satisfies $\CADCUnrelName$, and we use the variant of the bar recursion operator defined in section~\ref{barrecursion} to realize the axiom of dependent choice $\CADCName$. First, we need to make some assumptions on our category of continuations. The first one is a continuity requirement:
\begin{defi}[Continuity]
\label{continuity}
If $\CatRCHomA\in\CatRC\left(\CatTerm,\CatInterpSort{\left(\CALmnSort\LmSortTo\LmSortA\right)\LmSortTo\LmSortBot}\CatPar\CatInterpSort{\CALmnSort}\right)$, $\CatRCHomB\in\CatRC\left(\CatTerm,\CatInterpSort{\CALmnSort\LmSortTo\LmSortA}\CatPar\CatInterpSort{\CALmnSort}\right)$, then there exists $m\in\mathbb{N}$ such that for any $\CatRCHomC\in\CatRC\left(\CatTerm,\CatInterpSort{\CALmnSort\LmSortTo\LmSortA}\CatPar\CatInterpSort{\CALmnSort}\right)$:
$$\left(\forall m'<m,\,\CatRCHomC\,\CALmn{m'}=\CatRCHomB\,\CALmn{m'}\right)\LogImp\mu\kappa.\CatRCHomA\,\CatRCHomC=\mu\kappa.\CatRCHomA\,\CatRCHomB$$
\end{defi}\medskip

\noindent This requirement is satisfied in particular by games models since in that case $\CatRC$ is a cpo-enriched category in which the base types $\CALmnSort$ and $\LmSortBot$ are interpreted as flat domains. Note however that the full set-theoretic model doesn't satisfy this, since we can for example consider a function which gives $0$ if the input sequence is the constant $0$ sequence, and $1$ otherwise.\par
The second requirement states that we can construct a function from any sequence of elements:
\begin{defi}[Sequence internalization]
\label{sequence}
If $\left(\CatRCHomA_n\right)_{n\in\mathbb{N}}$ is a sequence of morphisms in the homset $\CatRC\left(\CatTerm,\CatInterpSort{\LmSortA}\CatPar\CatInterpSort{\CALmnSort}\right)$, then there exists a morphism $\CatRCHomA\in\CatRC\left(\CatTerm,\CatInterpSort{\CALmnSort\LmSortTo\LmSortA}\CatPar\CatInterpSort{\CALmnSort}\right)$ such that for any $n\in\mathbb{N}$, $\CatRCHomA\,\CALmn{n}=\CatRCHomA_n$.
\end{defi}
In particular, all functions on natural numbers must exist in the model, even the uncomputable ones. It is in particular true in the games models, but it is of course not satisfied by the term model of $\mu$PCF, and this is the main motivation for working in a model of $\mu$PCF rather than directly with the syntactic language.\par
Assuming $\CatRC$ satisfies these two assumptions, we now prove that the interpretation of $\CADCName$ realizes $\CADCName$:
\begin{lem}
Let $\RealBot\subseteq\CatRC\left(\CatTerm,\CatInterpSort{\CALmnSort}\right)$.
\begin{multline*}
\LmInterpAxiom{\CADCName}=\CADCTerm\\*
\in\RealVal{\CADC{\SortA}{\LogFormA}{\SortB}}
\end{multline*}
where $\LogFormA$ denotes a formula over $\Sigma_\LogRelForm{\CAom}=\Sigma_\LogRelForm{\PAom}$ with free variables among $\LogSortedTerm{\LogVarA}{\CASort},\LogSortedTerm{\LogVarB}{\SortA},\LogSortedTerm{\LogVarC}{\SortA},\LogSortedTerm{\vec{\LogVarD}}{\vec{\SortB}}$ and which is of the shape $\LogRel{\LogVarC}\LogAnd\LogFormB$. For clarity, we write $\LogFormA\LogSubst{\LogTermA,\LogTermB,\LogTermC}$ instead of $\LogFormA\LogSubst{\LogTermA/\LogVarA,\LogTermB/\LogVarB,\LogTermC/\LogVarC}$.
\end{lem}
\proof
Recall that we use the bar recursor:
$$\LmTerm{\CALmbarrec}{\CALmbarrecSort{\LmInterpForm{\LogFormA}}{\LmSortBot}}$$
To simplify notations we define:
$$\LmConst{last}\Def\lambda\LmVarE.\CALmifz\,\CALmlen{\LmVarE}\,\CALmn{0}\left(\LmProj_1\,\CALmind{\LmVarE}{\CALmpred\,\CALmlen{\LmVarE}}\right)$$
so $\LmInterpAxiom{\CADCName}$ is:
\begin{multline*}\LmTerm{\LmInterpAxiom{\CADCName}=\\\lambda\LmVarA\LmVarB.\CALmbarrec\left(\lambda\LmVarE.\LmVarA\,\CALmlen{\LmVarE}\left(\LmConst{last}\,\LmVarE\right)\right)\LmVarB\,\CALmnil}{\CADCSort{\LmSortA}{\LogFormA}}\end{multline*}
Let now $\vec{\ModElemA}\in\ModMInterp{\vec{\SortB}}$ and write $\LogFormA'\equiv\LogFormA\LogSubst{\vec{\ModElemA}/\vec{\LogVarD}}$. We use again the notation $\LogFormA'\LogSubst{\LogTermA,\LogTermB,\LogTermC}$. Let:
$$\CatRCHomA\in\CatRC\left(\CatTerm,\CatInterpSort{\CALmnSort\LmSortTo\LmSortA\LmSortTo\left(\LmInterpForm{\LogFormA}\LmSortTo\LmSortBot\right)\LmSortTo\LmInterpForm{\LogFormA}}\CatPar\CatInterpSort{\CALmnSort}\right)\qquad\CatRCHomB\in\CatRC\left(\CatTerm,\CatInterpSort{\left(\CALmnSort\LmSortTo\LmInterpForm{\LogFormA}\right)\LmSortTo\LmSortBot}\CatPar\CatInterpSort{\CALmnSort}\right)$$
be such that:
$$\CatRCHomA\in\RealVal{\LogForallRel\LogSortedTerm{\LogVarA}{\CASort}\,\LogForallRel\LogSortedTerm{\LogVarB}{\SortA}\left(\forall\LogSortedTerm{\LogVarC}{\SortA}\,\neg\LogFormA'\LogSubst{\LogVarA,\LogVarB,\LogVarC}\LogImp\forall\LogSortedTerm{\LogVarA'}{\CASort}\,\LogFormA'\LogSubst{\LogVarA',\LogVarB,\LogVarB}\right)}\;\;\;\CatRCHomB\in\RealVal{\forall\LogSortedTerm{\LogVarE}{\CASort\to\SortA}\,\neg\LogForallRel\LogSortedTerm{\LogVarA}{\CASort}\,\LogFormA'\LogSubst{\LogVarA,\LogVarE\,\LogVarA,\LogVarE\left(\CALogS\,\LogVarA\right)}}$$
We then have to prove:
$$\CALmbarrec\left(\lambda\LmVarE.\CatRCHomA\,\CALmlen{\LmVarE}\left(\LmConst{last}\,\LmVarE\right)\right)\CatRCHomB\,\CALmnil\in\RealVal{\LogBot}$$
For conciseness we write:
$$\CatRCHomC=\CALmbarrec\left(\lambda\LmVarE.\CatRCHomA\,\CALmlen{\LmVarE}\left(\LmConst{last}\,\LmVarE\right)\right)\CatRCHomB\in\CatRC\left(\CatTerm,\CatInterpSort{\CASortList{\LmInterpForm{\LogFormA}}\LmSortTo\LmSortBot}\CatPar\CatInterpSort{\CALmnSort}\right)$$
so we must prove that $\CatRCHomC\,\CALmnil\in\RealVal{\LogBot}$. With our notations, for any $\CatRCHomD\in\CatRC\left(\CatTerm,\CatInterpSort{\CASortList{\LmInterpForm{\LogFormA}}}\CatPar\CatInterpSort{\CALmnSort}\right)$ we have:
\begin{align*}
\CatRCHomC\,\CatRCHomD&=\CALmbarrec\left(\lambda\LmVarE.\CatRCHomA\,\CALmlen{\LmVarE}\left(\LmConst{last}\,\LmVarE\right)\right)\CatRCHomB\,\CatRCHomD\\
&=\CatRCHomB\left(\CatRCHomD\CALmconcat\CatRCHomA\,\CALmlen{\CatRCHomD}\left(\LmConst{last}\,\CatRCHomD\right)\left(\lambda\LmVarA.\CALmbarrec\,\left(\lambda\LmVarE.\CatRCHomA\,\CALmlen{\LmVarE}\left(\LmConst{last}\,\LmVarE\right)\right)\,\CatRCHomB\left(\CatRCHomD\CALmextend\LmVarA\right)\right)\right)\\
&=\CatRCHomB\left(\CatRCHomD\CALmconcat\CatRCHomA\,\CALmlen{\CatRCHomD}\left(\LmConst{last}\,\CatRCHomD\right)\left(\lambda\LmVarA.\CatRCHomC\left(\CatRCHomD\CALmextend\LmVarA\right)\right)\right)
\end{align*}
The following iteration lemma (the proof of which is deferred to the end of the section) is the heart of the adequacy:
\begin{lem}
Let $\ModElemB_0,\ldots,\ModElemB_n\in\ModMInterp{\SortA}$ and $\CatRCHomE_0,\ldots,\CatRCHomE_{n-1}\in\CatRC\left(\CatTerm,\CatInterpSort{\LmInterpForm{\LogFormA}}\CatPar\CatInterpSort{\CALmnSort}\right)$ be such that:
$$\ModElemB_0=\ModMInterp{\LogSortedTerm{\CALogZ}{\SortA}}\qquad\forall 0\leq i<n,\ \CatRCHomE_i\in\RealVal{\LogFormA'\LogSubst{i,\ModElemB_i,\ModElemB_{i+1}}}\qquad\CatRCHomC\left(\CatRCHomE_0\CALmextend\ldots\CALmextend\CatRCHomE_{n-1}\right)\notin\RealVal{\LogBot}$$
where $\LogSortedTerm{\CALogZ}{\SortA}$ is the term from section~\ref{PeanoTheory}. There exists $\ModElemB_{n+1}\in\ModMInterp{\SortA}$ and $\CatRCHomE_n\in\CatRC\left(\CatTerm,\CatInterpSort{\LmInterpForm{\LogFormA}}\CatPar\CatInterpSort{\CALmnSort}\right)$ such that:
$$\CatRCHomE_n\in\RealVal{\LogFormA'\LogSubst{n,\ModElemB_n,\ModElemB_{n+1}}}\qquad\CatRCHomC\left(\CatRCHomE_0\CALmextend\ldots\CALmextend\CatRCHomE_{n-1}\CALmextend\CatRCHomE_n\right)\notin\RealVal{\LogBot}$$\smallskip
\end{lem}

\noindent In order to prove $\CatRCHomC\,\CALmnil\in\RealVal{\LogBot}$, we use a reductio-ad-absurdum and suppose $\CatRCHomC\,\CALmnil\notin\RealVal{\LogBot}$. By iterating the lemma, we build sequences $\left(\ModElemB_n\right)_{n\in\mathbb{N}}$ in $\ModMInterp{\SortA}$ and $\left(\CatRCHomE_n\right)_{n\in\mathbb{N}}$ in $\CatRC\left(\CatTerm,\CatInterpSort{\LmInterpForm{\LogFormB}}\CatPar\CatInterpSort{\CALmnSort}\right)$ such that:
$$\ModElemB_0=\ModMInterp{\LogSortedTerm{\CALogZ}{\SortA}}\qquad\forall n\in\mathbb{N},\ \CatRCHomE_n\in\RealVal{\LogFormA'\LogSubst{n,\ModElemB_n,\ModElemB_{n+1}}}\text{ and }\CatRCHomC\left(\CatRCHomE_0\CALmextend\ldots\CALmextend\CatRCHomE_{n-1}\right)\notin\RealVal{\LogBot}$$
Since $\ModM$ satisfies $\CADCUnrelName$, we can build $\ModElemB\in\ModMInterp{\left(\CASort\SortTo\SortA\right)}$ such that for every $n\in\mathbb{N}$, $\ModElemB\,n=\ModElemB_n$, and using the second assumption on $\CatRC$ (definition~\ref{sequence}) we also build $\CatRCHomE\in\CatRC\left(\CatTerm,\CatInterpSort{\CALmnSort\LmSortTo\LmInterpForm{\LogFormA}}\CatPar\CatInterpSort{\CALmnSort}\right)$ such that for every $n\in\mathbb{N}$, $\CatRCHomE\,\CALmn{n}=\CatRCHomE_n$. We now prove that:
$$\CatRCHomE\in\RealVal{\LogForallRel\LogSortedTerm{\LogVarA}{\CASort}\,\LogFormA'\LogSubst{\LogVarA,\ModElemB\,\LogVarA,\ModElemB\left(\CALogS\,\LogVarA\right)}}$$
Since $\RealVal{\LogRel{\ModElemC}}=\emptyset$ if $\ModElemC\notin\mathbb{N}$, it is sufficient to prove that for $n\in\mathbb{N}$, $\CatRCHomE\,\CALmn{n}\in\RealVal{\LogFormA'\LogSubst{n,\ModElemB\,n,\ModElemB\left(n+1\right)}}$, but since $\CatRCHomE\,\CALmn{n}=\CatRCHomE_n$, $\ModElemB\,n=\ModElemB_n$ and $\ModElemB\left(n+1\right)=\ModElemB_{n+1}$, this is immediate. Now, since:
$$\CatRCHomB\in\RealVal{\forall\LogSortedTerm{\LogVarE}{\CASort\to\SortA}\,\neg\LogForallRel\LogSortedTerm{\LogVarA}{\CASort}\,\LogFormA'\LogSubst{\LogVarA,\LogVarE\,\LogVarA,\LogVarE\left(\CALogS\,\LogVarA\right)}}\subseteq\RealVal{\neg\LogForallRel\LogSortedTerm{\LogVarA}{\CASort}\,\LogFormA'\LogSubst{\LogVarA,\ModElemB\,\LogVarA,\ModElemB\left(\CALogS\,\LogVarA\right)}}$$
we get $\CatRCHomB\,\CatRCHomE\in\RealVal{\LogBot}$, so $\mu\kappa.\CatRCHomB\,\CatRCHomE\in\RealBot$. We use now the first assumption on $\CatRC$ (definition~\ref{continuity}), so there is some $m\in\mathbb{N}$ such that any morphism $\CatRCHomE'\in\CatRC\left(\CatTerm,\CatInterpSort{\CALmnSort\LmSortTo\LmInterpForm{\LogFormB}}\CatPar\CatInterpSort{\CALmnSort}\right)$ which satisfies:
$$\forall m'<m,\,\CatRCHomE'\,\CALmn{m'}=\CatRCHomE\,\CALmn{m'}$$
is such that $\mu\kappa.\CatRCHomB\,\CatRCHomE'=\mu\kappa.\CatRCHomB\,\CatRCHomE\in\RealBot$. If we write $\CatRCHomD=\CatRCHomE_0\CALmextend\ldots\CALmextend\CatRCHomE_{m-1}$, this is verified in particular for:
$$\CatRCHomD\CALmconcat\CatRCHomA\,\CALmlen{\CatRCHomD}\left(\LmConst{last}\,\CatRCHomD\right)\left(\lambda\LmVarA.\CatRCHomC\left(\CatRCHomD\CALmextend\LmVarA\right)\right)$$
so we get:
$$\mu\kappa.\CatRCHomB\left(\CatRCHomD\CALmconcat\CatRCHomA\,\CALmlen{\CatRCHomD}\left(\LmConst{last}\,\CatRCHomD\right)\left(\lambda\LmVarA.\CatRCHomC\left(\CatRCHomD\CALmextend\LmVarA\right)\right)\right)\in\RealBot$$
and finally:
$$\CatRCHomC\left(\CatRCHomE_0\CALmextend\ldots\CALmextend\CatRCHomE_{m-1}\right)=\CatRCHomC\,\CatRCHomD=\CatRCHomB\left(\CatRCHomD\CALmconcat\CatRCHomA\,\CALmlen{\CatRCHomD}\left(\LmConst{last}\,\CatRCHomD\right)\left(\lambda\LmVarA.\CatRCHomC\left(\CatRCHomD\CALmextend\LmVarA\right)\right)\right)\in\RealVal{\LogBot}$$
from which we get our contradiction.
\qed
Here is the proof of the iteration lemma:
\proof
Write $\CatRCHomD=\CatRCHomE_0\CALmextend\ldots\CALmextend\CatRCHomE_{n-1}$. We have:
$$\CatRCHomB\left(\CatRCHomD\CALmconcat\CatRCHomA\,\CALmlen{\CatRCHomD}\left(\LmConst{last}\,\CatRCHomD\right)\left(\lambda\LmVarA.\CatRCHomC\left(\CatRCHomD\CALmextend\LmVarA\right)\right)\right)=\CatRCHomC\,\CatRCHomD=\CatRCHomC\left(\CatRCHomE_0\CALmextend\ldots\CALmextend\CatRCHomE_{n-1}\right)\notin\RealVal{\LogBot}$$
Using $\ModMInterp{\CALogk}$, $\ModMInterp{\CALogs}$ and $\ModMInterp{\CALogrec}$ we can build some $\ModElemB\in\ModMInterp{\left(\CASort\SortTo\SortA\right)}$ such that for any $0\leq i\leq n$, $\ModElemB\,i=\ModElemB_i$, and for any $i>n$, $\ModElemB\,i=\ModElemB_n$. Since $\CatRCHomB\in\RealVal{\forall\LogSortedTerm{\LogVarE}{\CASort\SortTo\SortA}\,\neg\LogForallRel\LogSortedTerm{\LogVarA}{\CASort}\,\LogFormA'\LogSubst{\LogVarA,\LogVarE\,\LogVarA,\LogVarE\left(\CALogS\,\LogVarA\right)}}$, we have in particular:
$$\CatRCHomB\in\RealVal{\neg\LogForallRel\LogSortedTerm{\LogVarA}{\CASort}\,\LogFormA'\LogSubst{\LogVarA,\ModElemB\,\LogVarA,\ModElemB\left(\CALogS\,\LogVarA\right)}}$$
therefore:
$$\CatRCHomD\CALmconcat\CatRCHomA\,\CALmlen{\CatRCHomD}\left(\LmConst{last}\,\CatRCHomD\right)\left(\lambda\LmVarA.\CatRCHomC\left(\CatRCHomD\CALmextend\LmVarA\right)\right)\notin\RealVal{\LogForallRel\LogSortedTerm{\LogVarA}{\CASort}\,\LogFormA'\LogSubst{\LogVarA,\ModElemB\,\LogVarA,\ModElemB\left(\CALogS\,\LogVarA\right)}}$$
Since $\RealVal{\LogRel{\ModElemC}}=\emptyset$ if $\ModElemC\notin\mathbb{N}$, there must be some $i\in\mathbb{N}$ such that:
$$\left(\CatRCHomD\CALmconcat\CatRCHomA\,\CALmlen{\CatRCHomD}\left(\LmConst{last}\,\CatRCHomD\right)\left(\lambda\LmVarA.\CatRCHomC\left(\CatRCHomD\CALmextend\LmVarA\right)\right)\right)\CALmn{i}\notin\RealVal{\LogFormA'\LogSubst{i,\ModElemB_i,\ModElemB_{i+1}}}$$
If $i<n$ then we get:
$$\CALmind{\CatRCHomD}{\CALmn{i}}=\CatRCHomE_i\notin\RealVal{\LogFormA'\LogSubst{i,\ModElemB_i,\ModElemB_{i+1}}}$$
which contradicts the hypothesis of the lemma. Therefore, $i\geq n$ and:
$$\CatRCHomA\,\CALmlen{\CatRCHomD}\left(\LmConst{last}\,\CatRCHomD\right)\left(\lambda\LmVarA.\CatRCHomC\left(\CatRCHomD\CALmextend\LmVarA\right)\right)\notin\RealVal{\LogFormA'\LogSubst{i,\ModElemB_i,\ModElemB_{i+1}}}=\RealVal{\LogFormA'\LogSubst{i,\ModElemB_n,\ModElemB_n}}$$
First, since $\CatRCHomD=\CatRCHomE_0\CALmextend\ldots\CALmextend\CatRCHomE_{n-1}$ we have $\CALmlen{\CatRCHomD}=\CALmn{n}\in\RealVal{\LogRel{n}}$, and therefore:
$$\CatRCHomA\,\CALmlen{\CatRCHomD}\in\RealVal{\LogForallRel\LogSortedTerm{\LogVarB}{\SortA}\left(\forall\LogSortedTerm{\LogVarC}{\SortA}\,\neg\LogFormA'\LogSubst{n,\LogVarB,\LogVarC}\LogImp\forall\LogSortedTerm{\LogVarA'}{\CASort}\,\LogFormA'\LogSubst{\LogVarA',\LogVarB,\LogVarB}\right)}$$
We prove now that $\LmConst{last}\,\CatRCHomD\in\RealVal{\LogRel{\ModElemB_n}}$ by distinguishing cases:
\begin{itemize}
\item$n=0$: $\LmConst{last}\,\CatRCHomD=\CALmifz\,\CALmlen{\CatRCHomD}\,\CALmn{0}\left(\LmProj_1\,\CALmind{\CatRCHomD}{\CALmpred\,\CALmlen{\CatRCHomD}}\right)=\CALmn{0}\in\RealVal{\LogRel{\ModMInterp{\LogSortedTerm{\CALogZ}{\SortA}}}}=\RealVal{\LogRel{\ModElemB_0}}$ (indeed, $\CALmn{0}\in\RealVal{\LogRel{\ModMInterp{\LogSortedTerm{\CALogZ}{\SortA}}}}$ by an easy induction on $\SortA$)
\item$n\neq0$: $\LmConst{last}\,\CatRCHomD=\CALmifz\,\CALmlen{\CatRCHomD}\,\CALmn{0}\left(\LmProj_1\,\CALmind{\CatRCHomD}{\CALmpred\,\CALmlen{\CatRCHomD}}\right)=\LmProj_1\,\CatRCHomE_{n-1}\in\RealVal{\LogRel{\ModElemB_n}}$ since:
$$\CatRCHomE_{n-1}\in\RealVal{\LogFormA'\LogSubst{n-1,\ModElemB_{n-1},\ModElemB_n}}=\RealVal{\LogRel{\ModElemB_n}\LogAnd\LogFormB\LogSubst{\vec{\ModElemA}/\vec{\LogVarD},n-1/\LogVarA,\ModElemB_{n-1}/\LogVarB,\ModElemB_n/\LogVarC}}$$
\end{itemize}
Therefore we have:
$$\CatRCHomA\,\CALmlen{\CatRCHomD}\left(\LmConst{last}\,\CatRCHomD\right)\in\RealVal{\forall\LogSortedTerm{\LogVarC}{\SortA}\,\neg\LogFormA'\LogSubst{n,\ModElemB_n,\LogVarC}\LogImp\forall\LogSortedTerm{\LogVarA'}{\CASort}\,\LogFormA'\LogSubst{\LogVarA',\ModElemB_n,\ModElemB_n}}$$
and so:
$$\lambda\LmVarA.\CatRCHomC\left(\CatRCHomD\CALmextend\LmVarA\right)\notin\RealVal{\forall\LogSortedTerm{\LogVarC}{\SortA}\,\neg\LogFormA'\LogSubst{n,\ModElemB_n,\LogVarC}}$$
which means that there exists some $\ModElemB_{n+1}\in\ModMInterp{\SortA}$ such that:
$$\lambda\LmVarA.\CatRCHomC\left(\CatRCHomD\CALmextend\LmVarA\right)\notin\RealVal{\neg\LogFormA'\LogSubst{n,\ModElemB_n,\ModElemB_{n+1}}}$$
and so there exists $\CatRCHomE_n\in\CatRC\left(\CatTerm,\CatInterpSort{\LmInterpForm{\LogFormA}}\CatPar\CatInterpSort{\CALmnSort}\right)$ such that $\CatRCHomE_n\in\RealVal{\LogFormA'\LogSubst{n,\ModElemB_n,\ModElemB_{n+1}}}$ and:
$$\CatRCHomC\left(\CatRCHomE_0\CALmextend\ldots\CALmextend\CatRCHomE_{n-1}\CALmextend\CatRCHomE_n\right)=\CatRCHomC\left(\CatRCHomD\CALmextend\CatRCHomE_n\right)=\left(\lambda\LmVarA.\CatRCHomC\left(\CatRCHomD\CALmextend\LmVarA\right)\right)\CatRCHomE_n\notin\RealVal{\LogBot}\eqno{\qEd}$$

\subsection{Extraction}
Since the Friedman translation is directly built in the realizability interpretation, the extraction result is an easy consequence of the definitions. We show that from any $\Pi^0_2$-formula provable in $\CAom$ we can extract a computable witness in $\CatRC$. Note that in the extraction lemma, the equality $\LogTermA=_\SortB\LogTermB$ is at any type:
\begin{lem}
From a proof of $\Sequent{}{\forall\LogSortedTerm{\LogVarA}{\SortA}\exists\LogSortedTerm{\LogVarB}{\CASort}\left(\LogTermA=_\SortB\LogTermB\right)}{}$ in $\CAom$, one can extract a $\lambda\mu$-term $\LmTerm{\LmTermA}{\LmSortA\LmSortTo\CALmnSort}$ such that for any $\ModElemA\in\ModMInterp{\SortA}$ and $\CatRCHomA\in\RealVal{\LogRel{\ModElemA}}$, there is some $n\in\mathbb{N}$ such that:
$$\ModMInterp{\left(\LogTermA\LogSubst{\ModElemA/\LogVarA,n/\LogVarB}\right)}=\ModMInterp{\left(\LogTermB\LogSubst{\ModElemA/\LogVarA,n/\LogVarB}\right)}\text{ and }\LmTermA\,\CatRCHomA=\CALmn{n}\text{ in }\CatRC$$
\end{lem}
\proof
Since $\exists$ and $=$ are just encodings, we actually have a proof of $\Sequent{}{\forall\LogSortedTerm{\LogVarA}{\SortA}\neg\forall\LogSortedTerm{\LogVarB}{\CASort}\neg\neg\left(\LogTermA\neq_\SortB\LogTermB\right)}{}$. First, we can easily turn it into a proof of $\Sequent{}{\forall\LogSortedTerm{\LogVarA}{\SortA}\neg\forall\LogSortedTerm{\LogVarB}{\CASort}\left(\LogTermA\neq_\SortB\LogTermB\right)}{}$, and we get by relativization a proof $\LogProofA$ in $\LogRelForm{\CAom}$ of $\Sequent{}{\LogForallRel\LogSortedTerm{\LogVarA}{\SortA}\neg\LogForallRel\LogSortedTerm{\LogVarB}{\CASort}\left(\LogTermA\neq_\SortB\LogTermB\right)}{}$. The adequacy lemma then tells us:
$$\LmInterpProof{\LogProofA}\in\RealVal{\LogForallRel\LogSortedTerm{\LogVarA}{\SortA}\neg\LogForallRel\LogSortedTerm{\LogVarB}{\CASort}\left(\LogTermA\neq_\SortB\LogTermB\right)}$$
If $\ModElemA\in\ModMInterp{\SortA}$ and $\CatRCHomA\in\RealVal{\LogRel{\ModElemA}}$, we get $\LmInterpProof{\LogProofA}\,\CatRCHomA\in\RealVal{\neg\LogForallRel\LogSortedTerm{\LogVarB}{\CASort}\left(\LogTermA\LogSubst{\ModElemA/\LogVarA}\neq_\SortB\LogTermB\LogSubst{\ModElemA/\LogVarA}\right)}$. Let fix now:
$$\RealBot=\SetSuch{\CALmn{n}\in\CatRC\left(\CatTerm,\CatInterpSort{\CALmnSort}\right)}{\ModMInterp{\left(\LogTermA\LogSubst{\ModElemA/\LogVarA,n/\LogVarB}\right)}=\ModMInterp{\left(\LogTermB\LogSubst{\ModElemA/\LogVarA,n/\LogVarB}\right)}}$$
We prove that $\lambda\LmVarA.\left[\kappa\right]\,\LmVarA\in\RealVal{\LogForallRel\LogSortedTerm{\LogVarB}{\CASort}\left(\LogTermA\LogSubst{\ModElemA/\LogVarA}\neq_\SortB\LogTermB\LogSubst{\ModElemA/\LogVarA}\right)}$. For that let $n\in\ModMInterp{\CASort}$, and let prove that $\left(\lambda\LmVarA.\left[\kappa\right]\,\LmVarA\right)\CALmn{n}=\left[\kappa\right]\,\CALmn{n}\in\RealVal{\LogTermA\LogSubst{\ModElemA/\LogVarA,n/\LogVarB}\neq\LogTermB\LogSubst{\ModElemA/\LogVarA,n/\LogVarB}}$. There are two cases:
\begin{itemize}
\item$\CALmn{n}\in\RealBot$: in that case, $\ModMInterp{\left(\LogTermA\LogSubst{\ModElemA/\LogVarA,n/\LogVarB}\right)}=\ModMInterp{\left(\LogTermB\LogSubst{\ModElemA/\LogVarA,n/\LogVarB}\right)}$, so:
$$\RealVal{\LogTermA\LogSubst{\ModElemA/\LogVarA,n/\LogVarB}\neq\LogTermB\LogSubst{\ModElemA/\LogVarA,n/\LogVarB}}=\RealVal{\LogBot}$$
and $\mu\kappa.\left[\kappa\right]\,\CALmn{n}=\CALmn{n}\in\RealBot$ so $\left[\kappa\right]\,\CALmn{n}\in\RealVal{\LogBot}$
\item$\CALmn{n}\notin\RealBot$: in that case, $\ModMInterp{\left(\LogTermA\LogSubst{\ModElemA/\LogVarA,n/\LogVarB}\right)}\neq\ModMInterp{\left(\LogTermB\LogSubst{\ModElemA/\LogVarA,n/\LogVarB}\right)}$, so:
$$\RealVal{\LogTermA\LogSubst{\ModElemA/\LogVarA,n/\LogVarB}\neq\LogTermB\LogSubst{\ModElemA/\LogVarA,n/\LogVarB}}=\CatRC\left(\CatTerm,\CatInterpSort{\LmSortBot}\CatPar\CatInterpSort{\CALmnSort}\right)$$
and therefore $\left[\kappa\right]\,\CALmn{n}\in\RealVal{\LogTermA\LogSubst{\ModElemA/\LogVarA,n/\LogVarB}\neq\LogTermB\LogSubst{\ModElemA/\LogVarA,n/\LogVarB}}$
\end{itemize}
We get $\LmInterpProof{\LogProofA}\,\CatRCHomA\,\left(\lambda\LmVarA.\left[\kappa\right]\,\LmVarA\right)\in\RealVal{\LogBot}$, so $\mu\kappa.\LmInterpProof{\LogProofA}\,\CatRCHomA\,\left(\lambda\LmVarA.\left[\kappa\right]\,\LmVarA\right)\in\RealBot$. This means that $\mu\kappa.\LmInterpProof{\LogProofA}\,\CatRCHomA\,\left(\lambda\LmVarA.\left[\kappa\right]\,\LmVarA\right)$ is some $\CALmn{n}\in\RealBot$ (so $\ModMInterp{\left(\LogTermA\LogSubst{\ModElemA/\LogVarA,n/\LogVarB}\right)}=\ModMInterp{\left(\LogTermB\LogSubst{\ModElemA/\LogVarA,n/\LogVarB}\right)}$). Finally, we have the claimed result with $\LmTermA\Def\lambda\LmVarD.\mu\kappa.\LmInterpProof{\LogProofA}\,\LmVarD\,\left(\lambda\LmVarA.\left[\kappa\right]\,\LmVarA\right)$.
\qed
Similarly to what was done in~\cite{BlotRibaBarRec}, we can define an operatonal semantics for $\mu$PCF and adapt the techniques of e.g.~\cite{AmadioCurien} to prove computational adequacy with respect to a non-degenerated $\CatRC$, i.e. if $\LmTerm{\LmTermA}{\CASort}$ in $\mu$PCF, then for any $n\in\mathbb{N}$:
$$\LmTermA\rightarrow^*\CALmn{n}\quad\Longleftrightarrow\quad\LmTermA=\CALmn{n}\text{ in }\CatRC$$
Using computational adequacy and the extraction lemma in the particular case of $\SortA=\CASort$, we get that for any $m\in\mathbb{N}$, there is some $n\in\mathbb{N}$ such that $\LmTermA\,\CALmn{m}\rightarrow^*\CALmn{n}$ and $\ModMInterp{\left(\LogTermA\LogSubst{m/\LogVarA,n/\LogVarB}\right)}=\ModMInterp{\left(\LogTermB\LogSubst{m/\LogVarA,n/\LogVarB}\right)}$.
\section*{Conclusion}
Realizability interpretation of classical logic with control operators is a recent field in which much of the contributions take place in Krivine's untyped setting. We defined here a realizability model in which proofs are interpreted in a category of continuations, which is the universal model of typed $\lambda\mu$-calculus. The duality between terms and contexts, which appears in Krivine's work as a duality between terms and stacks, is reflected here by the duality in a category of continuations $\CatRC$ between $\CatC$ and $\CatRC$.\par
The direct interpretations of classical logic can be seen as a way to avoid G\"odel's negative translation on formulas, by using programming languages with an operational semantics which corresponds to that of their CPS-translation to $\lambda$-calculus. Similarly, the choice of having a free $\mu$-variable $\kappa$ in our realizers may be seen as a way to avoid Friedman's $A$-translation. Indeed, in Friedman's original work, the translation is obtained by replacing each basic predicate $\LogPredA$ with the disjunction $\LogPredA\vee A$. In classical sequent calculus, the right-hand context is meant to be interpreted as a disjunction, so adding a fixed $\mu$-variable to this context corresponds to applying Friedman's translation on programs instead of proofs. In many implementations of Friedman's trick, the formula $A$ is not added as a disjunction to every predicate, but instead the $\LogBot$ formula is replaced everywhere with $A$. Our implementation of Friedman's translation however allows simpler interpretations of the realizers in some models. Indeed, the continuation passing hidden behind the usual implementations is abstracted with the help of $\lambda\mu$-calculus constructs. Rather than adding a continuation on top of every interpretation of $\LogBot$, we only add one continuation on top of the whole realizer. This simplicity of the interpretation requires some properties on the model, and in particular it rules out Scott domains. However, all these properties are satisfied in unbracketed games, and it would be interesting to find other models satisfying these. The natural candidates for this are bistable biorders~\cite{LairdBistable} and sequential algorithms~\cite{BerryCurienSequential}, but these are not the only ones. Coherent spaces may also satisfy these properties if we choose carefully the object of natural numbers.\par
The $\mu$-variable $\kappa$ is also exploited in the definition of the realizability interpretation, to parameterize the orthogonality relation between realizers and counter-realizers. However, contrary to usual interpretations of classical logic, we consider positive predicates, for which no counter-realizers are defined. Through a suitable restriction on proofs we preserve the adequacy of the interpretation. Roughly, it amounts to forbid classical reasoning on the positive formulas, while keeping it in the general case. This introduction of positive predicates in a classical, negative setting is mainly motivated by the decomposition of the relativized universal quantifier into a uniform quantifier and a relativization predicate, the relativization predicate being a positive one.\par
We proved that the usual terms of G\"odel's system T realize the axioms of Peano arithmetic in our direct setting, without CPS-translation. This is not a surprise since all these axioms do imply their negative translation intuitionistically. However, when it comes to the axiom of choice, this is not as easy and we prove, as was done in~\cite{BlotRibaBarRec}, that the bar recursion operator realizes it in a direct interpretation. Interestingly, this bar recursor also gives computational content to the double-negation shift principle in an intuitionistic setting, and that same principle allows to derive intuitionistically the negative translation of the axiom of choice from the axiom of choice.\par
Finally, we validate our model by proving an extraction result for $\Pi^0_2$ formulas, relying once again on the $\mu$-variable $\kappa$ which allows the orthogonality relation to be a parameter of the model.
\subsection*{Acknowledgments}I thank the reviewers for their careful reading of my work, which led to constructive comments and suggestions.

\end{document}